\documentstyle[draft, psfig]{mn}
\newcommand{\mnref}[1]{\hangindent=0.5in \hangafter=1 #1 \par}

\newcommand{\mn}{MNRAS}

\newcommand{\apj}{ApJ}
\newcommand{\apjs}{ApJS}
\newcommand{\aaa}{A\&A}
\newcommand{\aas}{A\&AS}
\newcommand{\pasp}{PASP}

\newcommand{\Lsolar}{\mbox{\,$\rm L_{\odot}$}}
\renewcommand{\vec}[1]{\underline{\mbox{\bf{\rm #1}}}}

\title[Infrared Helium-Hydrogen Line Ratios]
{Infrared Helium-Hydrogen Line Ratios as a Measure of Stellar 
Effective Temperature}
\author[S.L. Lumsden, P.J. Puxley \& M.G. Hoare]
{S.L. Lumsden$^{1,3}$, P.J. Puxley$^{2}$ and M.G. Hoare$^{1}$\\
{}$^1$ {\em Department of Physics and Astronomy, University of Leeds, 
Leeds LS2 9JT, UK -- sll@ast.leeds.ac.uk, mgh@ast.leeds.ac.uk}\\
{}$^2$ {\em Gemini Observatory, 670 N. A'ohuku Place, Hilo, Hawaii 96720, USA -- ppuxley@gemini.edu}\\
{}$^3$ {\em Anglo-Australian Observatory, PO Box 296, Epping, NSW 1710,
Australia}\\
}

\begin{document}

\maketitle

\begin{abstract}
We have observed a large sample of compact planetary nebulae in the
near-infrared to determine how the 2$^1$P--2$^1$S HeI line at 2.058$\mu$m
varies as a function of stellar effective temperature, $T_{eff}$.  The ratio of
this line with HI Br$\gamma$ at 2.166$\mu$m has often been used as a measure of
the highest $T_{eff}$ present in a stellar cluster, and hence on whether there
is a cut-off in the stellar initial mass function at high masses.  However,
recent photoionisation modelling has revealed that the behaviour of this line
is more complex than previously anticipated.  Our work shows that in most
aspects the photoionisation models are correct.  In particular, we confirm the
weakening of the 2$^1$P--2$^1$S as $T_{eff}$ increases beyond 40000K.  However,
in many cases the model underpredicts the observed ratio when we consider the
detailed physical conditions in the individual planetary nebulae.  Furthermore,
there is evidence that there is still significant 2$^1$P--2$^1$S HeI line
emission even in the planetary nebulae with very hot ($T_{eff}>$100000K)
central stars.  It is clear from our work that this ratio cannot be considered
as a reliable measure of effective temperature on its own.
\end{abstract}

\begin{keywords}{infrared: ISM: lines and bands -- ISM: planetary nebulae:
general}
\end{keywords}

\section{Introduction}
One of the fundamental parameters characterising any stellar population is its
mass function, and in particular its {\em initial} mass function (IMF), 
defined simply as the number of stars per unit mass that are actually formed
in any system.  Determining the form of the IMF is vital to our understanding
of star formation in our own and other star-forming galaxies, since it allows
us to model the chemical and physical evolution of these systems.  

Despite ongoing debate (eg.\ Leitherer 1998) it is not clear whether the IMF
varies widely between galaxies or even within different star formation regions
of our own galaxy.  Since starburst galaxies are generally dusty, IR indicators
of the IMF are preferable since they provide considerable improvement in
sensitivity since the extinction is lower than in the optical, and perhaps as
important, any error in the derived extinction value has a correspondingly
smaller effect on any observed line ratio when the lines are close in
wavelength.  Indeed any method that relies on absolute flux values or widely
separated line pairs is likely to lead to significant errors in the derived
results when the extinction is large.

There are limited methods for directly measuring the current IMF from stellar
features at these wavelengths, so indirect methods are required.  Most work has
relied on inferring the nature of the stars present by comparing the ionisation
structure of the enveloping nebular gas with the predictions of detailed
photoionisation models.  There has been extensive use of strong mid and far-IR
forbidden line data in modelling HII regions in this fashion.  However, such
data is difficult to obtain for a large sample of galaxies, the atomic data is
typically less certain for forbidden transitions and extinction corrections are
complicated by the presence of the silicate absorption feature at 9.7$\mu$m and
our poor understanding of the extinction law at longer wavelengths.
Consequently there has not been widespread application of that method.

Given the problems inherent with the longer wavelength forbidden line data, it
was natural that attention should turn instead to the near infrared.  The
brightest lines in general below 2.5$\mu$m are permitted HI and HeI lines.  In
principle, the comparison of HeI and HI lines tests the relative volumes
occupied by He$^0$ and He$^+$ relative to H$^+$, and hence places constraints
on the form of the emergent UV radiation field shortwards of the HI Lyman limit
given the higher ionisation potential of helium.  This is certainly true for
any lines that arise solely from recombination.

The brightest HeI line longward of 1.1$\mu$m is the 2.058$\mu$m HeI
2$^1$P--2$^1$S transition.  This has led to considerable attention being given
to the use of the ratio of HeI 2$^1$P--2$^1$S with HI Br$\gamma$ at 2.166$\mu$m
(eg.\ Doyon, Puxley \& Joseph 1992 and references therein).  However, the
2.058$\mu$m HeI 2$^1$P--2$^1$S transition is determined by the population in
the 2$^1$P state.  This is largely driven by resonance scattering of the
584\AA\ HeI 2$^1$P--1$^1$S transition rather than a simple recombination
cascade.  These resonance photons can be absorbed by dust or hydrogen reducing
the population in the 2$^1$P state.  They also depend implicitly on the neutral
helium fraction present since they are a resonance from the ground state.  In
addition, microscopic velocity structure can lead to a sufficient shift in the
wavelength emitted so that the photon cannot be reabsorbed by helium at all.

Initially it was thought that these secondary effects were relatively small and
easily calculable.  However, Shields (1993) showed from a full photoionisation
treatment that the strength of the 2$^1$P--2$^1$S line was much more dependent
than previously thought on the neutral helium fraction, and indeed predicted
that as the stellar effective temperature, T$_{eff}$, increases much beyond
40000K that this line actually gets weaker as most helium in the HII region is
turned into He$^+$.  More accurate atomic data and photoionisation models have
confirmed this general picture (Ferland 1999).  Observational confirmation of
this prediction was evident from the extragalactic work of Lumsden, Puxley \&
Doherty (1994) and Doherty et al.\ (1995), and the study of planetary
nebulae by DePoy \& Shields (1994).  However, the former papers deal with
complex clusters of OB stars for which we have no accurate knowledge of the
effective temperature.  The latter used slitless spectroscopy of extended
planetary nebulae, resulting in very low resolution spectra in which only the
brightest lines were visible.  Therefore, although we know the trend in
the observed data appears to be in accord with the photoionisation models,
there have been no attempts to really constrain the model results in any
detail. 

Crucially, it was still unclear whether or not the HeI 2$^1$P--2$^1$S to HI
Br$\gamma$ ratio can be used as a tracer of the high mass end of the IMF.  We
therefore decided to carry out observations of objects whose stellar
temperature could be defined more accurately than is the case for typical
obscured compact HII regions.  We observed a sample of 23 planetary nebulae
(PN) to derive as full a set as possible of the observable line parameters for
both the HeI 2$^1$P--2$^1$S line and HI Br$\gamma$.  PN are ideal for this test
since they are predominantly excited by only a single central star (unlike HII
regions), and a sample covering a wider range of effective temperature is much
easier to construct.

\section{The Planetary Nebula Sample}
Our sample of PN was chosen (i) to be compact on the sky, (ii) to have good
quality line measurements in the ESO planetary nebula catalogue (Acker et al.\
1992) and (iii) to have a spread in RA to make observation easier.  Where
possible we chose those objects with predicted Stoy temperatures less than
60000K (Kaler \& Jacoby 1991) in preference to hotter objects, but the final
sample contains a selection of PN with predicted temperatures well above this
value as well.  The sample is therefore largely heterogeneous, since our
selection criteria do not really distinguish, for example, between nearby low
temperature and luminosity PN, and more distant high excitation and luminosity
PN.  In addition, we did not impose any morphological selection on the sample,
so there are objects ranging from spherical to bipolar present.  For most of
our sources there are no published Peimbert types, and even those with
published types are often contradictory.  It is likely therefore that at least
some of our objects are Type I PN, with the consequence that they will probably
have enhanced helium abundances (eg Peimbert \& Torres-Peimbert 1983).  Since
we cannot distinguish these PN accurately however, we do not split our sample
into Peimbert types.  This is not a particular problem for this project
however, since all we require is a set of PN with a reasonable spread in
effective temperature, which are bright enough that we can derive line ratio
measurements easily.  We can test for abundance effects by examining the
behaviour of the HeI 2$^1$P--2$^1$S line with other helium recombination lines.
The only real problem with the inclusion of PN which are not spherical is that
the photoionisation modelling reported in Section 5 will not be an ideal match.
We do not expect that this will significantly effect the conclusions we reach
however.

In addition to our own near infrared spectra we also sought high quality
optical data in order to provide more accurate constraints on the
photoionisation models.  For the most part this comes from a search of the
literature, and from the electronic catalogue of Kaler, Shaw \& Browning
(1997).  For some of the PN in our sample we obtained low resolution
4300--7400\AA\ CCD spectra with the red arm of ISIS on the William Herschel
Telescope on the night of 30 July 1996 in poor weather.  This allowed us to
check the reliability of the published data in the literature as well.  In
particular, we have chosen not to rely on the fluxes from the Acker et al.\
catalogue in our work beyond the aspect of sample selection wherever possible.
The bright lines are often saturated, and the faint HeI lines we require in our
analysis are often undetected making the catalogue largely unsuitable if there
is an alternative.  We also excluded a large amount of the oldest (pre-1980)
data since there were clear indications that it disagreed completely with our
own CCD data where there were objects in common.  Where necessary we placed all
data onto a common extinction corrected intensity scale using the extinction
law given in Seaton (1979).  Since all of the objects (with the exception of
CRL~618) have low--moderate optical extinction, we also made the implicit
assumption that $A_K=0$ in our analyses.  Since there are no direct means of
determining the extinction from the near infrared data alone, we felt this was
the safest option.  Assuming $A_K=0 $leads to errors in the observed HeI
2$^1$P--2$^1$S to HI Br$\gamma$ ratio of $0.1A_K$ or $\sim0.008A_V$, which are
essentially negligible for this sample.

The complete sample is listed in Table 1.  
We give the optical references we
have used there.  Also given there are the averaged fluxes of [OIII] 5007\AA,
HeI 6678\AA\ and HeII 4686\AA\ relative to H$\beta$ taken from those
references.  We attribute 10\% errors to the relatively weak 6678\AA\
line, and 5\% errors to the others, to account for potential differences
in optical and infrared beamsizes when making comparisons of the actual
lines.  These errors are not included in the analysis of the optical
data on its own, since there is no aperture effect to correct for.
From these data we have calculated the Stoy temperatures (from the
[OIII]/H$\beta$ ratio and equation 1  of Kaler \& Jacoby 1991), electron
densities (from the [SII] 6717/6731\AA\ line ratio) and electron temperature
(from either the [OIII] 4363/5007\AA\ ratio or the [NII] 5755/6548\AA\ ratio
where these exist).  These are also given in the Table.  Although the density
derived from the [SII] doublet is potentially unrepresentative of the density
in the He$^+$ region of the nebula, it is the only indicator that is available
for the majority of our sample, and we therefore adopt it throughout.

Unfortunately, as is clear from Table 1, we are lacking in PN with Stoy
temperatures between 40000 and 60000K.  This was not clear at the time of
observation since the predicted values from the ESO Planetary Nebula Catalogue
(Acker et al.) did not show this effect.  It is only with the adoption of
higher quality, and more strictly global, flux values from the literature that
it became obvious.  

We have included two objects whose identification as PN is uncertain, M~1-78
and CRL~618.  The first of these is sometimes listed as a possible HII region,
largely on grounds of excitation and extinction.  However, there are other
confirmed PN with similar characteristics so we have included it in our sample
anyway.  CRL~618 is more strictly a proto-planetary nebula, since the central
source is still obscured, and there is a clear bipolar outflow.  Since it
will eventually become a PN however, we include it in our sample.  It should
be noted that the optical data for this object is generally from the bipolar
lobes, so may not reflect the actual excitation of the core.

\section{Observations}
We obtained separate low resolution K band spectra ($R\sim400$) spanning
$\sim$1.9--2.5$\mu$m, and high resolution echelle spectra around the 2.16$\mu$m
HI Br$\gamma$ and 2.058$\mu$m HeI 2$^1$P--2$^1$S lines ($R\sim20000$) using
CGS4, the facility infrared spectrograph on the United Kingdom Infrared
Telescope.  The data were mainly obtained on the nights of 8 and 9 January
1997, but additional data were also obtained on 15 July 1995, 27 and 28
September 1995, and 4, 5 and 6 July 1997.  In all cases we used a two pixel
wide slit for the grating spectra, corresponding to 3 arcseconds on the sky.
We used a one pixel wide slit for the echelle data to preserve the highest
possible velocity resolution.  Proper sampling of the spectrum is obtained by
stepping the array a fraction of a pixel between `sub-exposures' which are
later combined in software to provide the final spectrum.  Anamorphic
distortion within the spectrograph makes the width of the slit with the echelle
grating close to 1 arcsecond.

The separate echelle data of the two lines were obtained primarily for two
reasons: first, to derive velocity profiles for the PN to test for the effects
of turbulence on the HeI 2$^1$P--2$^1$S to HI Br$\gamma$ ratio as expected from
theory; second, as a means of correcting the low resolution fluxes obtained
around 2.058$\mu$m where there is a sequence of deep narrow atmospheric
absorption lines due to CO$_2$.  However, the echelle data alone cannot be used
to derive reliable estimates of the ratio of the HI Br$\gamma$ and HeI
2$^1$P--2$^1$S lines because we used the narrowest possible slit to maximize
the spectral resolution.  This means that any small inaccuracy in positioning
the telescope can have a significant impact on the derived line ratio given the
compact nature of the PN.  It is clearly preferable to derive ratios from lines
observed in the same spectrum, so wherever possible we use the low resolution
data for this.

The grating spectra were observed in a standard manner, nodding the telescope
to keep the source always on the spectrograph slit.  The beam separation was
between 20 and 30 arcseconds for all data.  The data were reduced by dividing
by an internal flat field, and coadding all the nod pairs.  The data obtained
during January 1997 suffered from a problem with the slit mechanism which was
tilted markedly relative to the axes of the array.  These data were
straightened to make the wavelength perpendicular to the columns of the
array. The spectra were then extracted, and the negative beam subtracted from
the initial positive one.  This procedure ensures that residual sky
contamination is minimal.  The low resolution data were wavelength calibrated
in the standard fashion using an argon arc lamp.

Flux calibration of the low resolution data was obtained using observations of
a set of A-type stars from the Bright Star Catalogue.  We adopted $V-K$
corrections from Johnson (1966) and used the $V$ magnitudes from the Bright
Star Catalogue itself.  This technique is reliable in providing approximate
absolute fluxes (accurate to $\sim$30\%) and very good ($<<5$\%) relative
fluxes within each spectra which are sufficient for our requirements.  The
Br$\gamma$ absorption feature present in the A-type standards was interpolated
over using a Lorenzian line profile.

The echelle data were largely obtained in a similar manner, with the exception
of the 1997 data.  The tilted slit in January 1997 meant that more than one
echelle order was present in the final image, so that only the central 20 rows
were actually useful for our observations.  Therefore we adopted the same
strategy as used for mapping extended HII regions described in Lumsden \& Hoare
(1996), and acquired only one `sky' frame for each set of up to 5 `object'
frames.  Given the lack of strong sky lines near the emission lines from the PN
this procedure provides a reasonable sky subtracted data-set.  We used the same
technique in July 1997 for simplicity.

For all echelle data there is also a residual distortion in the wavelength
calibration due to the fact that the projection of the slit onto the array is
actually curved.  We used OH night sky lines to `straighten' the spectra for
this and the more extreme distortion caused by the tilted slit.  We also
used the OH lines to provide wavelength calibration for the Br$\gamma$ data.
The HeI data were calibrated using an internal argon arc lamp.

We also acquired spectra of standard stars for the echelle to map the
atmospheric absorption features.  We used A type standards for the 2.058$\mu$m
data and K type for the 2.166$\mu$m data, since both are relatively featureless
in these wavelength ranges at high resolution.  In practice there are only weak
atmospheric absorption features around 2.166$\mu$m so there is little problem
with this data.  However, around 2.058$\mu$m the wavelength region is strongly
attenuated by a sequence of deep narrow (almost unresolved) atmospheric
absorption features due to CO$_2$ in the upper atmosphere.  Although the
echelle data can be corrected for this absorption using the standard star
observations it reveals a potential problem with the accuracy of the flux
calibration in this narrow wavelength range for the low resolution data.  The
basic argument is simple: given a narrow emission line as seen in these PN, the
actual data as observed through the telescope can range from almost unaffected
by the atmospheric absorption to almost completely absorbed.  However, the
spectra of the standard stars as measured at low resolution average over these
absorption features, so that the correction they give to the low resolution
data is wrong in most cases.

We used the basic procedures outlined by Doherty et al.\ (1994) to correct the
observed low resolution HeI 2$^1$P--2$^1$S fluxes for this atmospheric
attenuation.  However, rather than using a model of the atmospheric absorption
as they did we used the actual observed echelle standard star spectra.  First
we correct the echelle spectra for changes in the observed radial velocity due
to the time of observation.  The correction is then derived by smoothing the
object and standard echelle spectra to the resolution of the grating data.  The
smoothed standard star spectrum is used to correct the smoothed PN echelle
spectrum, and the flux of the HeI line measured.  The original PN echelle
spectrum is corrected by the original standard star spectrum as well, and the
flux measured of the full resolution data.  The ratio of the measured fluxes
from the full resolution data with the smoothed data then give the required
correction factor.  The actual values are given in Table 2.  It is worth noting
here that no such corrections were applied to the DePoy and Shields (1994)
data, potentially leading to errors in their results of up to 30\% from this
factor alone.

\section{The Observational Data}
\subsection{The Spectra}
The spectra acquired with the echelle grating are shown in Figures 1 and 2.
The full set of low resolution spectra will be published separately in a paper
that discusses the other spectral features present.  However, a representative
sample showing the trend with effective temperature are shown in Figure 3.  In
all cases we have summed up the data over the whole object 
in forming these spectra.

The observed HI Br$\gamma$ fluxes are listed in Table 2, along with the ratios
of the strongest helium lines with Br$\gamma$.  The tabulated values of HeI
2$^1$P--2$^1$S against HI Br$\gamma$ are all corrected for atmospheric
absorption using the method given in Section 3.  All ratios are taken from the
low resolution data with the exception of the HeI 7$^{1,3}$G--4$^{1,3}$F
transition at 2.16475$\mu$m, which was measured from the echelle spectra.  This
is the strongest of the HeI satellite lines to Br$\gamma$, and is visible in
most of our echelle spectra.  The ratio given for that line is against the
observed Br$\gamma$ flux from our echelle spectra.  This line can also be
compared with the other HeI lines, since they would all arise from the same
region of the PN.  It should not however be compared with the HeII line
strengths, since they will come from a different region in the PN, and the
possible differences in slit position and size again may come into play.  Also
visible in some of the Br$\gamma$ echelle spectra are the other HeI transitions
of 7$^{1}$F--4$^{1}$D at 2.16229$\mu$m and 7$^{3}$F--4$^{3}$D at 2.16137$\mu$m,
and the HeII 14--8 transition at 2.1653$\mu$m.  It should be noted that the low
resolution Br$\gamma$ data have not been corrected for any underlying HeII
emission (the bright satellite HeI lines are sufficiently well resolved not to
contribute to the measured fluxes).  Examination of the echelle spectra shows
that these lines are insignificant for most of the sample, and even for the
higher excitation PN contribute at most 5\% of the flux.  The one exception is
Hu~1-2, which shows evidence for a bipolar flow and enhanced helium abundance
indicating it may be a Type I PN.  For this case, the HeII lines may contribute
up to 10\% of the observed low resolution Br$\gamma$ flux.

We also used the echelle data to derive line widths for the HI Br$\gamma$ and
HeI 2$^1$P--2$^1$S lines.  We used Gaussian fits in all cases.  The results are
given in Table 3.  Where the fit deviated noticeably from a single Gaussian, we
note this below in Section 4.3, and in the Table by including the separate
Gaussian components required to achieve a good overall fit.  We also measured
the widths of isolated OH night sky lines to determine the instrumental
resolution, since these lines are significantly narrower than this.  The
instrumental widths changed by no more than 10\% between the different
observing sessions (the worst data being that obtained in January 1997 due to
the problems with the slit mechanism).  In all cases, the HeI data is higher
resolution than the equivalent Br$\gamma$ data.  The intrinsic linewidth for
each observation was derived assuming the intrinsic and instrumental widths
added in quadrature.  The values give in Table 3 have been corrected for the
instrumental resolution.

\subsection{Optical and Infrared Line Ratios}
The observed linear correlation between the [OIII] 5007\AA/H$\beta$ line ratio
and the Stoy temperature found by Kaler \& Jacoby (1991) implies that we can
look for trends in our data in an entirely model independent fashion by
examining the correlation between infrared line ratios and the [OIII]/H$\beta$
ratio.  The basic results of our survey then are shown in Figure 4.  Figure
4(a) shows the ratio of the HeI 7$^{1,3}$G--4$^{1,3}$F and Br$\gamma$ against
the [OIII] 5007\AA/H$\beta$ line ratio.  The result is the expected one in the
sense that the volume of the He$^+$ region rises rapidly with effective
temperature ([OIII] line strength) until it fills the entire PN.  In Figure
4(b) we show the ratio of HeI 2$^1$P--2$^1$S and Br$\gamma$ against the [OIII]
5007\AA/H$\beta$ line ratio.  Here there is a much less well defined trend with
increasing [OIII]/H$\beta$ ratio, as expected if the results from previous
photoionisation models are correct.  In particular we note two key points: this
ratio peaks at lower effective temperature than the HeI 7$^{1,3}$G--4$^{1,3}$F
to Br$\gamma$ ratio and it decreases with increasing [OIII]/H$\beta$ ratio
beyond this peak, both of which are in accord with the models presented
by Ferland (1999).

We can test whether we have introduced any bias in these results from combining
the optical data from the literature with our observed infrared data by further
examining the behaviour of helium-to-hydrogen line ratios in the optical and
infrared separately.  In Figure 5(a) we plot the ratio of 6678\AA\ HeI
3$^1$D--2$^1$P transition with H$\beta$ against HeI 7$^{1,3}$G--4$^{1,3}$F with
Br$\gamma$.  Smits (1996) determined values of expected line emissivities for
high-$n$ states such as the HeI 7$^{1,3}$G--4$^{1,3}$F lines, as well as the
more common optical lines in the general recombination case, including
collisional effects amongst the lowest levels.  Benjamin, Skillman \& Smits
(1999) expanded this treatment of the collisional effects to higher levels, and
detail the resultant emissivities of the better known lines, but otherwise
recover the same data for the pure recombination case.  The two solid lines in
Figure 5(a) show the full range of possible theoretical values for
$n_e=10^2-10^6$cm$^{-3}$ and T$_e=10000-20000$K from the Smits (1996) data.
All of the 3$^1$D--2$^1$P and 7$^{1,3}$G--4$^{1,3}$F lines should be dominated
by recombination processes.  Line opacity and collisional excitation, which can
affect the lower lying triplet states significantly, should be relatively
unimportant here.  Figure 5(a) clearly indicates that this is the case, since
there is a well defined linear relation between the two ratios which is in
reasonable agreement with the predicted theoretical values.

It is interesting to compare this with Figure 5(b), in which we plot the same
optical ratio against the ratio of HeI 4$^{1,3}$S--3$^{1,3}$P with Br$\gamma$.
Again the straight lines show the expected theoretical values for the range
$n_e=10^2-10^6$cm$^{-3}$ and T$_e=10000-20000$K (assuming no collisional
excitation).  There is a larger theoretical dispersion in this instance,
largely due to a stronger dependence of the HeI 4$^{1,3}$S--3$^{1,3}$P line
fluxes on electron temperature.  Despite this, there is a clear trend for some
points to lie well away from the theoretical values, in the sense that the
4$^{1,3}$S--3$^{1,3}$P blend is brighter than expected from recombination
alone.  These are clearly objects in which the population of the 4$^3$S level
of HeI is enhanced.  This can be due to either collisional excitation (largely
from 2$^3$S--4$^{3}$S) or opacity in the 2$^3$S--$n^3$P series (leading to
corresponding enhancements in the S--P series).  The relative contribution of
the two can be estimated from the atomic data (eg Lumsden \& Puxley 1996) and
for the typical conditions of the PN in this survey they contribute almost
equally since it is primarily density that drives both effects.  The role of
collisions have also been calculated more exactly for HeI emission in general
by Benjamin, Skillman \& Smits (1999) as noted above.  They find enhancement of
order 50\% in the 4$^{3}$S--3$^{3}$P line is possible for the physical
conditions being considered, in accord with our data.  The PN that lie away
from the linear trend are obvious from Table 2 from a comparison of the HeI
4$^{1,3}$S--3$^{1,3}$P and HeI 7$^{1,3}$G--4$^{1,3}$F fluxes, since these
should be similar in the presence of recombination alone.  However, there does
not appear to be a clear correlation between these objects and their electron
density as would be expected.  This suggests that the estimated densities from
the [SII] doublet are not entirely reliable.  In particular, all of the PN
showing HeII emission appear to show this effect.  In such objects of course
the S$^+$ zone will lie at the edge of the nebula, which may not be
representative of the nebula as a whole.  In many cases there is evidence for
higher densities in the literature, as estimated from [OII] lines.  This is
consistent with the simple model in which the [SII] emission arises at the edge
of the nebula since O$^+$ has a higher ionisation potential than S$^+$.

\subsection{Velocity Structure}
As noted in the introduction purely local velocity broadening of the 584\AA\
2$^1$S--1$^1$S HeI resonance line can reduce the likelihood of absorption by
neutral helium (cf Ferland 1999), and hence reduce the ratio of the HeI
2$^1$P--2$^1$S and HI Be$\gamma$ lines.  In the on-the-spot approximation for
emission and absorption of this resonance line, only local broadening processes
count, since no 584\AA\ photon will travel a sufficient distance for a global
velocity field to be an important factor.  We therefore need a measure
of the local broadening if we are to compare our data with the correct 
photoionisation model.

Therefore we have assumed a simple model to decompose the observed line width
into local and global velocity broadening with the PN.  We assume that the
intrinsic line width is determined by three possible factors: the expansion
velocity of the nebula, the thermal line width, and turbulence, the first
of these being the only purely global broadening mechanism.  We also assume
that the expansion velocity has a particularly simple form, namely a constant
value everywhere in the nebula.  Although unrealistic in practice, the
variation in expansion velocity is generally sufficiently small that we can
ignore it here.  The profile locally will appear Gaussian, with the broadening
as given by equation 2 below.  Along a line of sight directly through the
centre of the shell we would then see the classic split line profile
composed of two Gaussians offset from each other by twice the expansion 
velocity.

However since the the objects are compact (mostly unresolved) on the sky we see
total velocity field in the beam.  The velocity profile is `infilled' from
those regions of the nebula where the velocity field is more perpendicular to
the line of sight than tangential, so that the double profile will not be
evident in the actual observed data.  
If we integrate this velocity profile
over the whole extent of the PN, then we can derive the observed profile, which
is given by (for an unresolved source)
\[ I ({\rm v}) \propto \int_V j(\vec{r})
\exp(-((\vec{v}-\vec{v}_{exp}).\vec{n})^2/2\sigma^2)\,\, dV  \] 
\begin{equation} \hspace*{7mm}
\propto   {\rm erf}\left(
\frac{{\rm v} + {\rm v}_{exp}}{\sqrt{2}\sigma_{\rm v}}\right)
- {\rm erf}\left(
\frac{{\rm v} - {\rm v}_{exp}}{\sqrt{2}\sigma_{\rm v}}\right)
\end{equation}
where $j$ is the emissivity as a function of position in the nebula (which
we assume here is constant), $\vec{v}_{exp}$ is the expansion velocity
so that $\vec{v}_{exp}.\vec{n}=v_{exp}\cos\theta$ if $\vec{n}$ is the 
vector defining our line-of-sight, and
\begin{equation}
\sigma_{\rm v}^2 = \frac{1}{2}
\left(\frac{2kT}{m_{ion}}+\rm{v}^{2}_{turb}\right).
\end{equation}
For a nominal T$_e=10000$K, and no turbulence, $\sigma_{\rm
v}\simeq9.1$kms$^{-1}$ for HI and $\simeq4.6$kms$^{-1}$ for HeI.  In the
absence of expansion, the line profile will be Gaussian, and the broadening is
given solely by equation 2.

For most of the objects considered here, the actual derived profile is
sufficiently similar to a Gaussian that we cannot uniquely determine the split
between turbulence and global expansion.  We can estimate the thermal line
width directly from the derived electron temperature in each region to remove
that uncertainty.  In practice therefore the best discrimination comes from a
comparison of the differences in the observed widths for the HI and HeI lines.
We calculated the widths of the lines for varying values of v$_{exp}$ and
v$_{turb}$ and sought the `best' match from this grid of data.  An example for
T$_e=10000$K is shown in Figure 6.  For the HI and HeI separately we find the
set of curves that match the observed data.  These then give the derived
v$_{turb}$ and v$_{exp}$ values shown in Table 4.  Since a family of curves is
possible in almost all cases only upper limits to v$_{turb}$ and v$_{exp}$ can
be derived.  Clearly, however, these are not joint limits, since when
v$_{turb}$ is large v$_{exp}$ must be small and vice-versa.  For some of the
objects there is a discrepancy between the values derived from the HI and HeI
lines.  In some cases this is due to the fact that the He$^+$ and H$^+$ zones
do not overlap because the effective temperature of the exciting star is too
low.  For these objects we use the limits on the HeI line alone.  For a small
sample of objects there are no reliable solutions within the context of our
simple model.  Comments on these objects are given below, together with other
notes on individual PN.

BD+30$^\circ$3639: there is considerable structure in BD+30$^\circ$3639
(cf Sahai \& Trauger 1998).  The split in the two components seen in the
HeI line give v$_{exp}\sim30$kms$^{-1}$ in terms of our simple model, so
no turbulent broadening is required.  This expansion velocity is not however
compatible with the HI line data.  It seems likely that our simple model
is not a good approximation for this source.  However, even given this, there
appears little need to invoke a large turbulent velocity component.

CRL~618: The spectra acquired for this paper were centred on the brighter of
the two outflow lobes visible in the optical.  The actual HII region is in fact
completely obscured at optical wavelengths.  Our data agree with previous
published near infrared data (Hora, Latter \& Deutsch 1999), which also seems
to have been centred on the visible lobe rather than the actual core.  We have
further low resolution data that will be published separately of the core
showing strong HI emission, and stronger HeI emission than present in the
lobes.  However, since we have only obtained echelle spectra of the lobe
position we only quote the result for the HI and HeI line strengths there.  It
is possible that most of the emission we detect in the lobes is actually
scattered light from the obscured central regions, since CRL~618 is known to
have significant polarisation.  Trammell, Dinerstein \& Goodrich (1993) show
that, in the optical, all the recombination lines and high excitation forbidden
lines are scattered into our line of sight.  We therefore do not attempt to
determine turbulent or expansion components to the velocity from the observed
data (which would imply very large velocities), and for sake of the
photoionisation model simply assume v$_{turb}\sim$10kms$^{-1}$.

Hu~1-2: this object shows clear signs of a bipolar outflow in the echelle
spectra, rather than straightforward expansion, since the velocity field flips
from red to blue shifted moving south across the position of the central star.
The double peaked line profile that results is clearly visible in both echelle
spectra.  The difference in line width probably reflects a differing
acceleration in the He$^{++}$ and He$+$ zones in the outflow.  The observed
widths limit v$_{turb}\le$10kms$^{-1}$.

M~1-1: Only a two component Gaussian fit gave a good match to the data.  The
velocity separation of the two peaks is $\sim$58kms$^{-1}$, indicating an
expansion velocity of $\sim$30kms$^{-1}$.  The predicted turbulent velocity
component is then small, since most of the remaining line broadening must be
thermal.  Because of the weakness of the HeI 2$^1$P--2$^1$S line in the low
resolution data we did not observe this object with the echelle at 2.058$\mu$m.
There are no satellite HeI lines visible in the Br$\gamma$ data either.

NGC~7027: this is the only one of our sample to show spatially resolvable
velocity structure.  The structure most closely resembles an expanding shell in
agreement with the radio data of Roelfsema et al.\ (1991).  Given this we can
further limit the velocity components so that v$_{exp}\sim20$kms$^{-1}$ and
v$_{turb}\le$10kms$^{-1}$.

\section{Photoionisation Models}
\subsection{Basic Parameters}
We used Cloudy version 90.05 (Ferland 1996) to model the emission line spectra
for a grid of PN with differing physical parameters.  The main parameters we
considered were effective temperature, central star luminosity, electron
density and the velocity structure.  All models used the standard Cloudy
planetary nebula abundances with their corresponding dust grain abundance.

We considered effective temperatures in the range 25000--150000K.  We used two
sets of model atmospheres.  First, to ensure that we have the same model for
all temperatures, we used a simple blackbody approximation.  Then, to test
whether a different model might affect the results, we used the standard Kurucz
(1992) models with $\log g=5$ for the more restrictive temperature range of
25000-50000K.  The only alternative models of stellar atmospheres
potentially applicable to low effective temperature PN in the version
of Cloudy used are due to Mihalas (1972).  We did not consider these however
when it became clear that the blackbody models were a sufficiently good
match to our data (see Section 5.2).

There is observational and theoretical evidence that PN central stars with low
effective temperatures all have approximately the same bolometric luminosity.
Observational evidence comes from studies of PN in the Magellanic Clouds (cf
Figure 11 in Dopita et al.\ 1996).  The theoretical model tracks of Vassiliadis
\& Wood (1993) also show that below 50000K there is little dispersion in the
luminosity-effective temperature domain.  From Vassiliadis \& Wood (1993) we
adopt a value of 5000\Lsolar\ as a mean bolometric luminosity, but consider
3000 and 7000\Lsolar\ as well, since these match the spread in the range of
observed and theoretical values for PN central stars below about 100000K.  The
range of electron density and turbulent/expansion velocity considered are taken
directly from the observed data.  We computed models with $n_e=$3000, 6000,
12000, 24000 and 48000cm$^{-3}$, and v$_{turb}=$0, 5, 10 and 15kms$^{-1}$.

Lesser effects that need to be considered are due to metallicity/abundance and
the final radius chosen.  This latter parameter is really only important for
the PN with hot central stars, since, for small outer radii, the nebula can be
density rather than ionisation bounded, resulting in reduced He$^+$ zones.
This is apparent in some of the observed data for those PN with hot central
stars.  Since it only affects the high temperature end of the models, we adopt
$r_{out}=0.1$pc as an appropriate value for our sample (cf the values in Cahn,
Kaler \& Stanghellini 1992), and accept that this may mean that we overestimate
the model HeI line strength at high effective temperature.  The inner radius
has little effect on the strengths of the lines we are interested in, and we
adopt $r_{in}=0.01$pc.  We do not consider metallicity effects further here,
since there is no direct evidence from the observational data that it is a
significant effect. For example, DdDm~1 is a known halo PN, which has lower
metallicity than the other PN in the sample, but whose helium-to-hydrogen line
ratios show little deviance from the rest of the sample.

\subsection{Model Results}
Our selection used the relation between the [OIII] 5007\AA\ to H$\beta$ ratio
and Stoy temperatures for the PN central stars as derived by Kaler and Jacoby
(1991) to determine approximate stellar effective temperatures for our sample.
However, these temperatures are not completely reliable measures of the actual
underlying T$_{eff}$, since the method is only sensitive to effective
temperatures below 60000K.  This can be seen clearly in Figure 7, where we show
the original data used by Kaler \& Jacoby, together with their fit to the same
data, and the results from our photoionisation modelling.  It is encouraging
that the blackbody stellar models reproduce the spread in the observed data
reasonably well.  In this case it appears that the identification of the
underlying effective temperature with the Stoy temperature is sound up to
60000K.  This is in spite of the fact that a black body is not a good match to
the actual spectrum of a PN central star.  However, previous workers have also
found that their use gives a good match between model and observation (cf.\ the
discussion in Dopita \& Meatheringham 1991).  By comparison, the Kurucz model
atmospheres which are shown as the dashed line in Figure 7 do not reproduce the
observed relation at all, particularly at the lower end of our effective
temperature range (note also that the Kurucz models only exist for
T$_{eff}<$50000K).  We therefore did not consider the further use of these
models.  The primary mechanism giving rise to the spread in [OIII] 5007\AA\ to
H$\beta$ ratio at a fixed temperature is the electron density.  Collisional
enhancement at a fixed temperature can change the ratio by 30\% over the range
of densities considered (the largest value of the ratio occurring at
$n_e=20000$cm$^{-3}$; at higher values the ratio declines again), whereas the
central star luminosity contributes at most a 15\% change, and the velocity
structure is essentially irrelevant.

As can be seen, the [OIII]/H$\beta$ ratio rises almost linearly with
temperature, as the O$^{2+}$ region in the nebula increases in size, until
about T$_{eff}=$60000K.  However, beyond this temperature, the O$^{2+}$ region
saturates, and eventually decreases with increasing T$_{eff}$ as O$^{3+}$
becomes dominant.  In addition, for our sample of compact PN, it is possible
that the nebula is density rather than ionisation bounded.  This effect tends
to become significant for temperatures above that at which the He$^{++}$ region
first appears ($\sim50000K$).  For a compact nebula, the available gas
reservoir at this point may be less than that required to absorb all the
emitted radiation below 912\AA.  In this case, the flux from the lower
ionisation species near the edge of the nebula will clearly drop first.  The
effect is the same as that which leads to the well known difference between
hydrogen and helium Zanstra temperatures in PN at high temperature (cf Kaler \&
Jacoby 1991 and references therein).  Therefore for those PN showing evidence
of HeII emission we deduced a calibration between the 4686\AA\ HeII line
strength and T$_{eff}$ from the models.  First, we extended the effective
temperature range by calculating a small set of models spanning the range
150000--400000K, with L$_*=$5000\Lsolar, $n_e=12000$cm$^{-3}$ and
v$_{turb}=$5kms$^{-1}$.  Variations in the luminosity and density lead to
changes of less than 25\% in the 4686\AA\ HeII to H$\beta$ ratio.  We
interpolated these data to find effective temperatures for the real PN.  There
was one problem with this approach, since the model HeII 4686\AA\ to H$\beta$
ratio never increased beyond 0.55 (near 300000K: again this is a reflection of
the fact that the He$^{++}$ zone is now becoming density bounded).  Three of
our PN show values larger than this however.  This may be a reflection of
either a density bounded nebula or non spherical geometry.  For the sake of
this paper we simply ascribe to these PN a lower limit on the effective
temperature of 250000K.

In Figure 8 we show how the HeI 2$^1$P--2$^1$S to HI Br$\gamma$ ratio varies as
a function of effective temperature.  We have explicitly split the dependence
on (a) density, (b) turbulent velocity and (c) stellar luminosity to show how
the ratio is affected by each.  The results are as expected.  The ratio
increases with increasing density because of collisions from the metastable
2$^1$S state enhancing the population of the 2$^1$P state.  As the microscopic
velocity width increases, the 584\AA\ 2$^1$P--1$^1$S resonance line weakens,
leading to a decrease in the population of the 2$^1$P state.  As the stellar
luminosity increases the ionisation parameter at the inner edge of the cloud
also increases, leading to a marginal increase in the He$^{++}$ region compared
to a lower luminosity star, once that star is hot enough to actually excite
this state, and a decline in the fraction of He$^0$ and He$^+$.  Since He$^0$
is required to scatter the 584\AA\ photons, the direct consequence is a
reduction in the ratio for effective temperatures beyond about 40000K.  From
Figure 8 it is also clear that the models with the `extreme' high and low
ratios are those with high density, and low turbulent velocity and central star
luminosity and low density, and high turbulent velocity and central star
luminosity respectively.  Since the luminosity dependence is only weak however,
we will ignore it in the following, and consider only our mean central star
luminosity of L$_*=5000$\Lsolar.

\subsection{Comparison with Observations}
All of the strong HI and HeI lines are calculated directly by Cloudy, but the
two weaker K band HeI lines are not.  Since the 2.16475$\mu$m
7$^{1,3}$G--4$^{1,3}$F line must be dominated by recombination processes, and
the 6678\AA\ 3$^1$D--2$^1$P is known to be (eg Benjamin et al.\ 1999), the
ratio of these two lines must in theory be roughly constant.  Figure 5(a) shows
that this is true observationally.  We can therefore use the data from Smits
(1996) to correct the model 6678\AA\ flux into a model 2.16475$\mu$m flux.  The
actual correction for T$_e=10000$K and $n_e=10000$cm$^{-3}$ is that the model
HeI 7$^{1,3}$G--4$^{1,3}$F to HI Br$\gamma$ ratio is 1.23 times the model HeI
3$^1$D--2$^1$P to H$\beta$ ratio (where we have also used the data on HI from
Storey \& Hummer 1995).  This values rises to 1.31 for T$_e=20000$K and
$n_e=10000$cm$^{-3}$.  The value depends only slightly on electron density but
much more strongly on electron temperature.

Unfortunately, the behaviour of the 2.113$\mu$m 4$^{1,3}$S--3$^{1,3}$P line is
not as simple to translate from the Cloudy results, given the comments in
section 4.2.  We therefore chose not to compare it to the model, but rather to
use its ratio with the 7$^{1,3}$G--4$^{1,3}$F line as an indicator of
collisional or line opacity effects, and hence of high electron density
($>10^4$cm$^{-3}$).  As noted in Section 4.2, these objects can be found in
Table 2 by a comparison of the HeI 4$^{1,3}$S--3$^{1,3}$P and
7$^{1,3}$G--4$^{1,3}$F fluxes.  Objects with significantly larger
4$^{1,3}$S--3$^{1,3}$P fluxes than 7$^{1,3}$G--4$^{1,3}$F fluxes must have
enhanced 4$^{3}$S populations, and hence high electron density.

In Figure 9(a) we plot the observed HeI 7$^{1,3}$G--4$^{1,3}$F to HI Br$\gamma$
ratio and in Figure 9(b) the observed HeI 6678\AA\ 3$^1$D--2$^1$P to H$\beta$
ratio as taken from the literature.  We also plot a typical result from our
photoionisation models (the actual results change by less than 5\% as a
function of density, luminosity or velocity, and 10\% as a function of electron
temperature).  We have not labelled the individual PN here since clearly both
sets of data are in good agreement with the models.  Therefore we can be
confident that the ratio of a pure recombination HeI and HI line does behave in
the fashion expected, and that such a ratio could be used as a measure of
effective temperature in the 30--40000K range.

In Figure 10 we plot the observed HeI 2$^1$P--2$^1$S to HI Br$\gamma$ ratio
against effective temperature (taken to be the greater of T(Stoy) or T(HeII)
from Table 1).  The solid lines are the model results for the `extreme' models
for a central star with a luminosity of 5000\Lsolar\ as discussed in the
previous section.  The lower curve is a model with $n_e=3000$cm$^{-3}$ and
v$_{turb}=15$kms$^{-1}$ and the upper curve is a model with
$n_e=48000$cm$^{-3}$ and v$_{turb}=0$kms$^{-1}$.  We have labelled the
individual PN to make identification easy.  We have compressed the temperature
scale to show the detail at T$_{eff}<50000$K, but note that the high excitation
PN that are not plotted all lie above the predicted model curve.  Finally in
Figure 11 we plot the observed HeI 2$^1$P--2$^1$S to HeI 7$^{1,3}$G--4$^{1,3}$F
ratio, since this removes any metallicity dependence.  We have plotted the same
models as in Figure 10.

From Figure 10, it is clear that the models systematically underpredict most of
the high effective temperature PN.  These objects do show evidence for high
density (either from their [SII] ratios, or from the strength of the HeI
4$^{1,3}$S--3$^{1,3}$P line).  Some must have low turbulent velocities (M~1-20
for example), but there does not appear to be a clear trend of the ratio with
this parameter (remembering from Figure 8 that even turbulent velocities at the
10kms$^{-1}$ level will suppress the HeI 2$^1$P--2$^1$S line by 25\% at high
T$_{eff}$).  Indeed 3 of the sources shown in Figure 10 and 11 (M~1-20, K~3-62
and M~1-4) lie above the maximum predicted theoretical values for the ratios
shown, and the same trends also carry through to the higher effective
temperature PN not plotted in Figures 10 and 11.  From Figure 11 it is also
clear that enhanced helium abundance cannot explain the observed data for these
objects, since they also tend to lie above the model predictions there.  For
effective temperatures near and above 60000K it is also clear from Figure 8
that density has only a small role in determining the HeI 2$^1$P--2$^1$S line
strength.  Therefore higher density than that used in our modelling is also
unlikely to account for these data.  The only obvious explanation for all these
results is that the population of the HeI 2$^1$P level is not as small for
these PN as predicted by Cloudy.  It is not the ionisation structure that is
wrong though, since the model does predict the strength of the neutral [OI]
6300\AA\ line correctly (at a level between 1 and 5\% of H$\beta$ for these
PN).  Therefore it seems that the only alternative is that the HeI Ly$\alpha$
destruction probability as used in Cloudy is an overestimate (see Ferland
1999 for a fuller discussion of this factor).

The behaviour of the low temperature PN (those with T$_{eff}<$28000K) may be
easier to understand.  Three of the five objects sit within the model bounds,
though they again show deviations from the model since BD+30$^\circ$3639 should
have high density and no turbulence, yet is the lowest of the three.  More
importantly, two objects lie below the predicted ratio.  For all these objects
we must of course be cautious as to whether the temperature scale is accurate.
A shift of 3000K in temperature (the effective error on the Kaler \& Jacoby
relation), would be sufficient to explain all the data.  It should also be
borne in mind that we may have `errant' observational values for CRL 618 given
the comments in section 4.3.  Also these PN are unlikely to be well matched by
black body stellar atmospheres (BD+30$^\circ$3639 is a WC star after all).  The
predicted ratio from the Kurucz models we calculated dropped off more sharply
with decreasing effective temperature for example (in the same sense that the
[OIII]/H$\beta$ ratio declines more sharply than for a black body stellar
atmosphere, as shown in Figure 7).  

Finally we come to the group of objects between 28000 and 40000K.  Again there
seems to be no direct correlation between the limits on the turbulent
velocities and the observed line ratio in contradiction with the models.  The
two most discrepant points here though are M~1-6 and M~1-14.  At least for the
these there is observational evidence for moderately high density.  M~1-6 may
simply have the wrong assigned effective temperature as well, since it is the
only object for which we have had to rely on the Acker et al.\ (1992) catalogue
for the 5007\AA\ [OIII] flux.  In addition, there is some evidence here for an
enhanced helium abundance, since Figure 11 shows that some of the PN that sit
above the maximum model value in Figure 10 lie well within the model boundaries
of the HeI 2$^1$P--2$^1$S to HeI 7$^{1,3}$G--4$^{1,3}$F ratio.  M~1-6, M~1-9
and M~1-14 are possible examples where this is true (and we note that the first
two of these PN are listed as having large helium abundances, and being
Peimbert Type I objects, by Shibata \& Tamura 1985).  We also note however that
there are other objects with similar effective temperatures that have even
higher observed HeI 2$^1$P--2$^1$S to HI Br$\gamma$ ratios than M~1-14, such as
the ultracompact HII region, G45.12+0.13, which again has high density (Lumsden
\& Puxley 1996).  Therefore, two plausible arguments present themselves here:
either there is some enhancement to the helium abundance over the 10\% assumed
in our modelling (as might be expected for Peimbert Type I PN) or the
population of the HeI 2$^1$P level is collisionally enhanced above the maximum
values derived using Cloudy (perhaps due to the limited number of triplet
states considered in Cloudy).  At these values of the effective temperature
however there is no need to invoke a change in the HeI Ly$\alpha$ destruction
probability.

\section{Conclusions}
It is clear from the results we have presented here that the behaviour of the
HeI 2$^1$P--2$^1$S line is sufficiently complex that it is not a reliable
indicator of stellar effective temperature.  The variation of 30\% or more in
both observed and model predictions for a fixed temperature mean that only a
very general picture of the effective temperature can be gained from this
ratio.  It is possible that in certain restricted circumstances, such as
objects with known density and turbulent velocity fields, that it may provide
secondary information about the effective temperature however.  Figure 11
provides an example of how it might be used in combination with another HeI
recombination line to constrain the temperature of objects above 40000K, where
the HeI/HI recombination line ratio alone is no longer varying rapidly.
However, for this to be reliable the discrepancies between model and data seen
in our results will need to be explained.  There are factors that we have not
accounted for that may help explain some of observations.  A change in the
dust-to-gas ratio, clumping within the ionised gas or a density rather than
ionisation bounded nebula could plausibly explain some of the discrepancies we
see.  We would certainly not recommend the use of this line however as a
primary indicator of stellar effective temperature.

By contrast, the behaviour of the HeI 7$^{1,3}$G--4$^{1,3}$F line, which should
be due solely to recombination, is as expected, and in agreement with the
models.  Although this line is not in itself of particular value, since it is
blended with HI Br$\gamma$ at intermediate spectral resolution, it does show
that the use of such HeI lines is in accord with the original intent of Doyon
et al.\ (1992).  In particular, the 1.7007$\mu$m 4$^{3}$D--3$^{3}$P transition
is also expected to be due mostly to recombination, is in a clear part of the
atmospheric H band window, and is well separated from other emission lines.
This line has been discussed in the past in this context (eg, Lumsden \& Puxley
1996, Vanzi \& Rieke 1997).  We shall discuss it further in the paper
presenting our low resolution spectra for this PN sample.

\section*{Acknowledgments}
We would like to thank Derck Smits for providing the machine readable version
of his HeI predictions from his 1996 paper.  SLL acknowledges support from
PPARC through the award of an Advanced Research Fellowship.  SLL also thanks
the Access to Major Research Facilities Program, administered by the Australian
Nuclear Science and Technology Organisation on behalf of the Australian
Government, for travel support for the observations reported here.  The United
Kingdom Infrared Telescope is operated by the Joint Astronomy Centre on behalf
of PPARC.

\parindent=0pt

\noindent{\bf References}\par
\mnref{Acker, A., Ochsenbein, F., Stenholm, B., Tylenda, R., Marcout, J.,
	Schohn, C., 1992, Strasbourg-ESO Catalogue of Galactic 
	Planetary Nebulae, ESO}
\mnref{Aller, L.H., Keyes, C.D., 1987, \apjs, 65, 405}
\mnref{Aller, L.H., Czyzak, S.J., 1979, Ap\&SS, 62, 397}
\mnref{Aller, L.H., Czyzak, S.J., 1983, \apjs, 51, 211}
\mnref{Aller, L.H., Hyung, S., 1995, \mn, 276, 1101}
\mnref{Baessgen, M., Hopfensitz, W., Zweigle, J., 1997, \aaa, 325, 277}
\mnref{Benjamin, R.A., Skillman, E.D., Smits, D.P., 1999, \apj, 514, 307}
\mnref{Cahn, J.H., Kaler, J.B., Stanghellini, L., 1992, \aas, 94, 399}
\mnref{Costa, R.D.D., Chiappini, C., Maciel, W.J., de Freitas Pacheco, J.A., 
	1996, \aas, 116, 249}
\mnref{Cuisinier, F., Acker, A., Koppen, J., 1996, \aaa, 307, 215}
\mnref{Depoy, D.L., Shields, J.C., 1994, \apj, 422, 187}
\mnref{Doherty, R.M., Puxley, P.J., Doyon, R.,  Brand, P.W.J.L., 1994,
	\mn, 268, 821}
\mnref{Doherty, R.M., Puxley, P.J., Lumsden, S.L., Doyon, R., 1995, \mn,
      277, 577}
\mnref{Dopita, M.A., Meatheringham, S.J., 1991, \apj, 367, 115}
\mnref{Dopita, M.A., Vassiliadis, E., Meatheringham, S.J., Bohlin, R.C., 
	Ford, H.C., Harrington, J.P., Wood, P.R., Stecher, T.P., Maran, S.P.,
	1996, \apj, 460, 320}
\mnref{Dopita, M.A., Hua, C.T., 1997, \apjs, 108, 515}
\mnref{Doyon, R., Puxley, P.J., Joseph, R.D., 1992, \apj, 397, 117}
\mnref{Ferland, G.J., 1999, \apj, 512, 247}
\mnref{Ferland, G.J., 1996, Hazy, a Brief Introduction to Cloudy, 
	University of Kentucky Department of Physics and Astronomy 
	Internal Report.}
\mnref{Hora, J.L., Latter, W.B., Deutsch, L.K., 1999, \apjs, 124, 195}
\mnref{Johnson, H.L., 1966, ARA\&A, 4, 193}
\mnref{Kaler, J.B., 1985, ApJ, 290, 531}
\mnref{Kaler, J.B., Jacoby, G.H., 1991, \apj, 372, 215}
\mnref{Kaler, J.B., Bell, D., Hayes, J., Stanghellini, L., 1993, \pasp, 105,
	599}
\mnref{Kaler, J.B., Kwitter, K.B., Shaw, R.A., Browning, L., 1996,
	\pasp, 108, 980}
\mnref{Kaler, J.B., Shaw, R.A., Browning, L., 1997, \pasp, 109, 289}
\mnref{Kelly, D.M., Latter, W.B., Rieke, G.H., 1992, \apj, 395, 174}
\mnref{Keyes, C.D., Aller, L.H., Feibelman, W.A., 1990, \pasp, 102, 59}
\mnref{Kingsburgh, R.L., Barlow, M.J., 1994, \mn, 271, 257}
\mnref{Kurucz, R.L., 1992, IAU Symposium 149: The Stellar Populations of
Galaxies, eds Barbuy, B., Renzini, A., Kluwer, Dordrecht, Holland, p.225}
\mnref{Kwitter, K.B., Henry, R.B.C., 1998, \apj, 493, 247}
\mnref{Leitherer, C.,  1998, ASP Conf. Ser. 142: The Stellar Initial 
Mass Function (38th Herstmonceux Conference), 61}
\mnref{Lumsden, S.L., Puxley, P.J., Doherty, R.M., 1994, \mn, 268, 821}
\mnref{Lumsden, S.L., Hoare, M.G., 1996, \apj, 464, 272}
\mnref{Lumsden, S.L., Puxley, P.J., 1996, \mn, 281, 493}
\mnref{Mihalas, D., 1972, Non-LTE Model Atmospheres for B and O Stars, NCAR-TN/STR-76}
\mnref{Peimbert M., Torres-Peimbert S., 1983, IAU Symp.\ 103: 
	Planetary Nebulae,  103, 233}
\mnref{Sabbadin, F., Cappellaro, E., Turatto, M., 1987, \aaa, 182, 305}
\mnref{Sahai, R., Trauger, J.T., 1998, \apj, 116, 1357}
\mnref{Seaton, M.J., 1979, \mn, 187, 73P}
\mnref{Shaw, R.A., Kaler, J.B., 1989, \apjs, 69, 495}
\mnref{Shibata, K., Tamura, S., 1985, PASJ, 37, 325}
\mnref{Shields, J.C., 1993, \apj, 419, 181}
\mnref{Smits, D.P., 1996, \mn, 278, 683}
\mnref{Storey, P.J., Hummer, D.G., 1995, \mn, 272, 41}
\mnref{Tamura, S., Shaw, R.A., 1987, \pasp, 99, 1264}
\mnref{Trammell, S.R., Dinerstein, H.L., Goodrich, R.W., 1993, \apj,
	402, 249}
\mnref{Vanzi, L., Rieke, G.H., 1997, \apj, 479, 694}
\mnref{Vassiliadis, E., Wood, P.R., 1993, \apj, 413, 641}

\clearpage

\onecolumn

\tabcolsep=5.5pt
\begin{tabular}{lrrrrrrrrl}
Name & 
\multicolumn{1}{c}{Size}  & \multicolumn{1}{c}{T(Stoy)} & 
\multicolumn{1}{c}{T(HeII)} & 
\multicolumn{1}{c}{[OIII]/H$\beta$} & \multicolumn{1}{c}{HeI/H$\beta$} & 
\multicolumn{1}{c}{HeII/H$\beta$} & \multicolumn{1}{c}{T$_e$ (K)} & 
\multicolumn{1}{c}{$n_e$ (cm$^{-3}$)}&Ref.\\
BD+303639&8 &25600 &&0.07 &0.009 & &9000&22000$^{+20000}_{-2500}$&a  \\
CRL~618 & 10 & 25600&& 0.04 & 0.013 &&&5500$\pm1000$&bc  \\
DdDm~1 &   1 &40000 && 4.53 & 0.040 &&12000&4000$\pm1000$&d  \\
Hu~1-2 &  8 &51000 &$>$250000&7.86 & 0.029&0.89&16000&4500$\pm1000$&efg  \\
K~3-60 &  3 &73000 &220000&14.84 & 0.028 & 0.50&&4000$\pm1000$&h  \\
K~3-62 &  3 &57800 &&10.08 &0.039 &&12500&20000$^{+20000}_{-3000}$&i  \\
K~3-66 &  0 & 33100 &&  2.40 & 0.035& & 12000&6000$\pm1000$&j \\
K~3-67 &  0 & 59200 && 10.50 & 0.039&&13800&4500$\pm1000$&hjk  \\
K~4-48 &  2 & 65300 &$>$250000&12.40 & 0.031&0.57&12500&2500$\pm1000$&l  \\
M~1-1 &   6 & 37807 &$>$250000&   4.37 & 0.009 & 1.00 & 15000 & 5300$\pm1000$& e \\
M~1-4 &   4 & 66860 &83000&  12.90 & 0.032 &0.06 &  12000 &5000$^{+1000}_{-3000}$& hmn  \\
M~1-6  &  5 & 29330 && 1.22 & -- & & &20000$^{+20000}_{-3000}$&o  \\
M~1-9 &  12 & 39900 &&4.50 & 0.032 &&10000&4500$\pm1000$&l  \\
M~1-11 &  0 & 25600 && 0.09 & 0.008 &&11400&3000$\pm1000$& klp  \\
M~1-12 &  0 & 25890&& 0.15 & 0.008 & & 11000&20000$^{+20000}_{-3000}$&kl\\
M~1-14 &  0 & 34500&& 2.82 & 0.037 &&11500&12000$\pm1000$& q \\
M~1-20&   7 &57000 &&9.80 &0.038 & &11000&20000$^{+20000}_{-3000}$&pq  \\
M~1-74 &  5 & 60700 && 10.98 &0.031 & &10000&15000$^{+20000}_{-7500}$&egi \\
M~1-78 &  6 & 35500 &&3.23 &0.031 & &&3500$\pm1000$&hir \\
NGC~7027 &14 &70700&190000&14.81&0.029 &0.45 &13000&20000$^{+20000}_{-3000}$&hs  \\
PC~12 &   5 &35400 &&3.03 &0.035 &&10000&7500$^{+2500}_{-2500}$& lt  \\
SaSt~2-3 & 0 & 25500 && 0.02 &&&&2000$\pm1000$&u  \\
Vy~1-1 &  6 & 56200 && 9.60 & 0.034 & & &&i  \\
\end{tabular}

\noindent{\bf Table 1:} The sample of planetary nebula observed for this
project.  The Stoy temperatures given are derived from equation 1 of Kaler \&
Jacoby (1991).  T(HeII) is the temperature of the best fitting black-body
stellar atmosphere from our modelling (Section 5) for the observed HeII line
strength.  It is a more accurate estimate of the stellar effective temperature
for those objects with HeII emission than the Stoy temperature.  Where T$_e$ is
not given we assume T$_e=10000$K.  Where $n_e$ is not given we assume
$n_e=5000$cm$^{-3}$.  Errors in T$_e$ are estimated at approximately 1000K.
The optical line ratios are derived from the references given.  The ratios
tabulated are for 5007\AA\ [OIII]/H$\beta$, 6678\AA\ HeI/H$\beta$ and 4686\AA\
HeII/H$\beta$.  These are as follows: (a) Aller \& Hyung 1995; (b) Baessgen,
Hopfensitz \& Zweigle (1997); (c) Kelly, Latter \& Rieke (1992); (d) Kwitter \&
Henry (1998); (e) Aller \& Czyzak (1979); (f) Sabaddin, Cappellaro \& Turatto
(1987); (g) Aller \& Czyzak (1983); (h) Aller \& Keyes (1987); (i) our own
data; (j) Tamura \& Shaw (1987); (k) Kingsburgh \& Barlow (1994); (l)
Cuisinier, Acker \& Koppen (1996); (m) Aller (1984) as reported in Kaler, Shaw
\& Browning (1997); (n) Kaler (1985); (o) Acker et al.\ (1989); (p) Kaler,
Kwitter, Shaw \& Browning (1996); (q) Costa et al.\ (1996); (r) Kaler et al.\
(1993); (s) Keyes, Aller \& Feibelman 1990; (t) Shaw \& Kaler (1989); (u)
Dopita \& Hua (1997);

\newpage

\tabcolsep=3pt
\begin{tabular}{lcccccr}
Name & 
\multicolumn{1}{c}{F(Br$\gamma$ (2.166$\mu$m))}  & 
\multicolumn{1}{c}{I(HeI (2.06$\mu$m))/} & 
\multicolumn{1}{c}{I(HeI (2.11$\mu$m))/} & 
\multicolumn{1}{c}{I(HeI (2.166$\mu$m))/} & 
\multicolumn{1}{c}{I(HeII(2.189$\mu$m))/  } & Corrn.
\\
\multicolumn{1}{c}{} &
\multicolumn{1}{c}{} &
\multicolumn{1}{c}{I(Br$\gamma$)} & 
\multicolumn{1}{c}{I(Br$\gamma$)} & 
\multicolumn{1}{c}{I(Br$\gamma$)} & 
\multicolumn{1}{c}{I(Br$\gamma$)} \\
BD+30$^\circ$3639  & $449\pm0.2$& $0.294\pm0.001$& & $0.013\pm0.001$& & 0.9531\\
CRL~618 & $0.36\pm0.03$& $0.150\pm0.100$ & &&& 0.9785\\
DdDm~1 & $3.10\pm0.03$& $0.786\pm0.001$& $0.077\pm0.001$&  & & 1.1991\\
Hu 1-2 &$17.8\pm0.2$& $0.366\pm0.003$& $0.067\pm0.001$& &$0.140\pm0.006$  &1.0139\\
K~3-60 &$17.3\pm0.2$& $0.197\pm0.001$ &$0.055\pm0.001$& $0.029\pm0.004$ &$0.121\pm0.011$ &0.8490\\
K~3-62 & $67.3\pm0.1$& $0.723\pm0.001$& $0.054\pm0.001$& $0.047\pm0.002$ &  & 1.1341\\
K~3-66 & $7.09\pm.04$&$0.987\pm0.007$& $0.035\pm0.001$ &$0.030\pm0.008$ && 1.0963\\
K~3-67 & $11.1\pm0.01$& $0.508\pm0.001$& $0.071\pm0.001$& $0.044\pm0.004$& & 1.0544\\
K~4-48 & $6.80\pm0.03$  &$0.418\pm0.002$& $0.078\pm0.001$& &$0.037\pm0.004$& 1.0592\\
M~1-1 & $1.87\pm0.03$ & $0.050\pm0.001$ &$0.031\pm0.001$  &  &$0.300\pm0.017$ & \\
M~1-4 & $25.2\pm0.03$ & $0.440\pm0.001$ & $0.068\pm0.001$&$0.041\pm0.007$
 &$0.018\pm0.002$ & 1.0009\\
M~1-6  & $47.4\pm0.05$& $1.058\pm0.002$& $0.034\pm0.001$& $0.031\pm0.002$& & 1.0030\\
M~1-9  & $18.3\pm0.04$& $1.030\pm0.004$& $0.053\pm0.001$&$0.053\pm0.004$ & & 0.9970\\
M~1-11 & $113\pm0.1$& $0.418\pm0.001$& $0.010\pm0.001$& $0.014\pm0.002$& & 1.1427\\
M~1-12 & $38.4\pm0.05$& $0.347\pm0.001$& $0.014\pm0.001$&$0.017\pm0.003$ & & 0.8114\\
M~1-14 & $32.5\pm0.05$& $1.132\pm0.003$& $0.046\pm0.001$& $0.048\pm0.007$& & 1.0085\\
M~1-20 & $27.9\pm0.05$& $0.774\pm0.002$& $0.050\pm0.001$& $0.049\pm0.004$&  & 1.0415\\
M~1-74 & $39.2\pm0.2$& $0.562\pm0.003$& $0.046\pm0.001$&$0.049\pm0.003$ & & 0.9119\\
M~1-78 & $85.1\pm0.2$& $0.787\pm0.002$& $0.052\pm0.001$&$0.042\pm0.002$  & & 0.9586\\
NGC 7027 & $995\pm0.2$& $0.279\pm0.001$& $0.075\pm0.001$& $0.035\pm0.001$ & $0.138\pm0.001$&  0.9427\\
PC~12  & $18.8\pm0.08$& $0.966\pm0.006$& $0.038\pm0.001$&$0.037\pm0.004$ & & 1.2120\\
SaSt~2-3 & $3.21\pm.05$& $0.153\pm0.002$& & & & 1.1233\\
Vy~1-1 & $7.3\pm0.1$ & $0.332\pm0.005$& $0.049\pm0.001$ & $0.044\pm0.011$ & & 1.1965\\
\end{tabular}

\noindent{\bf Table 2:} Observed HI Br$\gamma$ fluxes, and ratios of the
various helium lines with Br$\gamma$ for the planetary nebula sample.  The HeI
lines tabulated are 2.058$\mu$m $2^1$S--$2^1$P, 2.113$\mu$m
$4^{3,1}$S--$3^{3,1}$P and 2.165$\mu$m $7^{1,3}$G--$4^{3,1}$F.  The HeII line
is the 2.189$\mu$m 10--7 transition.  The Br$\gamma$ fluxes are in units of
$10^{-17}$Wm$^{-2}$.

\vspace*{1cm}

\begin{tabular}{lcclcc}
Name & \multicolumn{2}{c}{Line Width (kms$^{-1}$)}
&Name & \multicolumn{2}{c}{Line Width (kms$^{-1}$)}
\\
 &
\multicolumn{1}{c}{HI Br$\gamma$}  & 
\multicolumn{1}{c}{HeI $2^1$S--$2^1$P } 
& &
\multicolumn{1}{c}{HI Br$\gamma$}  & 
\multicolumn{1}{c}{HeI $2^1$S--$2^1$P } 
\\
BD+30$^\circ$3639  &  $42.80\pm0.69$ & $51.17\pm0.29$ & M~1-4     &  $39.33\pm0.69$ & $30.47\pm0.58$ \\
BD+30$^\circ$3639  &  & $48.40\pm4.37$ &M~1-6     &  $26.73\pm0.69$ & $14.87\pm0.29$ \\
CRL~618   &  $65.09\pm4.15$ & $40.52\pm7.29$ &M~1-9     &  $31.30\pm0.69$ & $22.30\pm0.29$ \\
DdDm~1    &  $39.75\pm1.11$ & $34.11\pm2.92$ &M~1-11    &  $26.59\pm0.69$ & $10.64\pm0.29$ \\
Hu~1-2    &  $41.00\pm2.91$ & $29.30\pm1.31$ &M~1-12    &  $30.89\pm0.69$ & $23.76\pm0.44$\\
Hu~1-2    &  $39.33\pm2.91$ & $23.47\pm0.44$ &M~1-14    &  $33.10\pm0.69$ & $26.38\pm0.29$ \\
K~3-60    &  $42.52\pm0.14$ & $39.21\pm1.31$ & M~1-20    &  $27.28\pm0.14$ & $20.55\pm0.15$ \\
K~3-62    &  $35.87\pm0.14$ & $34.26\pm0.15$ &M~1-74    &  $39.47\pm0.14$ & $33.24\pm0.29$ \\
K~3-66    &  $47.23\pm0.83$ & $47.52\pm0.29$ &M~1-78    &  $49.72\pm0.14$ & $48.69\pm1.02$ \\
K~3-67    &  $44.60\pm0.69$ & $37.76\pm0.44$ &NGC~7027  &  $42.93\pm0.28$ & $41.25\pm0.15$ \\
K~4-48    &  $37.39\pm1.38$ & $31.78\pm2.92$ &PC~12      &  $33.79\pm0.14$ & $16.03\pm0.29$ \\
M~1-1     &  $53.32\pm6.92$ &  &SaSt~2-3  &  $26.87\pm1.38$ & $ 7.29\pm7.29$ \\
M~1-1     &  $45.70\pm2.46$ &  &Vy~1-1    &  $30.47\pm0.83$ & $25.66\pm0.73$ \\
\end{tabular}

\noindent{\bf Table 3:} Observed widths of the HI Br$\gamma$ and HeI
2$^1$P--2$^1$S lines.  For M~1-1 and BD+30$^\circ$3639, two component Gaussian
fits were required for the HI and HeI lines respectively.  The values in the
table are for the blue and red lines respectively.  Empty entries indicate that
an object was not observed with the echelle grating for that line, with the
exception of BD+30$^\circ$3639, where the HI line was well fit by a single
Gaussian.

\newpage

\begin{tabular}{lrrrr}
 & \multicolumn{2}{c}{HI} &  \multicolumn{2}{c}{HeI}\\
\multicolumn{1}{c}{Name} & \multicolumn{1}{c}{v$_{turb}$}
&\multicolumn{1}{c}{v$_{exp}$} &  \multicolumn{1}{c}{v$_{turb}$}
&\multicolumn{1}{c}{v$_{exp}$} \\
 & \multicolumn{1}{c}{(kms$^{-1}$)} & \multicolumn{1}{c}{(kms$^{-1}$)} &  
\multicolumn{1}{c}{(kms$^{-1}$)} & \multicolumn{1}{c}{(kms$^{-1}$)}\\
BD+30$^\circ$3639 & \multicolumn{4}{l}{see comments in text}\\
CRL~618 &\multicolumn{4}{l}{see comments in text} \\
DdDm~1 & 21 & 20 & 20 & 20 \\
Hu~1-2 & \multicolumn{4}{l}{see comments in text}\\
K~3-60 & 23 & 22 & 23 & 22 \\
K~3-62 & 18 & 18 & 20 & 18 \\
K~3-66 & 27 & 25 & 29 & 25 \\
K~3-67 & 25 & 23 & 23 & 21 \\
K~4-48 & 20 & 20 & 18 & 16 \\
M~1-1 & \multicolumn{4}{l}{see comments in text}\\
M~1-4 & 21 & 20 & 17 & 15 \\
M~1-6 & 14 & 14 & 9 & 9 \\
M~1-9 & 15 & 15 & 12 & 12 \\
M~1-11 & 13 & 13 & 5 & 5 \\
M~1-12 & 15 & 15 & 14 & 14 \\
M~1-14 & 18 & 18 & 15 & 15 \\
M~1-20 & 13 & 13 & 11 & 11 \\
M~1-74 & 28 & 25 & 28 & 25 \\
M~1-78 & 27 & 25 & 27 & 25 \\
NGC~7027 & 23 & 22 & 23 & 22 \\
PC~12 & 17 & 17 & 8 & 8 \\
SaSt~2-3 & 15 & 15 & 0 & 0 \\
Vy~1-1 & 15 & 15 & 15 & 15 \\
\end{tabular}

\noindent{\bf Table 3:} Derived upper limits on the turbulent and expansion
velocity components from our data.

\newpage

\begin{center}
\begin{minipage}{3.5in}{
\psfig{file=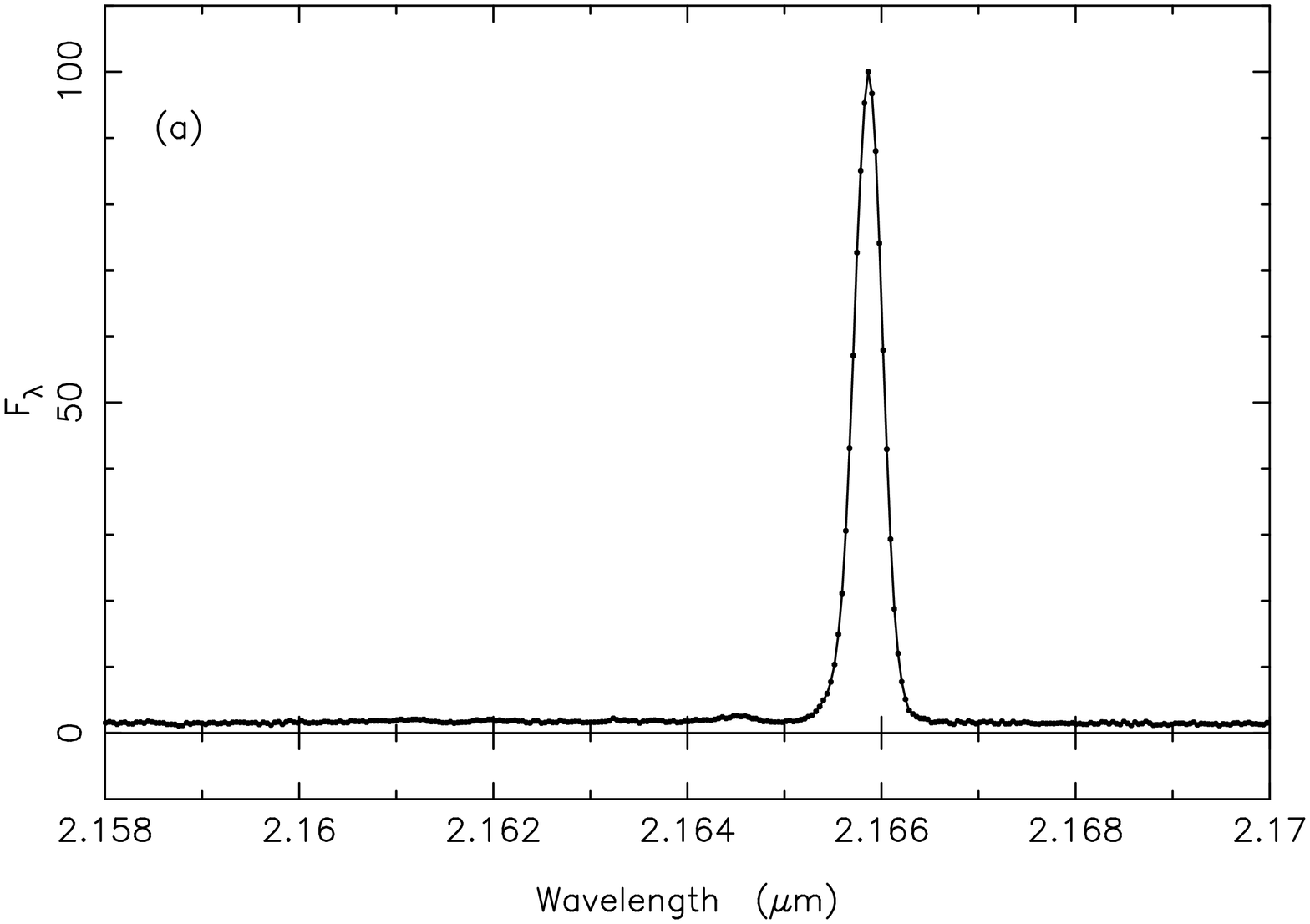,width=3in,angle=-90,clip=}
}\end{minipage}
\begin{minipage}{3.5in}{
\psfig{file=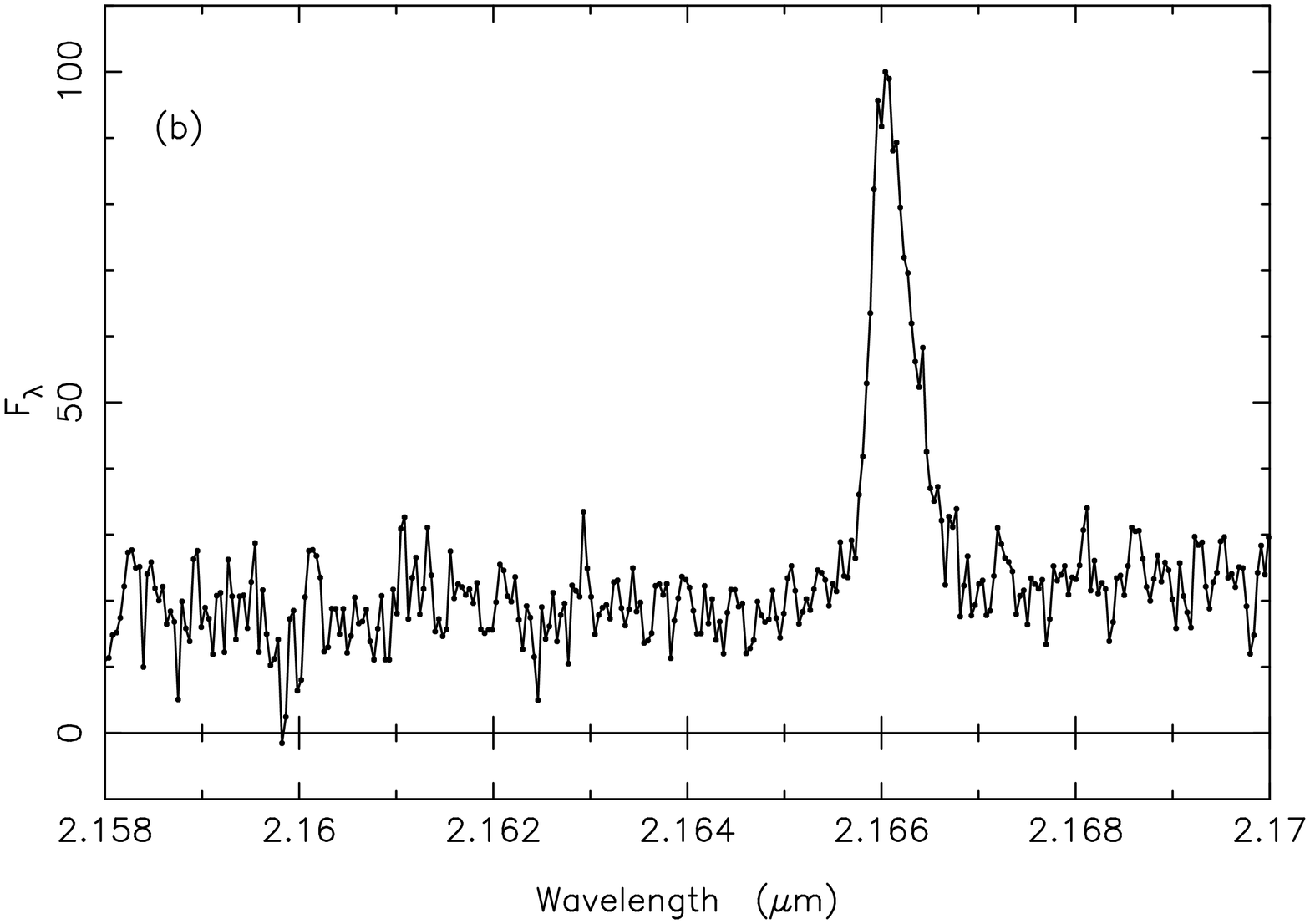,width=3in,angle=-90,clip=}
}\end{minipage}
\vspace*{-2mm}

\begin{minipage}{3.5in}{
\psfig{file=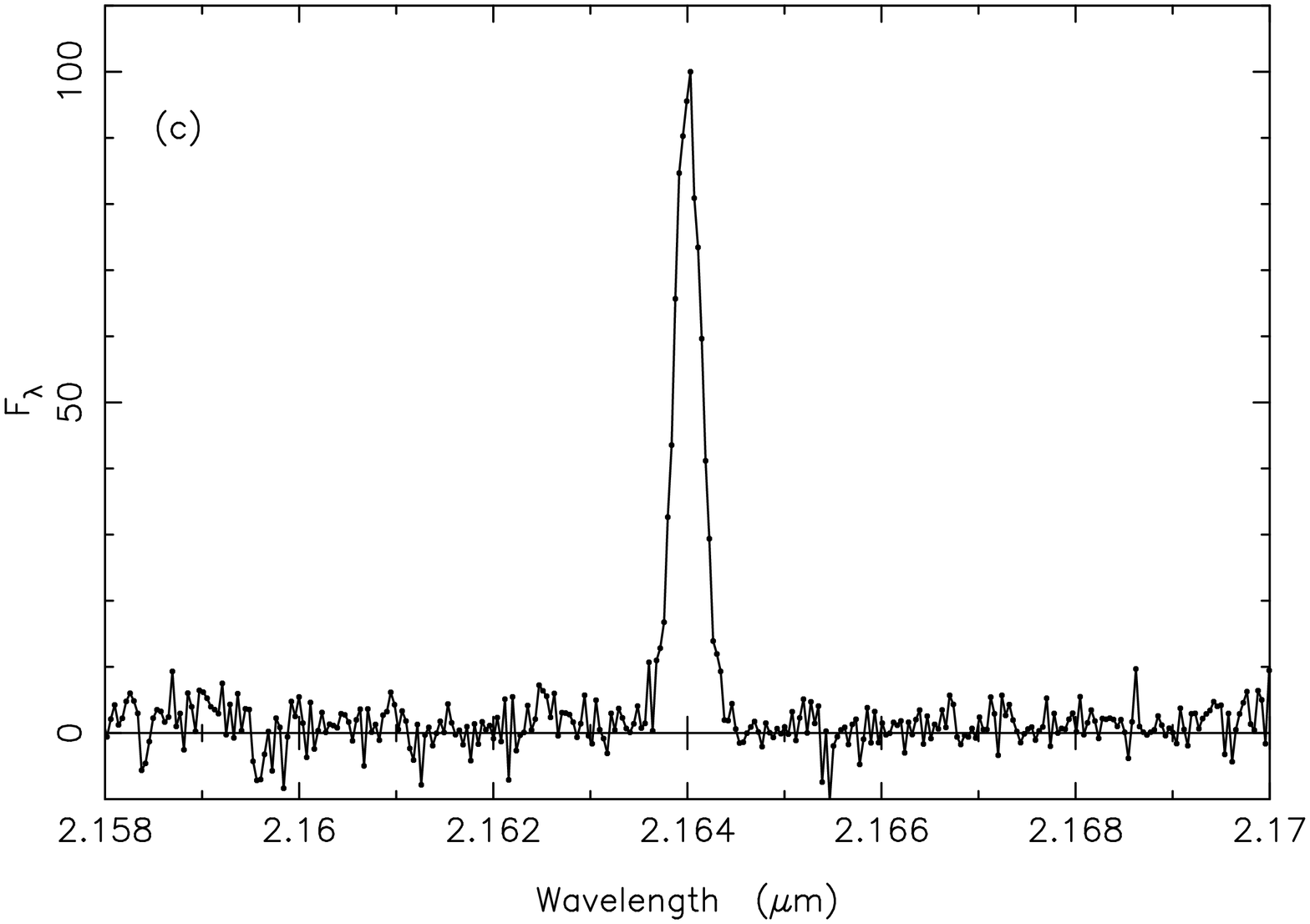,width=3in,angle=-90,clip=}
}\end{minipage}
\begin{minipage}{3.5in}{
\psfig{file=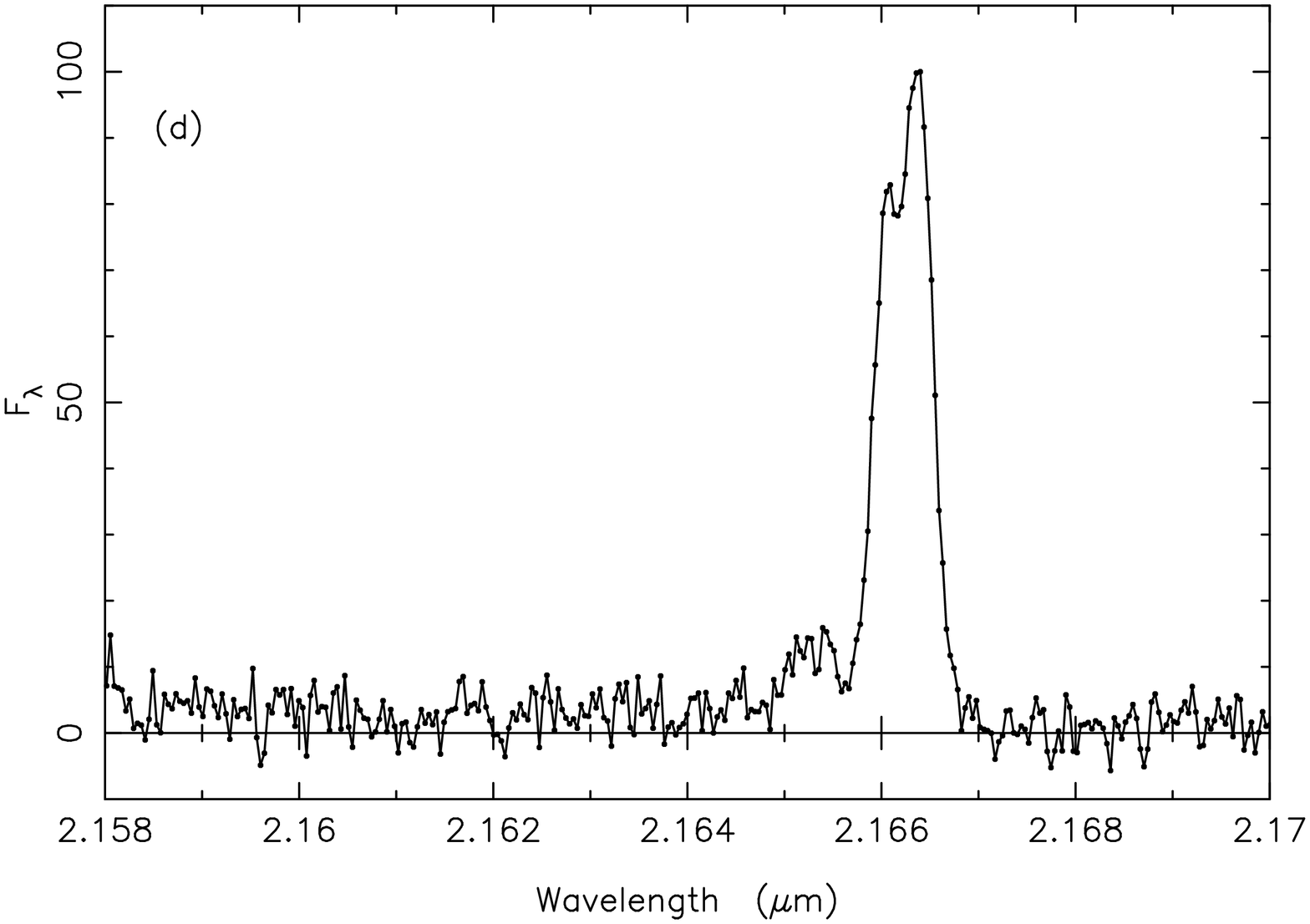,width=3in,angle=-90,clip=}
}\end{minipage}
\vspace*{-2mm}

\begin{minipage}{3.5in}{
\psfig{file=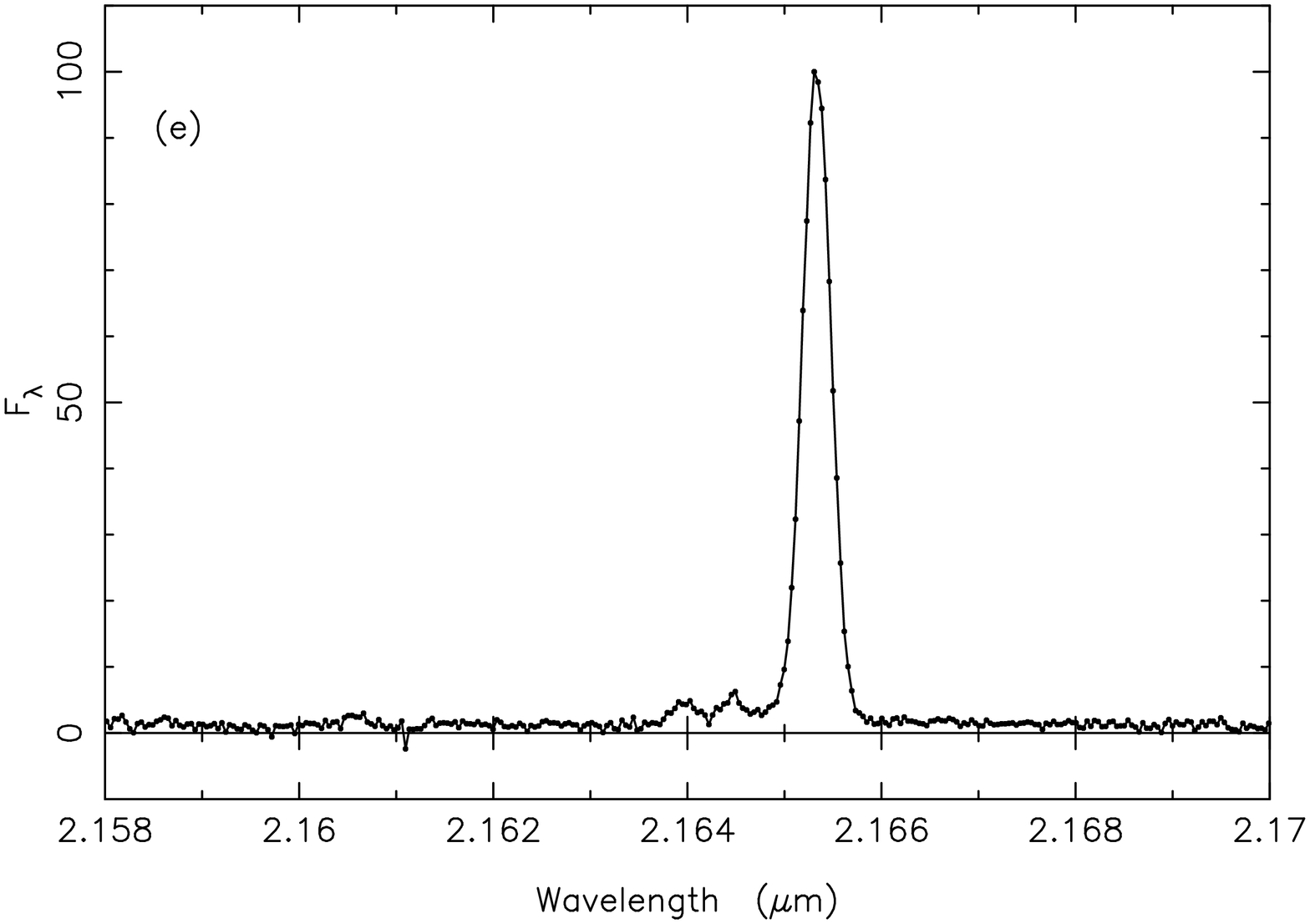,width=3in,angle=-90,clip=}
}\end{minipage}
\begin{minipage}{3.5in}{
\psfig{file=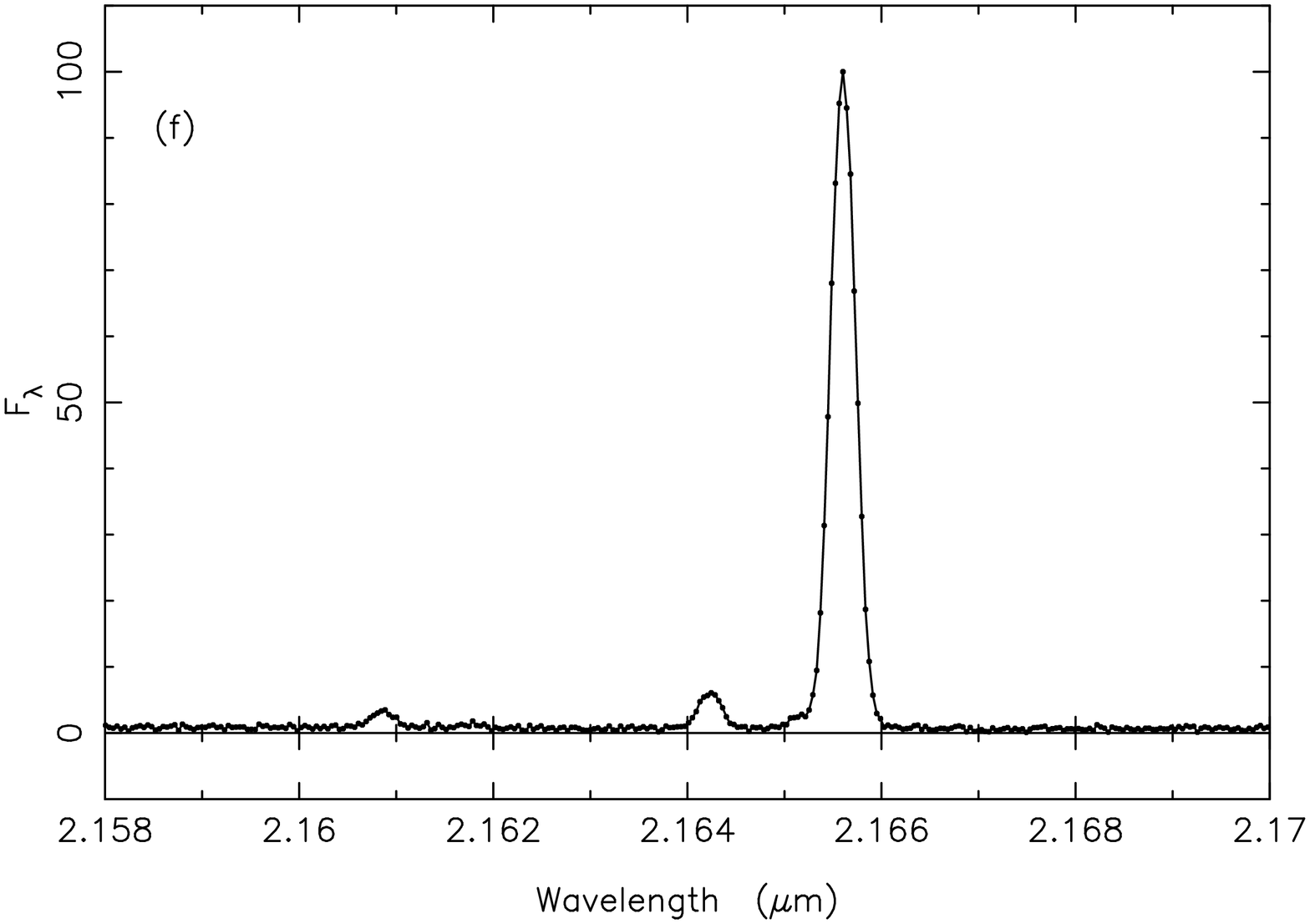,width=3in,angle=-90,clip=}
}\end{minipage}
\vspace*{-2mm}

\begin{minipage}{3.5in}{
\psfig{file=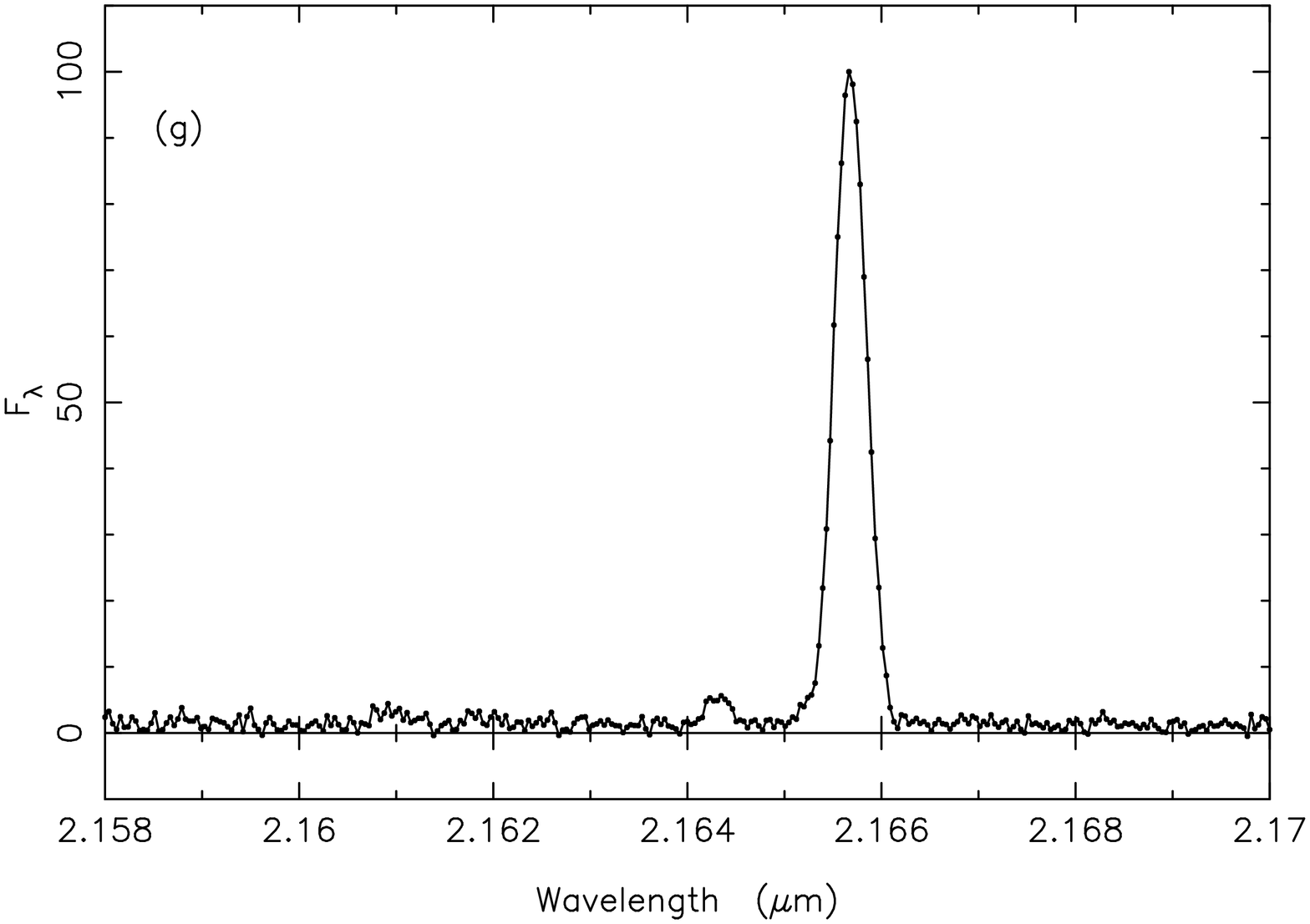,width=3in,angle=-90,clip=}
}\end{minipage}
\begin{minipage}{3.5in}{
\psfig{file=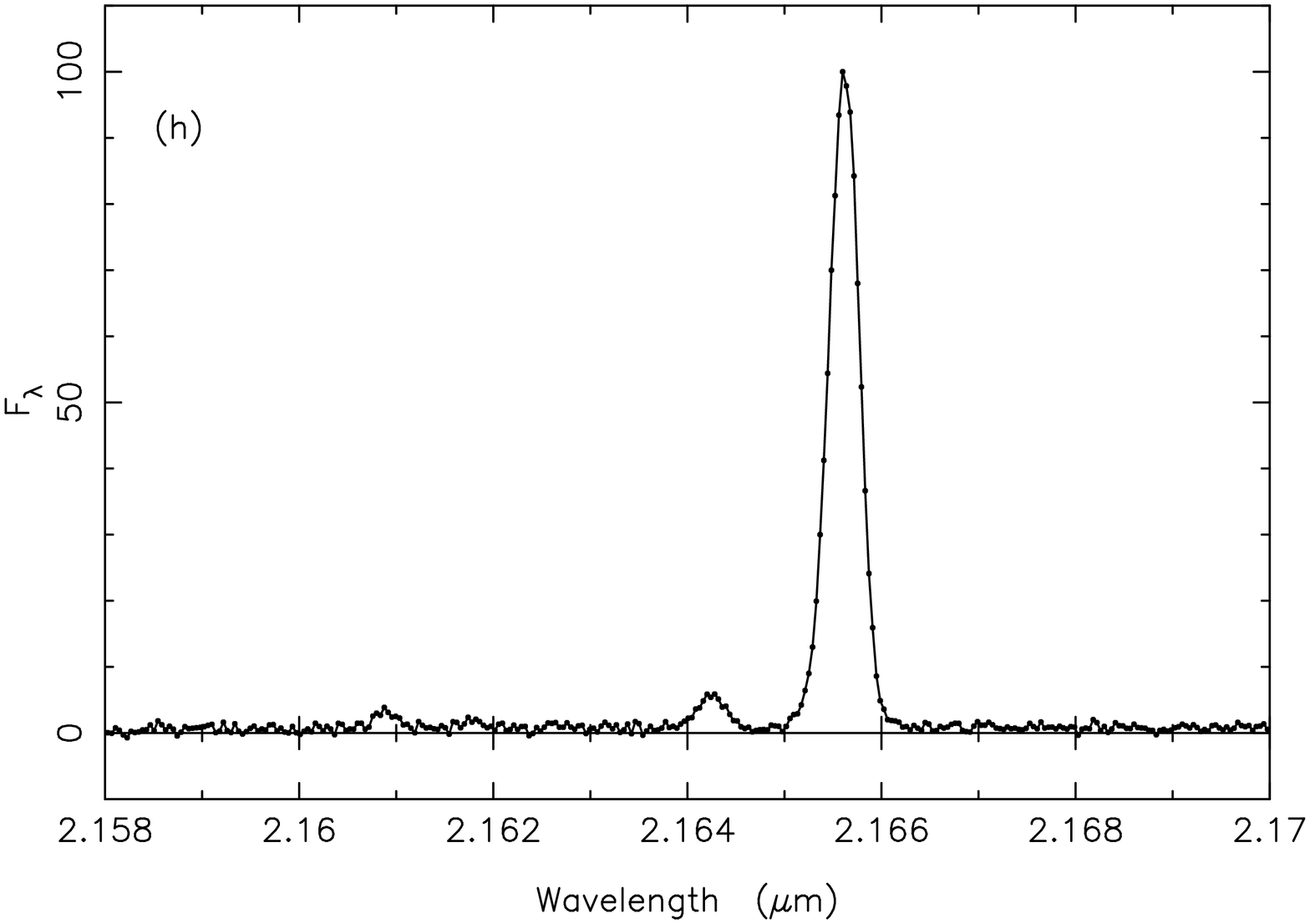,width=3in,angle=-90,clip=}
}\end{minipage}
\vspace*{-2mm}

\end{center}

\begin{center}

\begin{minipage}{3.5in}{
\psfig{file=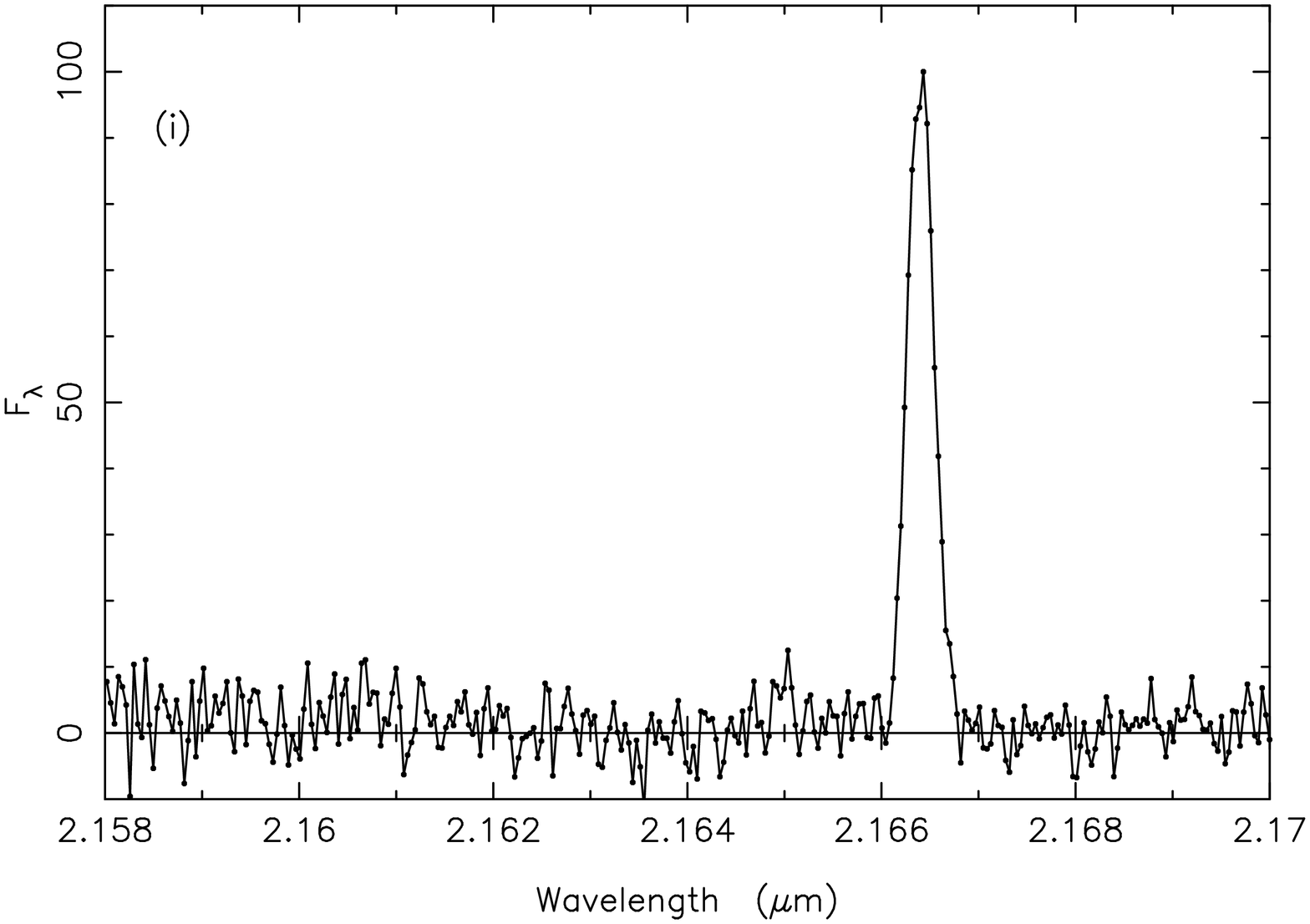,width=3in,angle=-90,clip=}
}\end{minipage}
\begin{minipage}{3.5in}{
\psfig{file=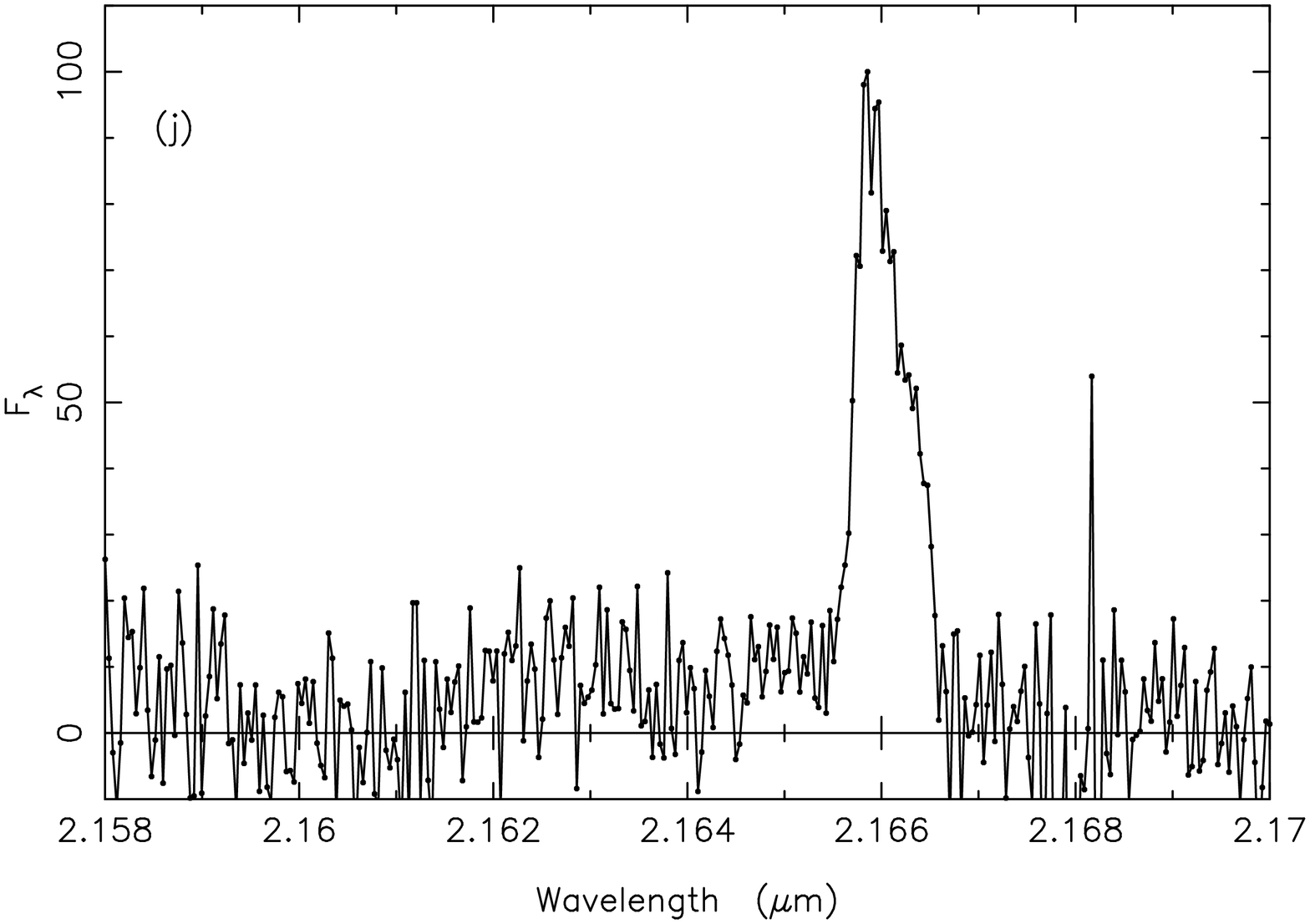,width=3in,angle=-90,clip=}
}\end{minipage}
\vspace*{-2mm}

\begin{minipage}{3.5in}{
\psfig{file=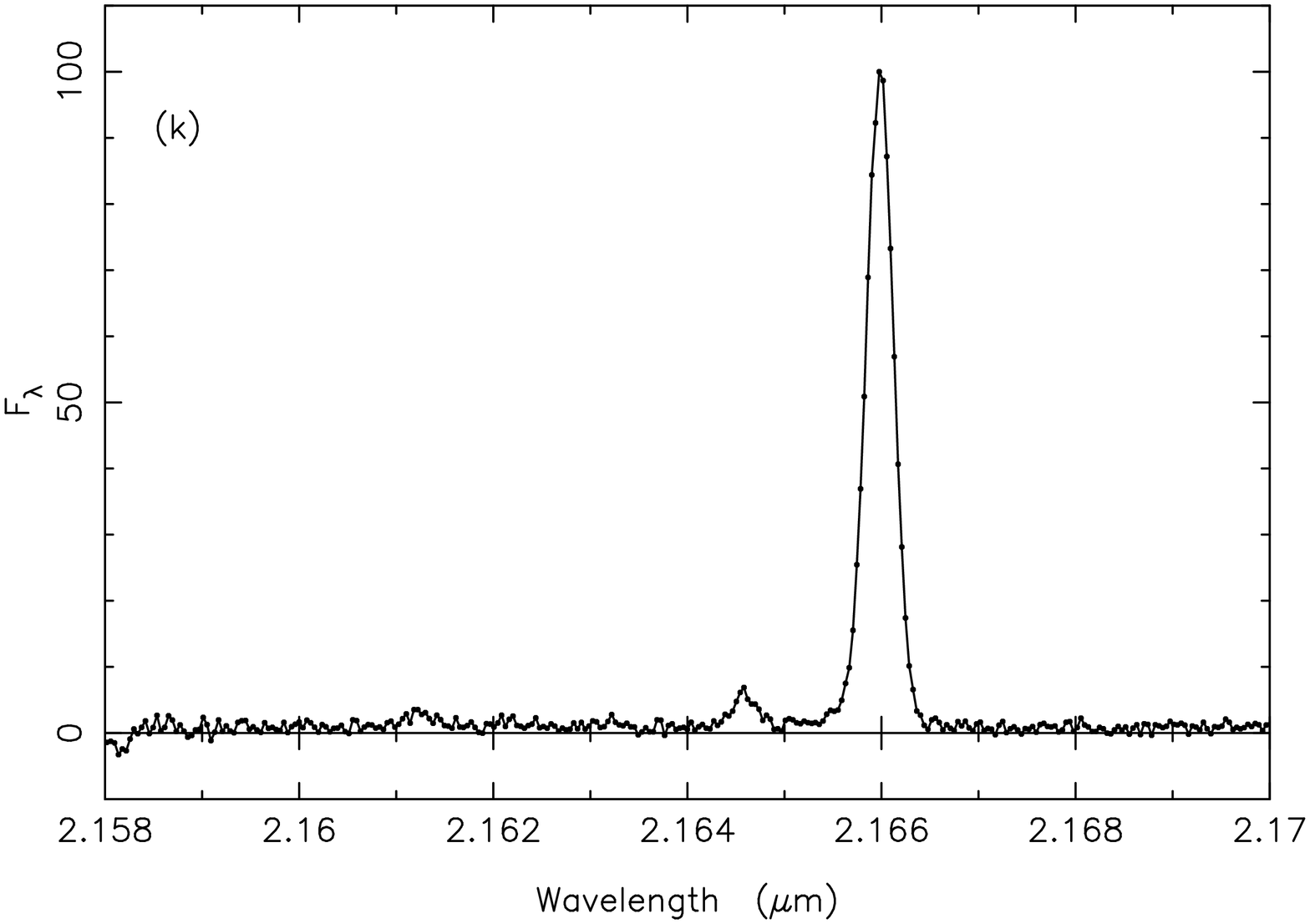,width=3in,angle=-90,clip=}
}\end{minipage}
\begin{minipage}{3.5in}{
\psfig{file=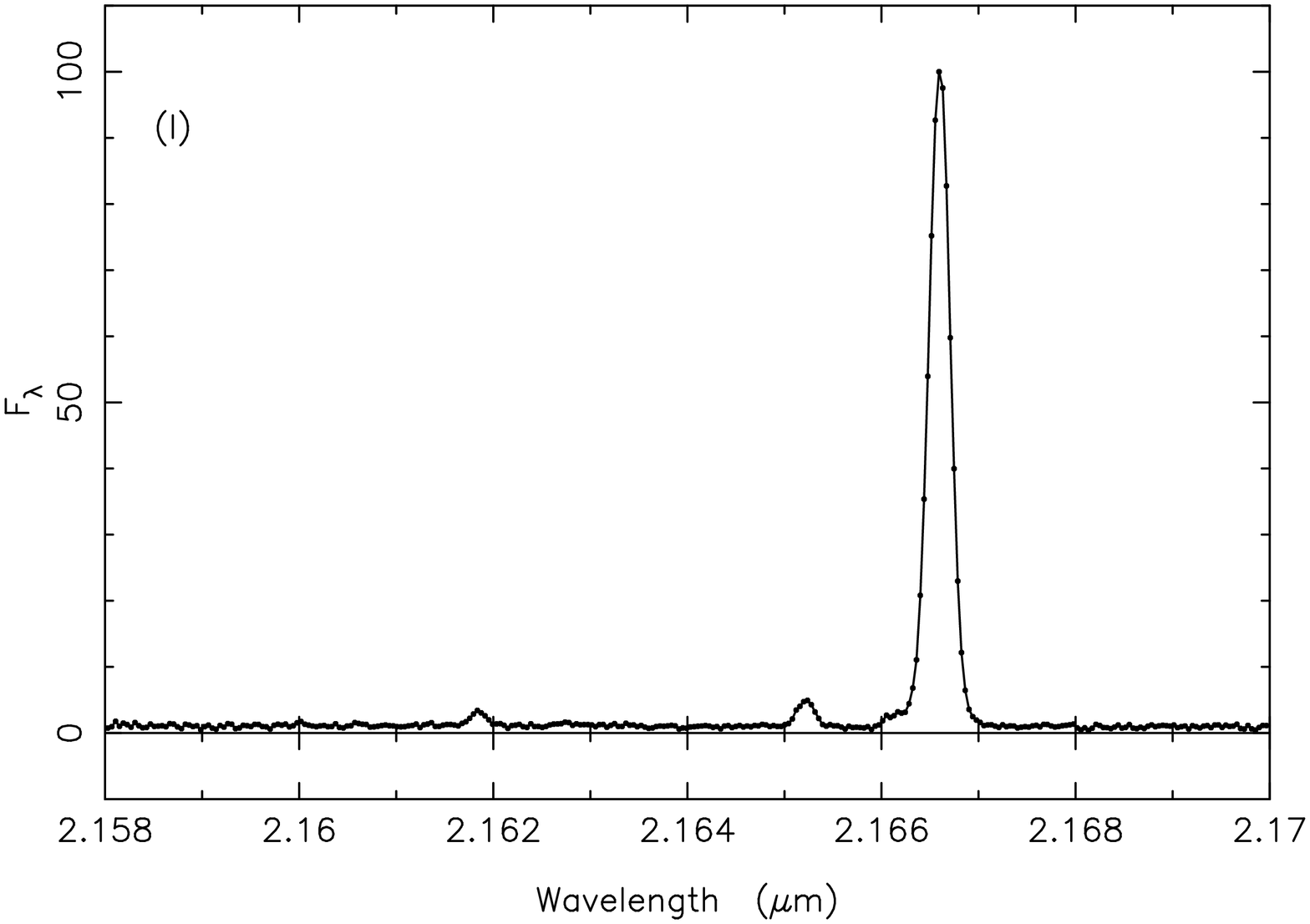,width=3in,angle=-90,clip=}
}\end{minipage}
\vspace*{-2mm}

\begin{minipage}{3.5in}{
\psfig{file=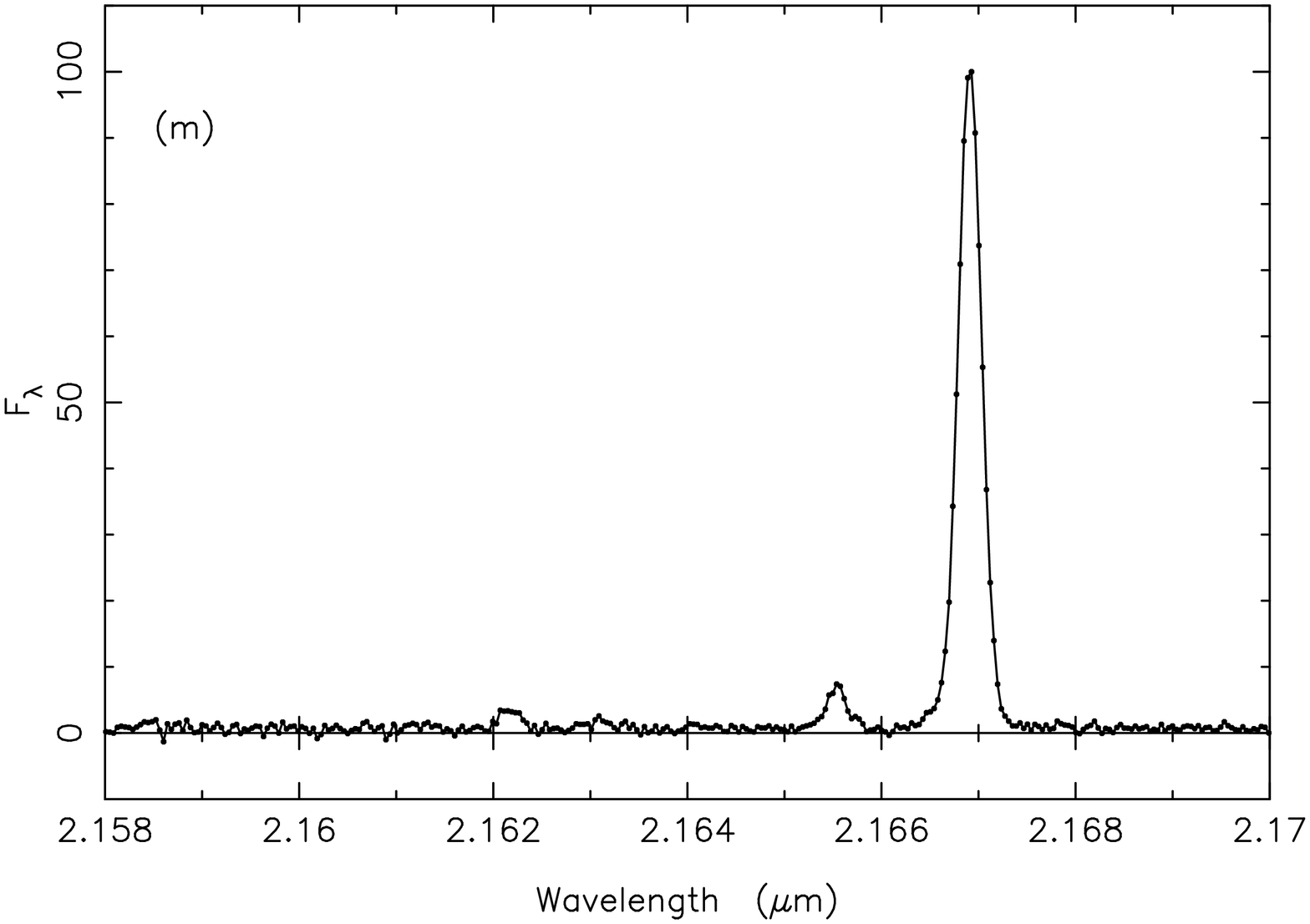,width=3in,angle=-90,clip=}
}\end{minipage}
\begin{minipage}{3.5in}{
\psfig{file=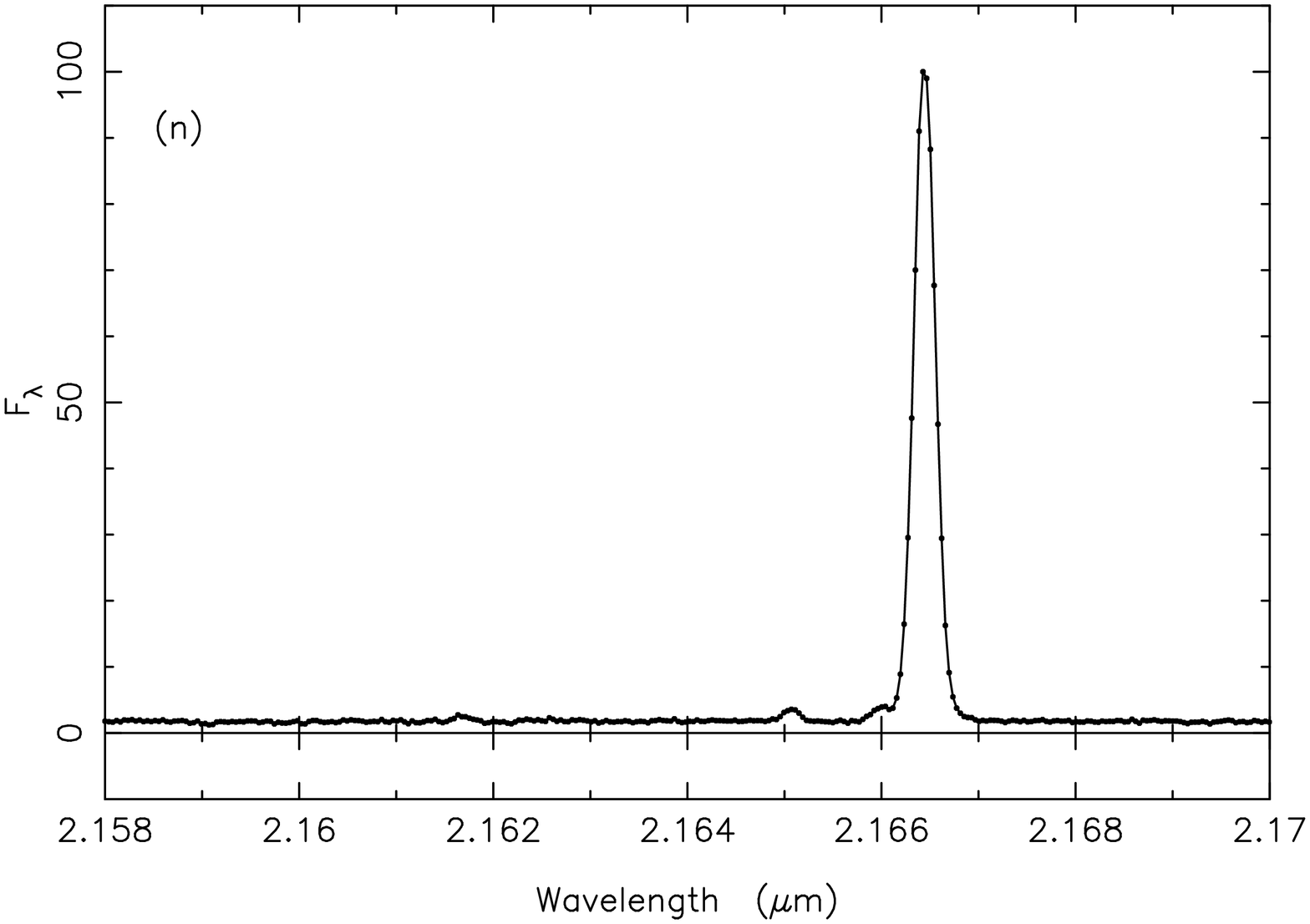,width=3in,angle=-90,clip=}
}\end{minipage}
\vspace*{-2mm}

\begin{minipage}{3.5in}{
\psfig{file=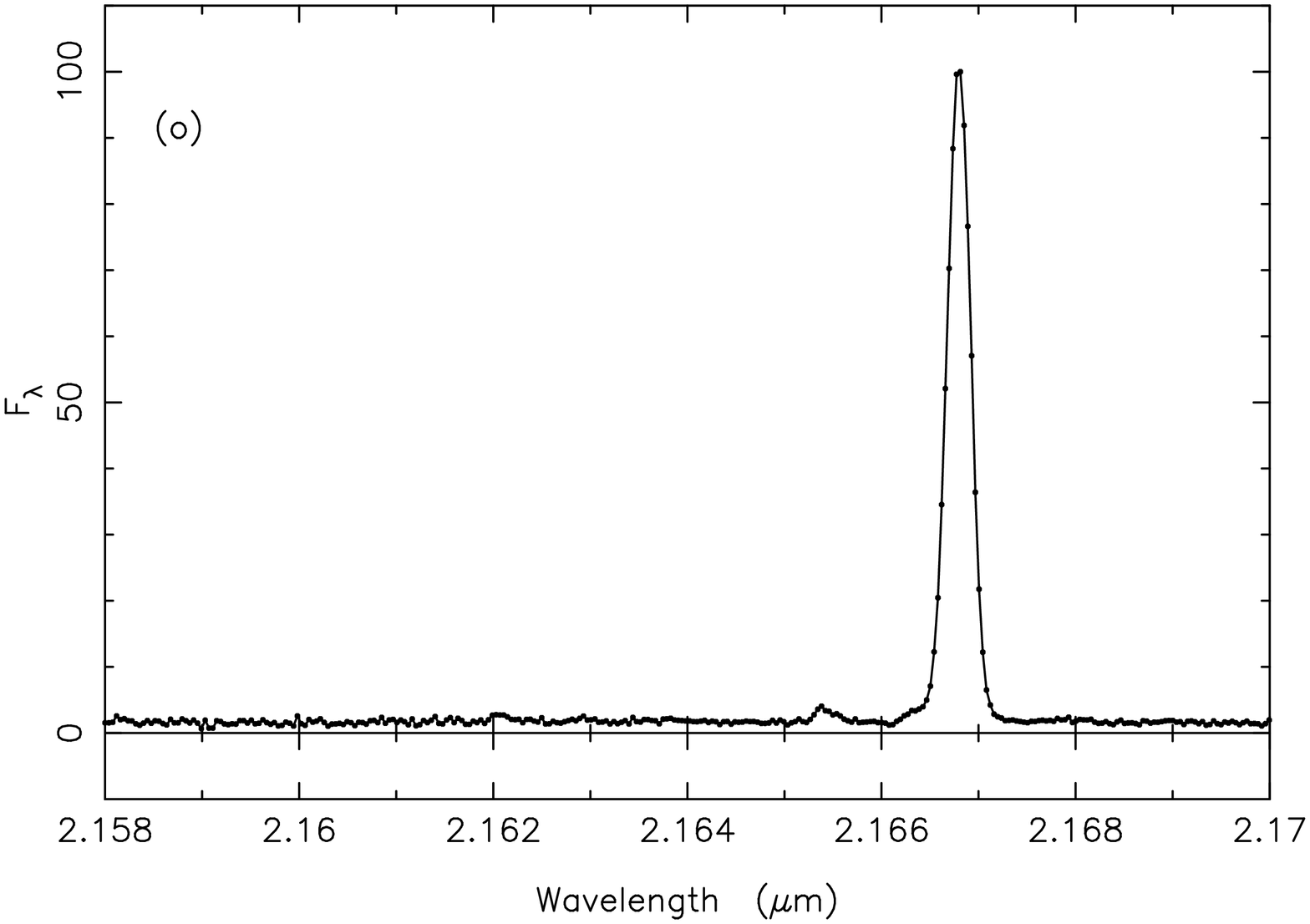,width=3in,angle=-90,clip=}
}\end{minipage}
\begin{minipage}{3.5in}{
\psfig{file=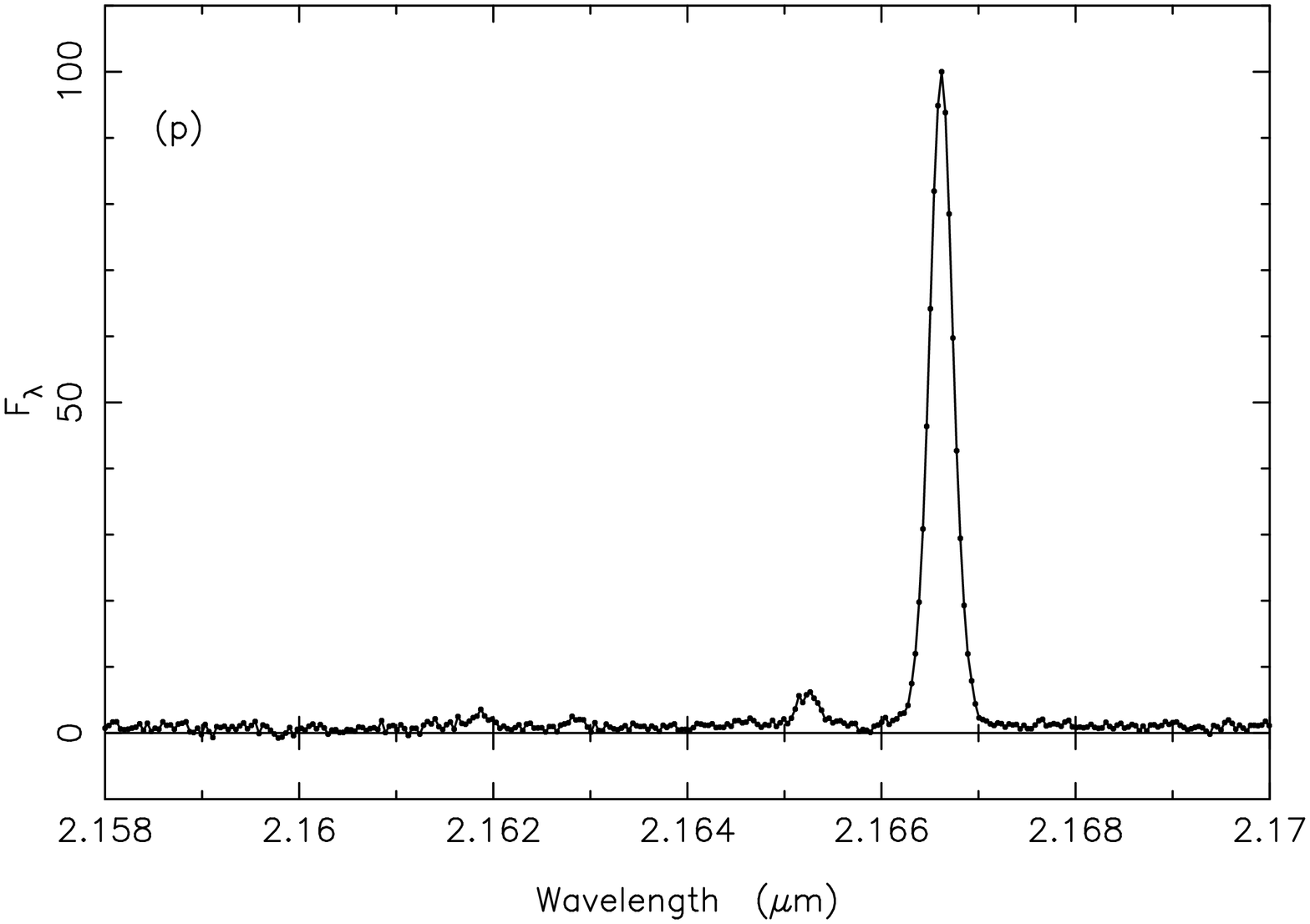,width=3in,angle=-90,clip=}
}\end{minipage}
\vspace*{-2mm}

\end{center}

\begin{center}

\begin{minipage}{3.5in}{
\psfig{file=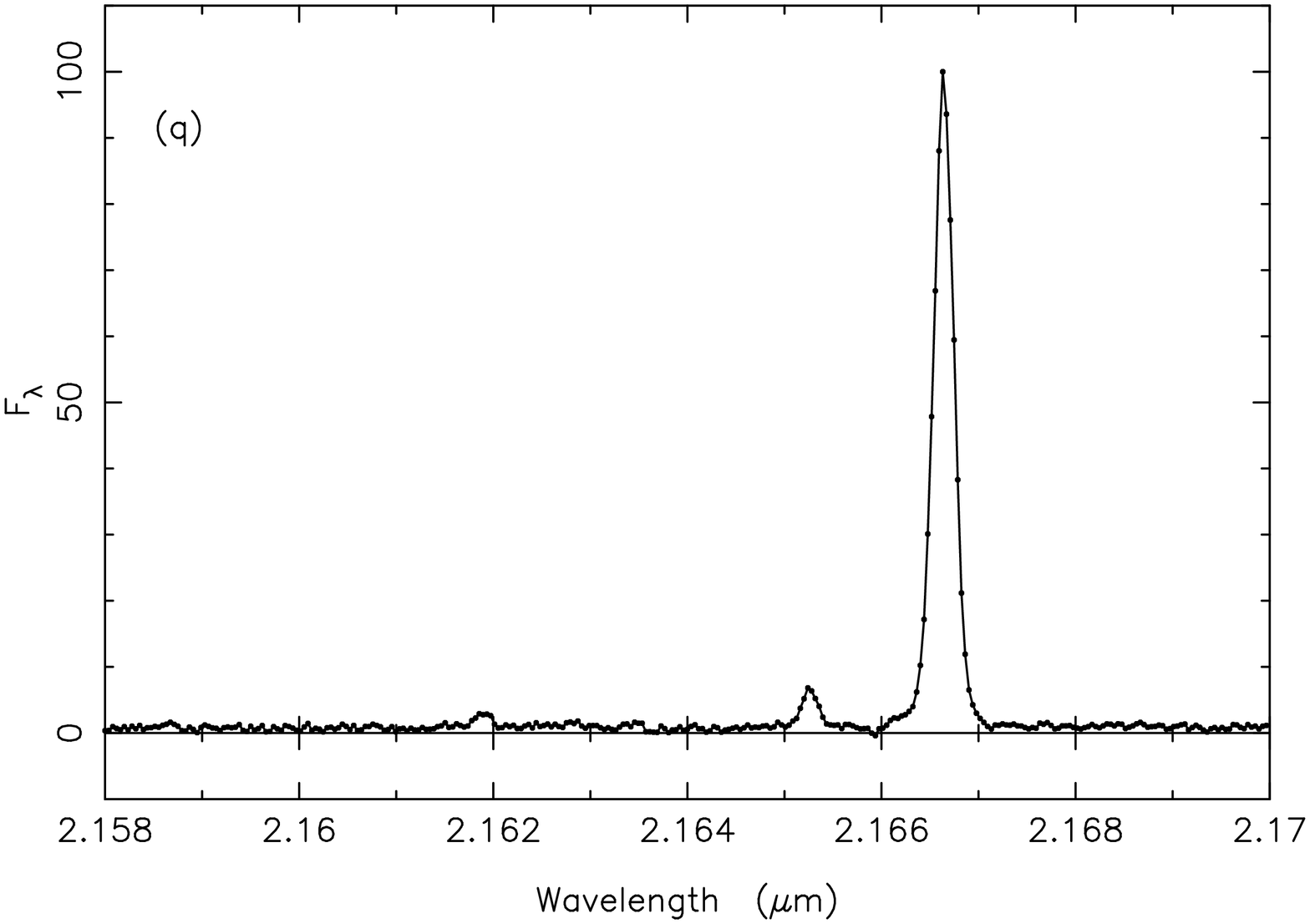,width=3in,angle=-90,clip=}
}\end{minipage}
\begin{minipage}{3.5in}{
\psfig{file=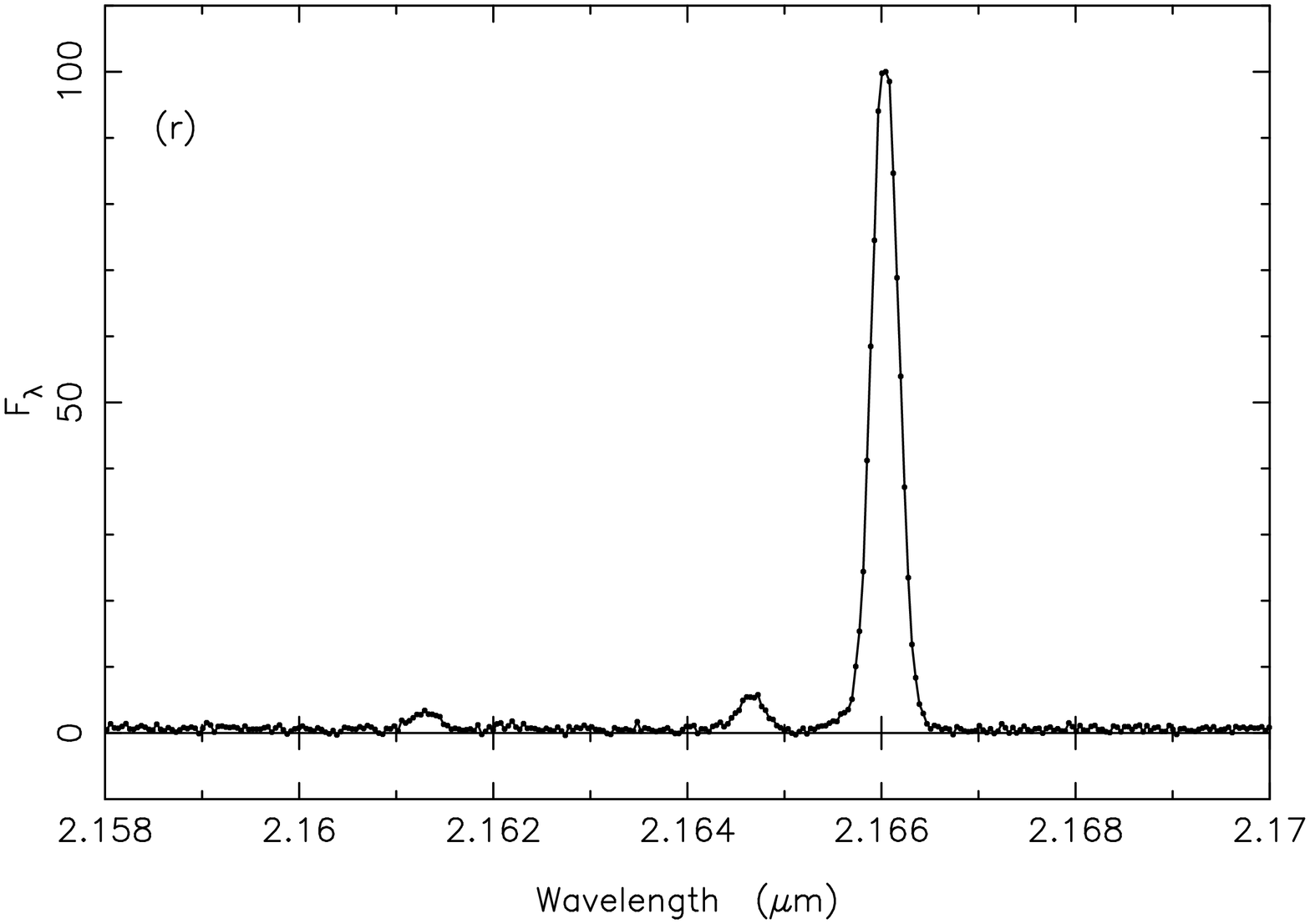,width=3in,angle=-90,clip=}
}\end{minipage}
\vspace*{-2mm}

\begin{minipage}{3.5in}{
\psfig{file=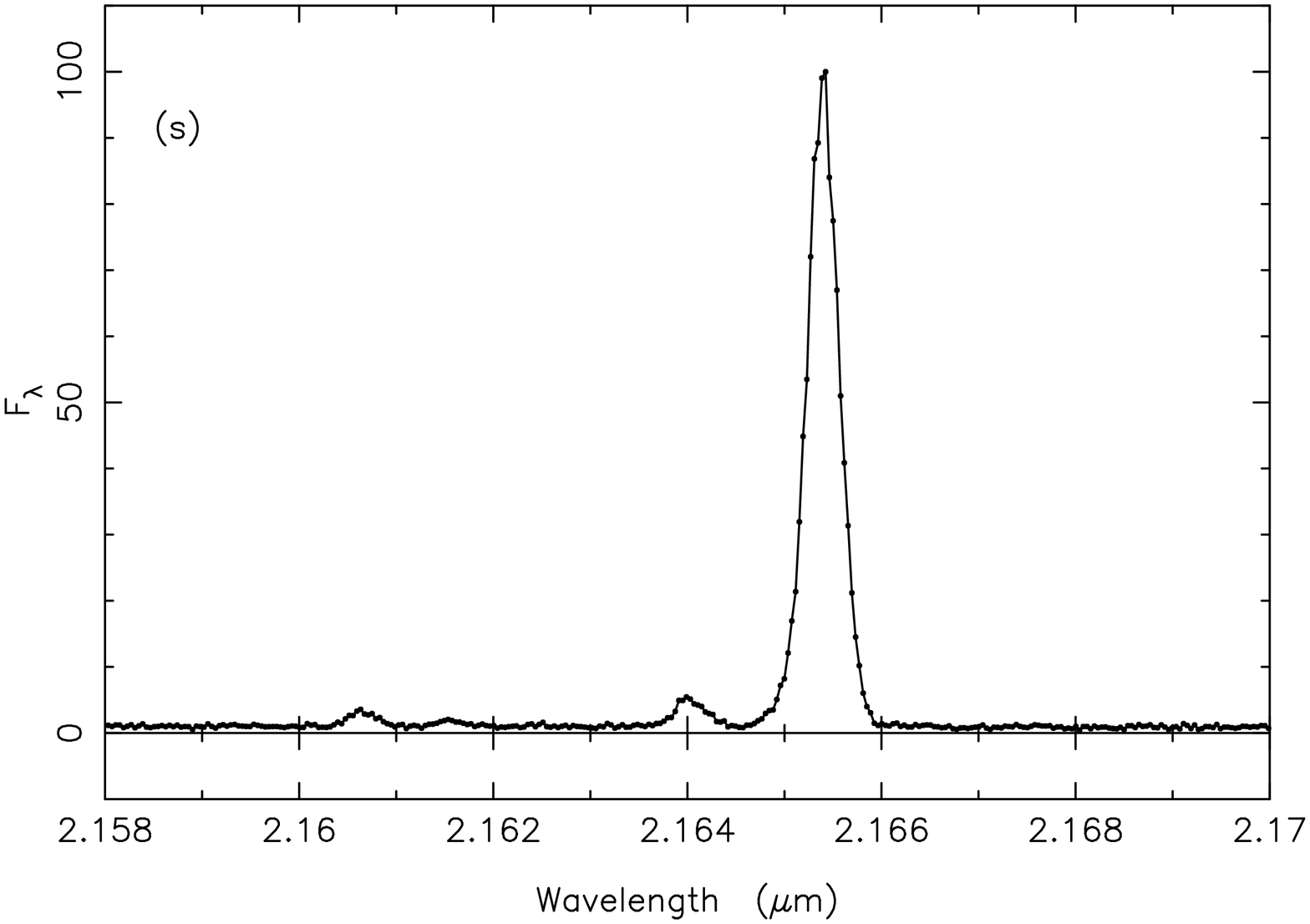,width=3in,angle=-90,clip=}
}\end{minipage}
\begin{minipage}{3.5in}{
\psfig{file=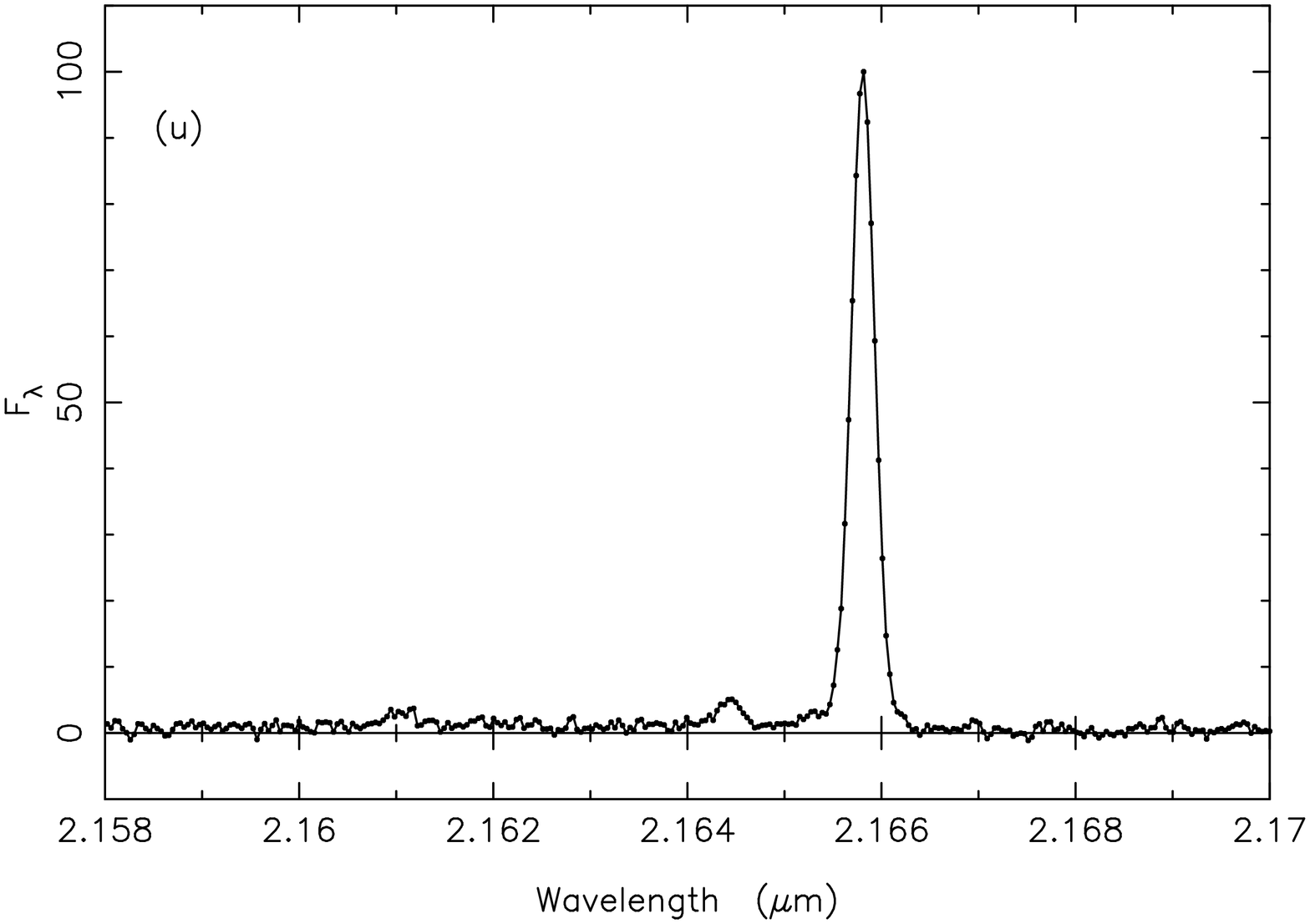,width=3in,angle=-90,clip=}
}\end{minipage}
\vspace*{-2mm}

\begin{minipage}{3.5in}{
\psfig{file=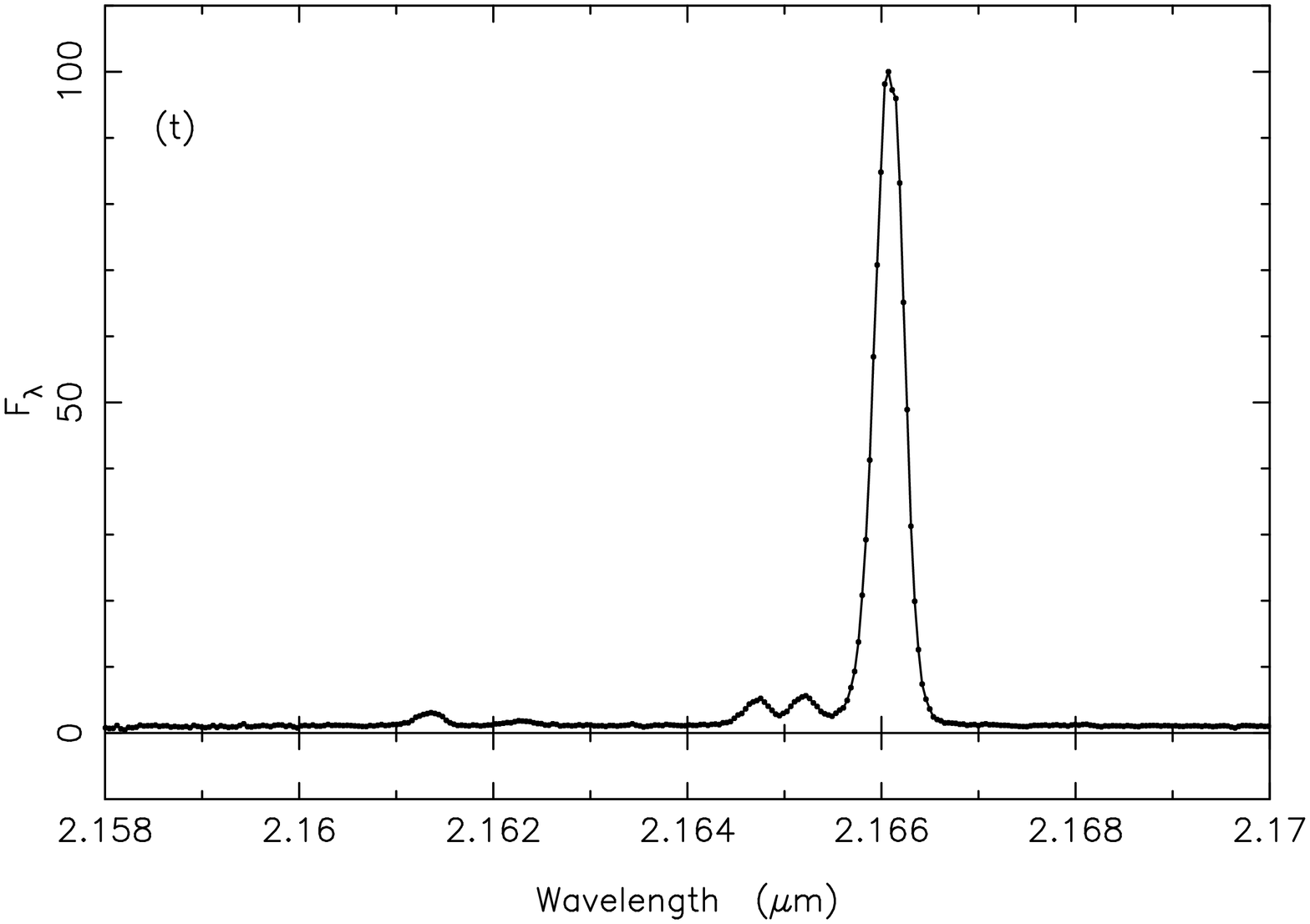,width=3in,angle=-90,clip=}
}\end{minipage}
\begin{minipage}{3.5in}{
\psfig{file=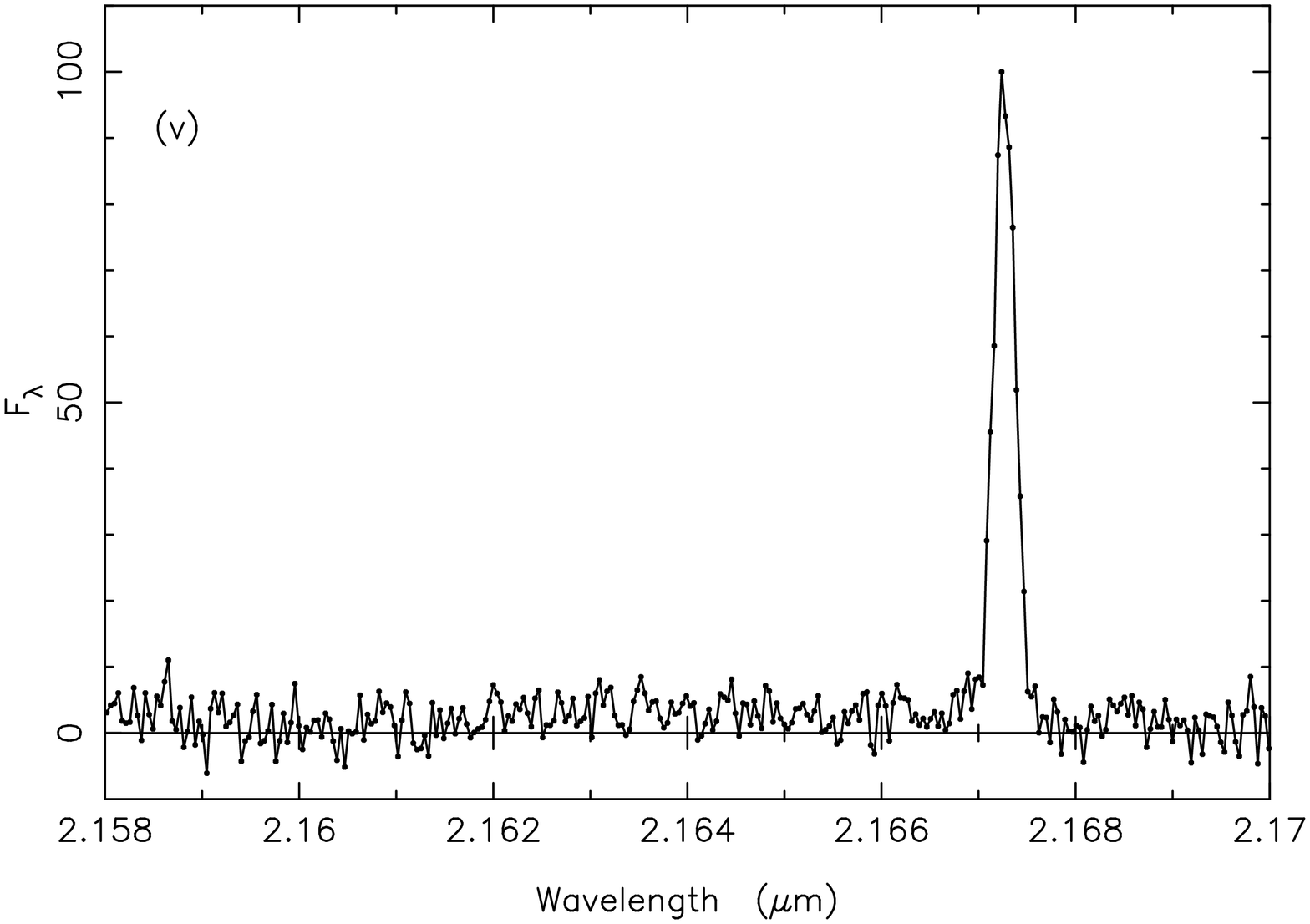,width=3in,angle=-90,clip=}
}\end{minipage}
\vspace*{-2mm}

\begin{minipage}{3.5in}{
\psfig{file=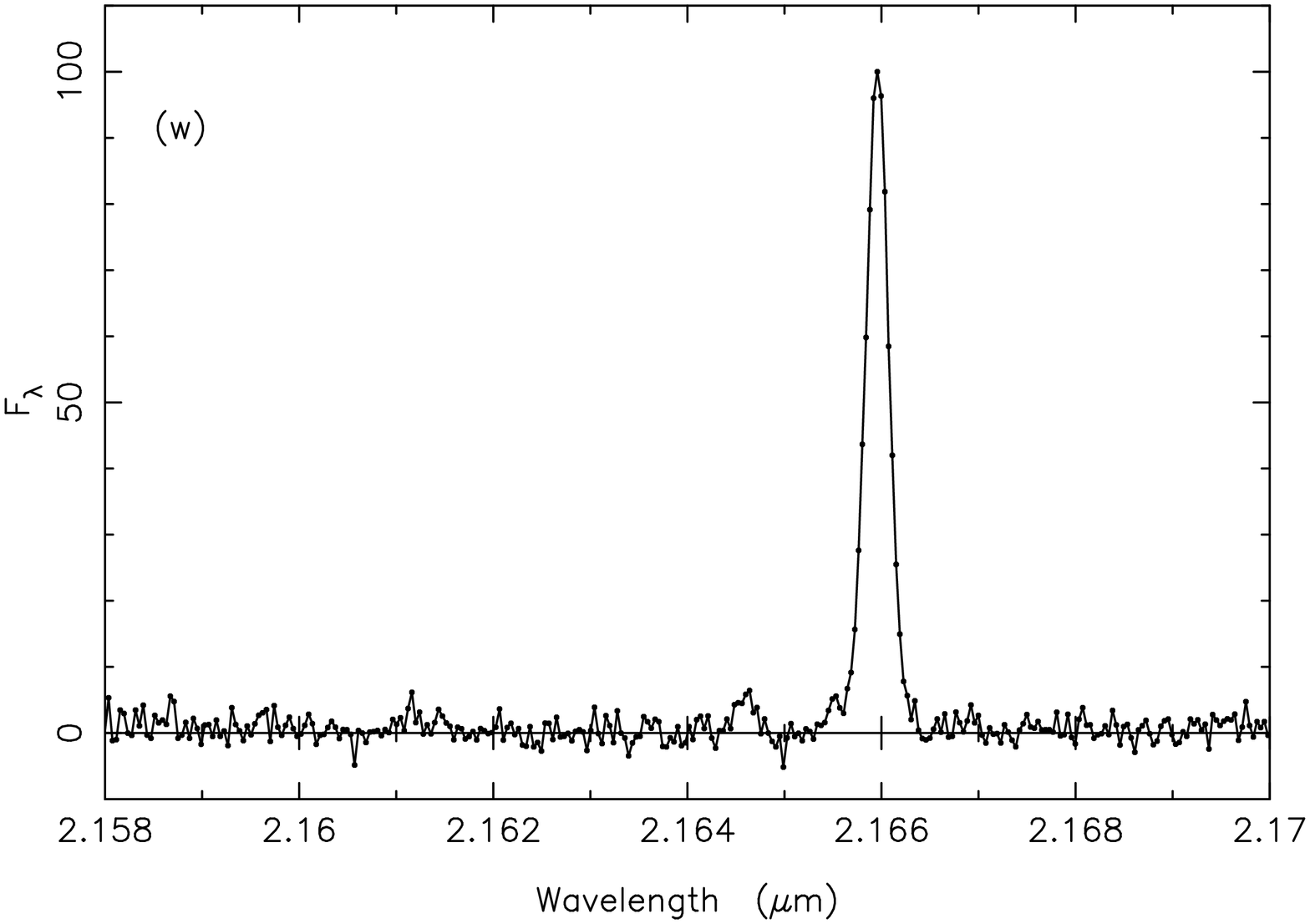,width=3in,angle=-90,clip=}
}\end{minipage}
\vspace*{-2mm}

\end{center}

{\bf Figure 1:} The HI Br$\gamma$ echelle spectra. The most common of
the satelite lines is the 2.16475$\mu$m HeI 7$^{1,3}$G--4$^{1,3}$F blend.
Also visible in some of the spectra are the other HeI
transitions of 7$^{1}$F--4$^{1}$D at 2.16229$\mu$m and 
7$^{3}$F--4$^{3}$D at 2.16137$\mu$m, and the HeII 14--8 transition at
2.1653$\mu$m.  
The spectra are of: (a) BD+303639; (b) CRL 618; (c) DdDm~1; (d) Hu~1-2; 
(e) K~3-60; (f) K~3-62; (g) K~3-66; (h) K~3-67; (i) K 4-48; 
(j) M~1-1; (k) M~1-4; (l) M~1-6; (m) M~1-9; (n) M~1-11;
(o) M~1-12; (p) M~1-14; (r) M~1-74; (s) M~1-78;
(t) NGC 7027; (u) PC~12; (v) SaSt~2-3; (w) Vy~1-1.   

\newpage

\begin{center}
\begin{minipage}{3.5in}{
\psfig{file=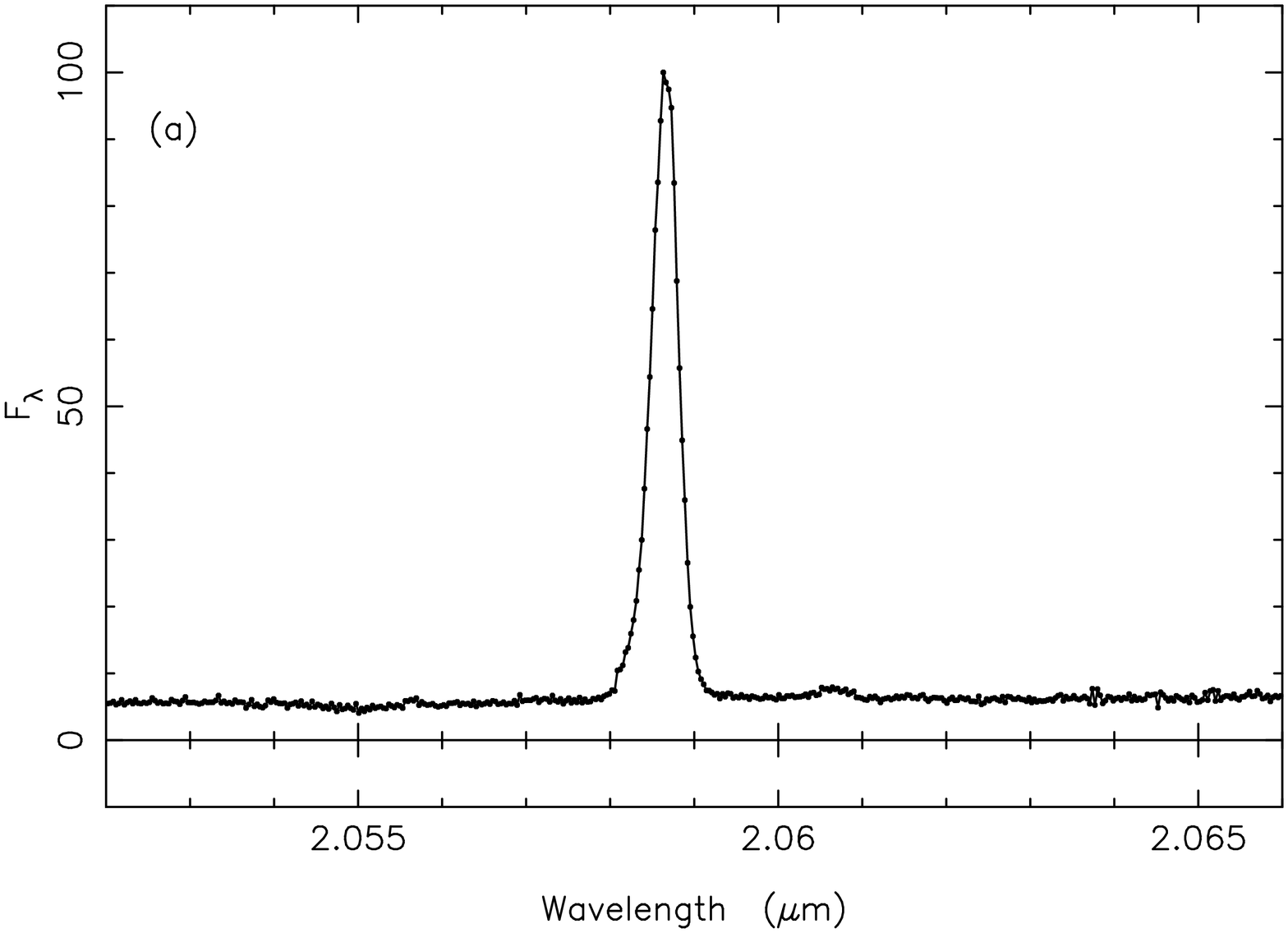,width=3in,angle=-90,clip=}
}\end{minipage}
\begin{minipage}{3.5in}{
\psfig{file=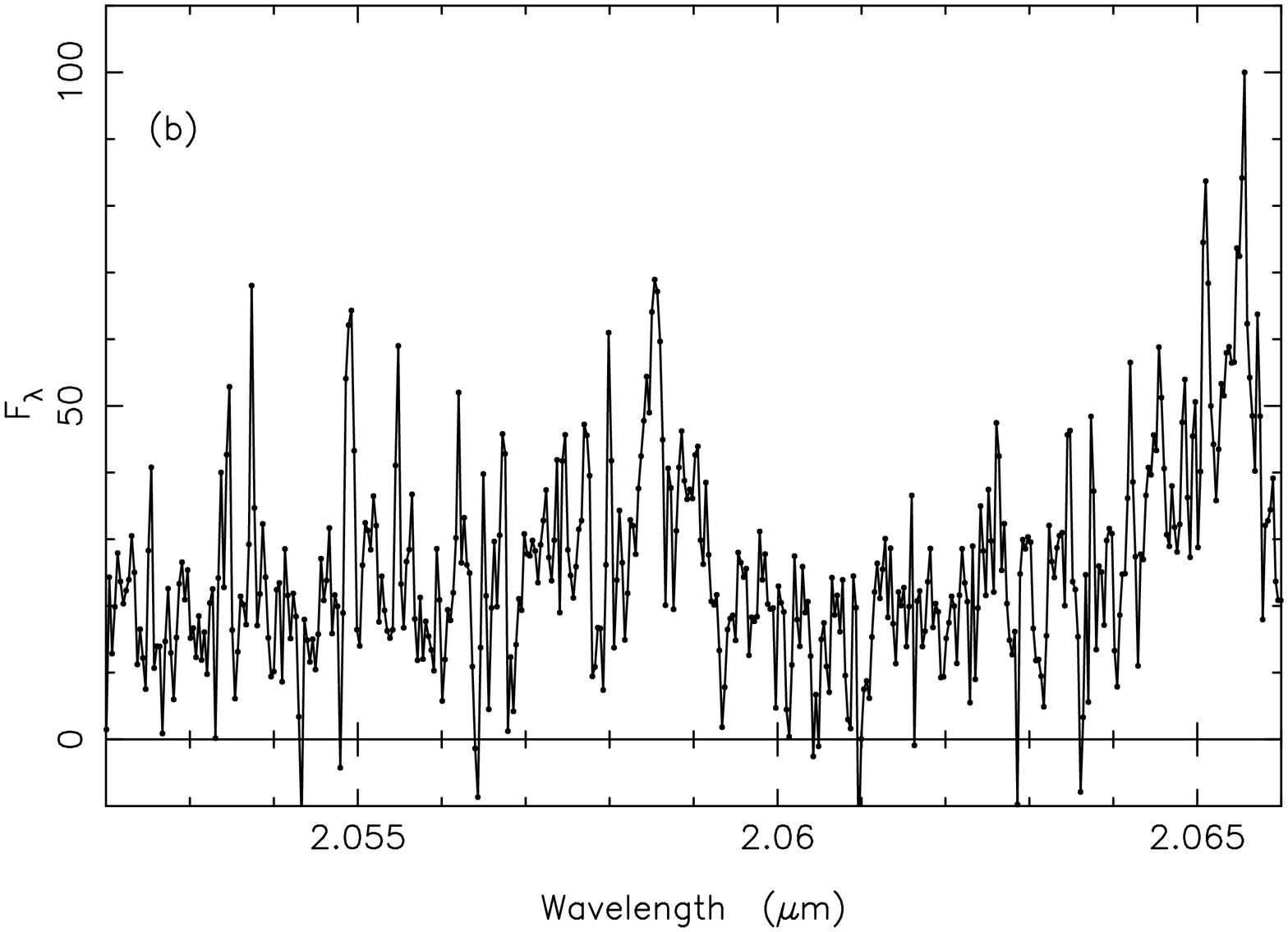,width=3in,angle=-90,clip=}
}\end{minipage}
\vspace*{-2mm}

\begin{minipage}{3.5in}{
\psfig{file=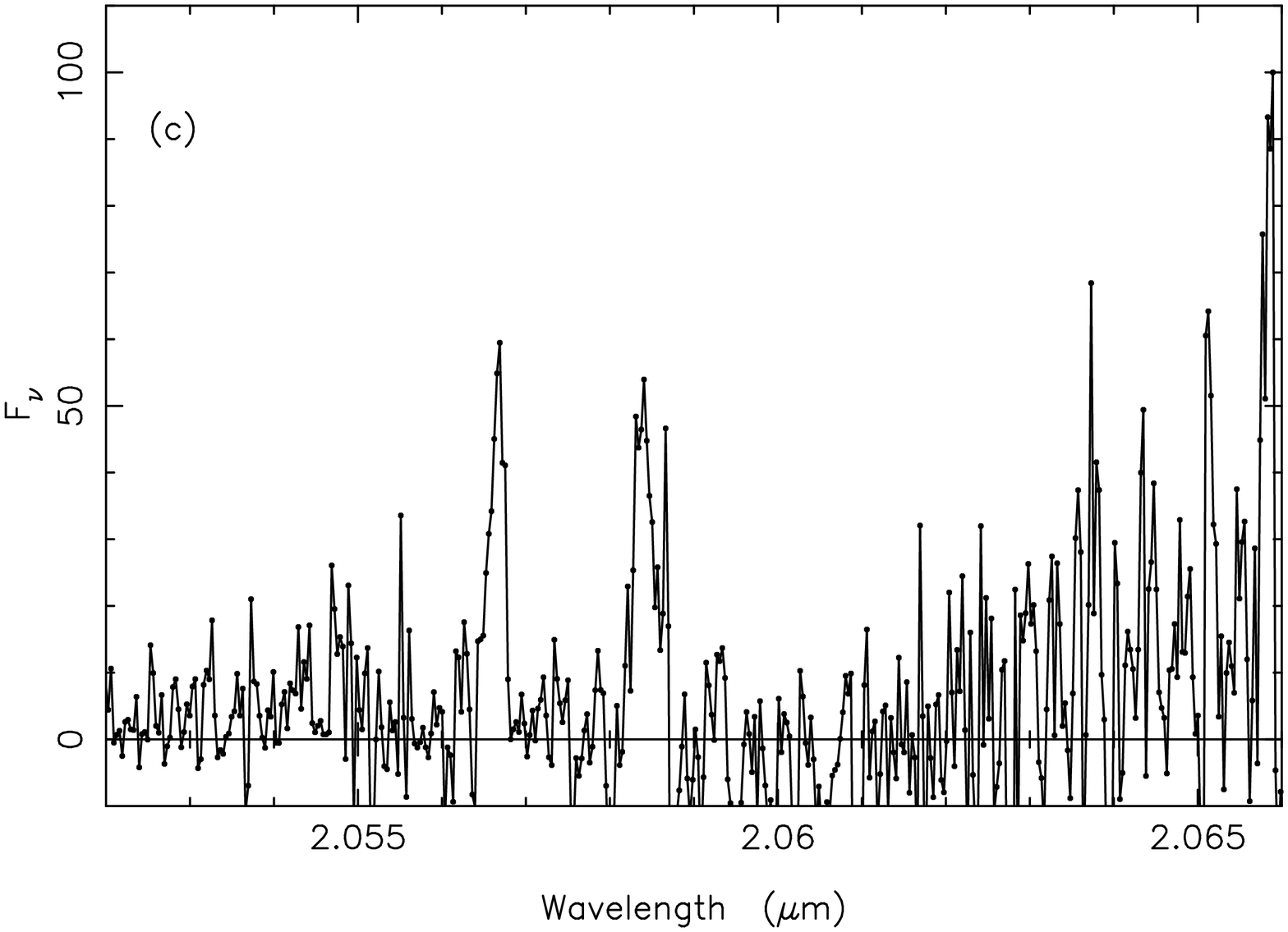,width=3in,angle=-90,clip=}
}\end{minipage}
\begin{minipage}{3.5in}{
\psfig{file=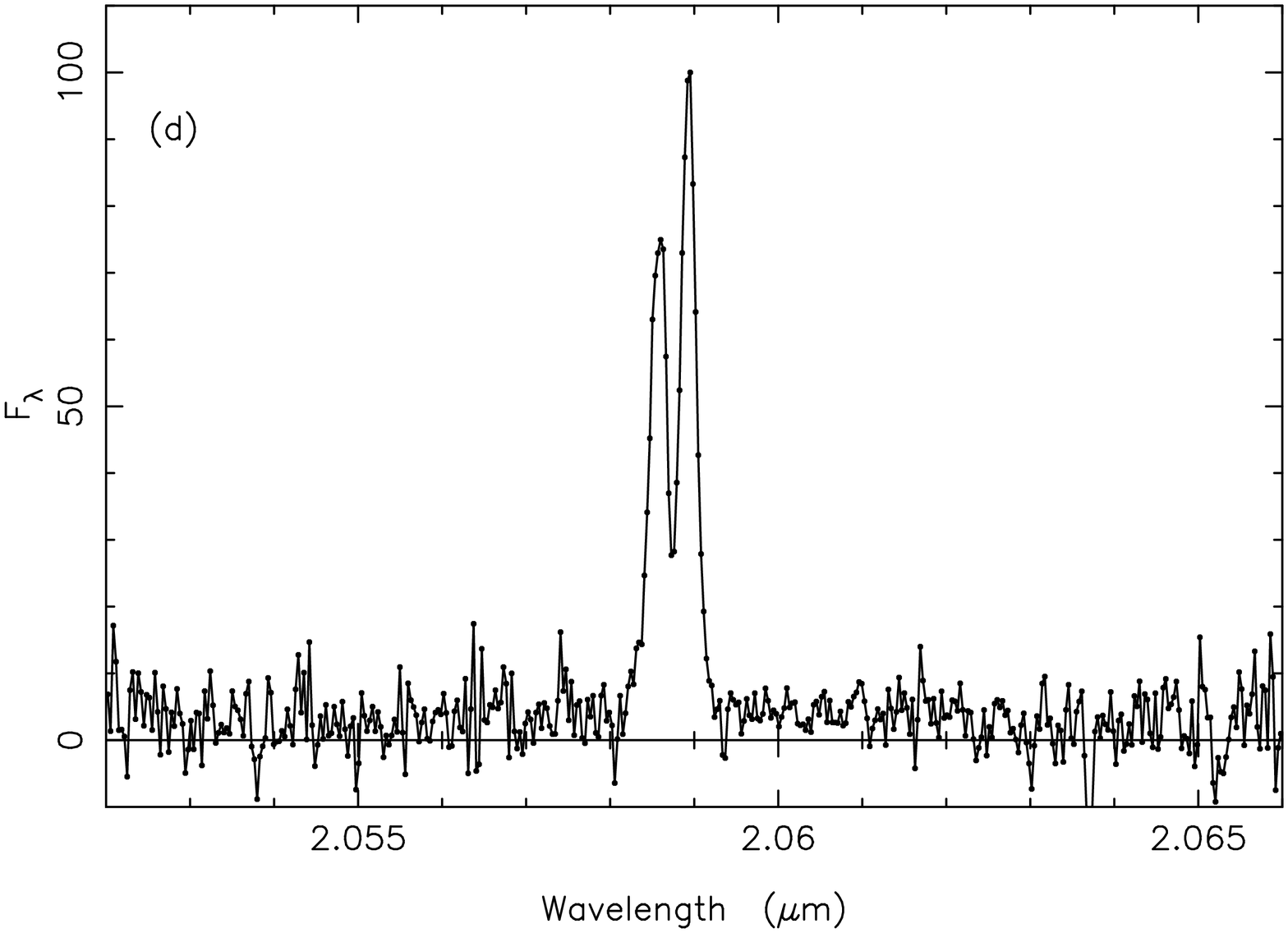,width=3in,angle=-90,clip=}
}\end{minipage}
\vspace*{-2mm}

\begin{minipage}{3.5in}{
\psfig{file=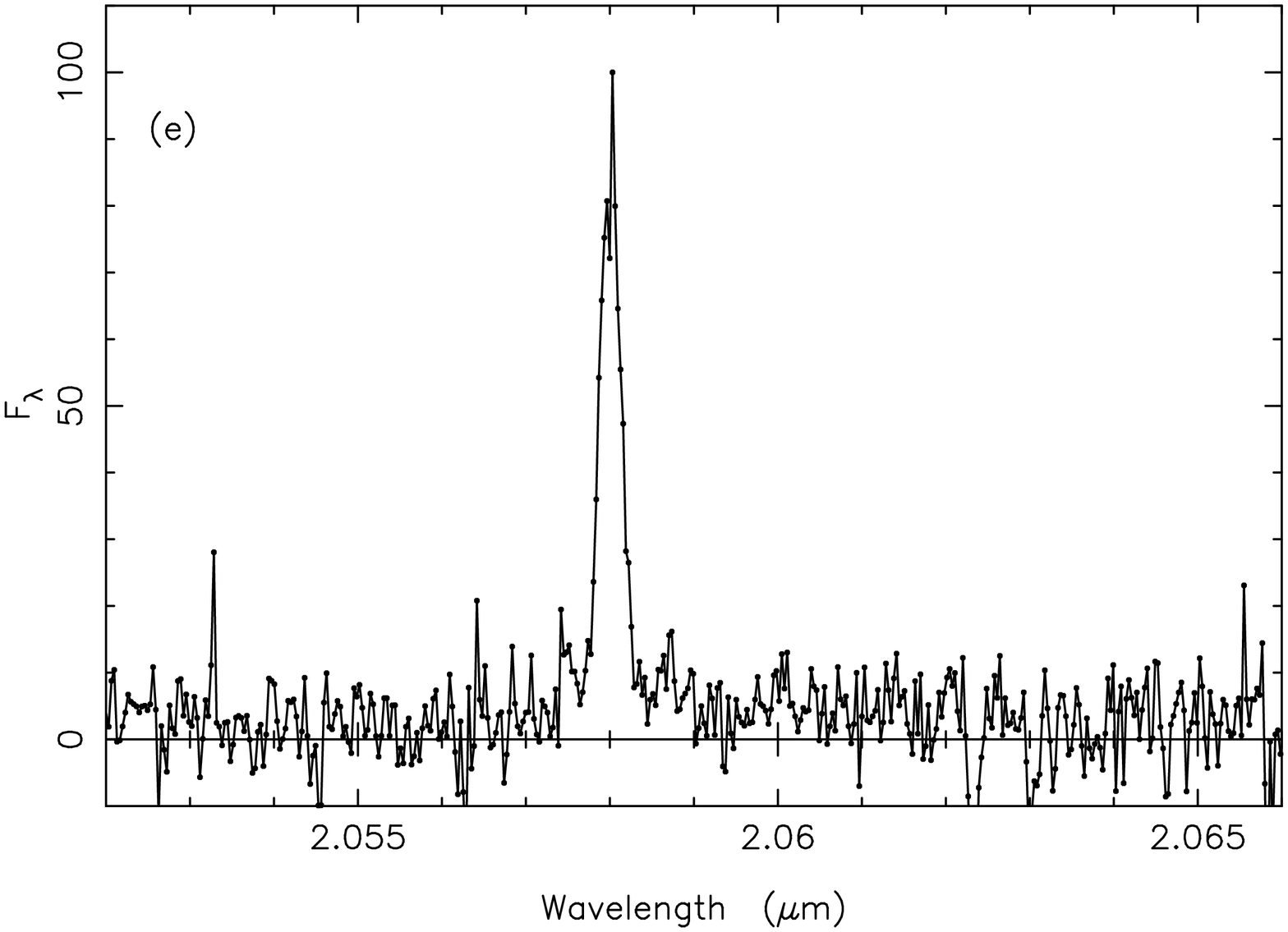,width=3in,angle=-90,clip=}
}\end{minipage}
\begin{minipage}{3.5in}{
\psfig{file=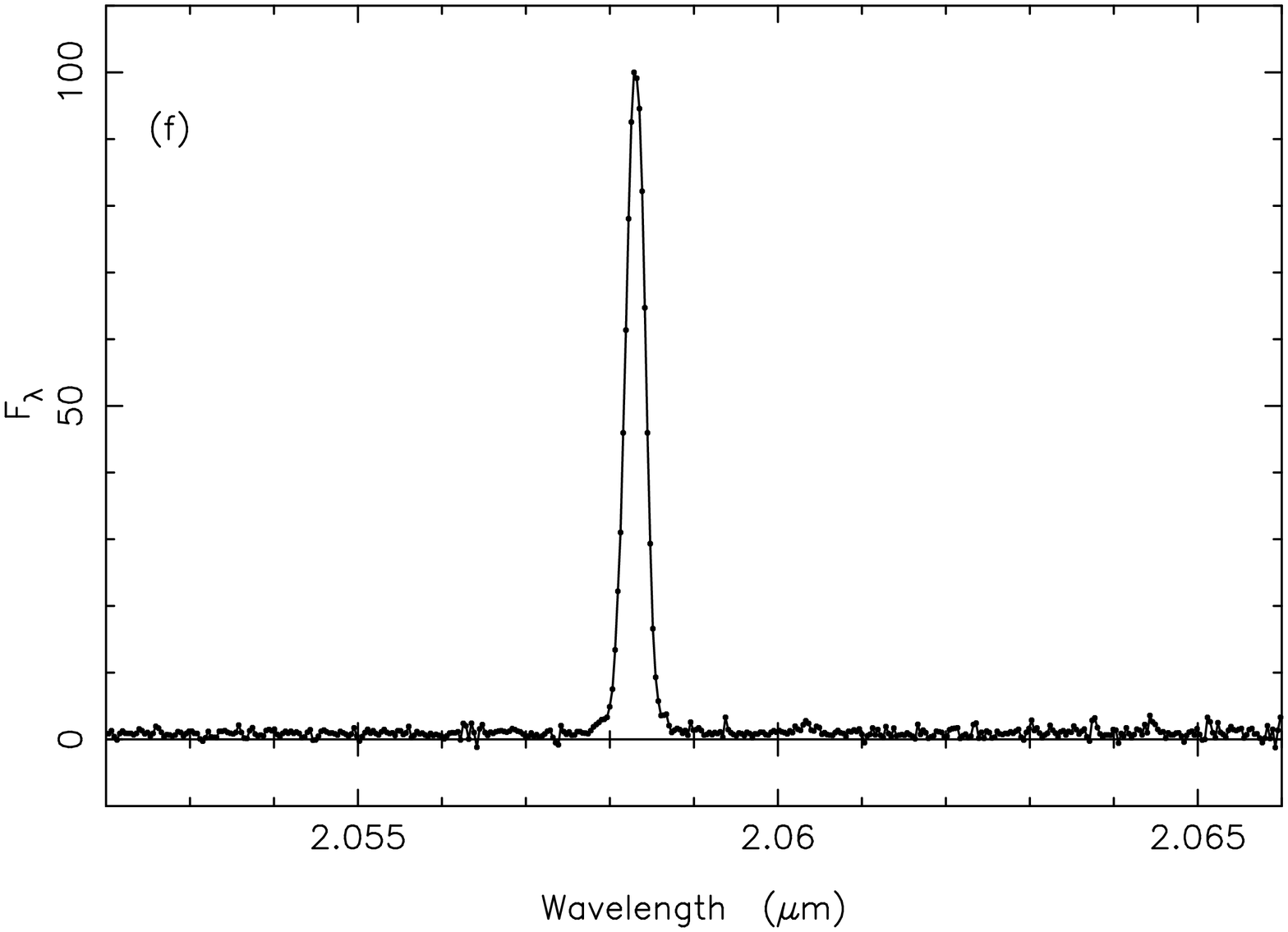,width=3in,angle=-90,clip=}
}\end{minipage}
\vspace*{-2mm}

\begin{minipage}{3.5in}{
\psfig{file=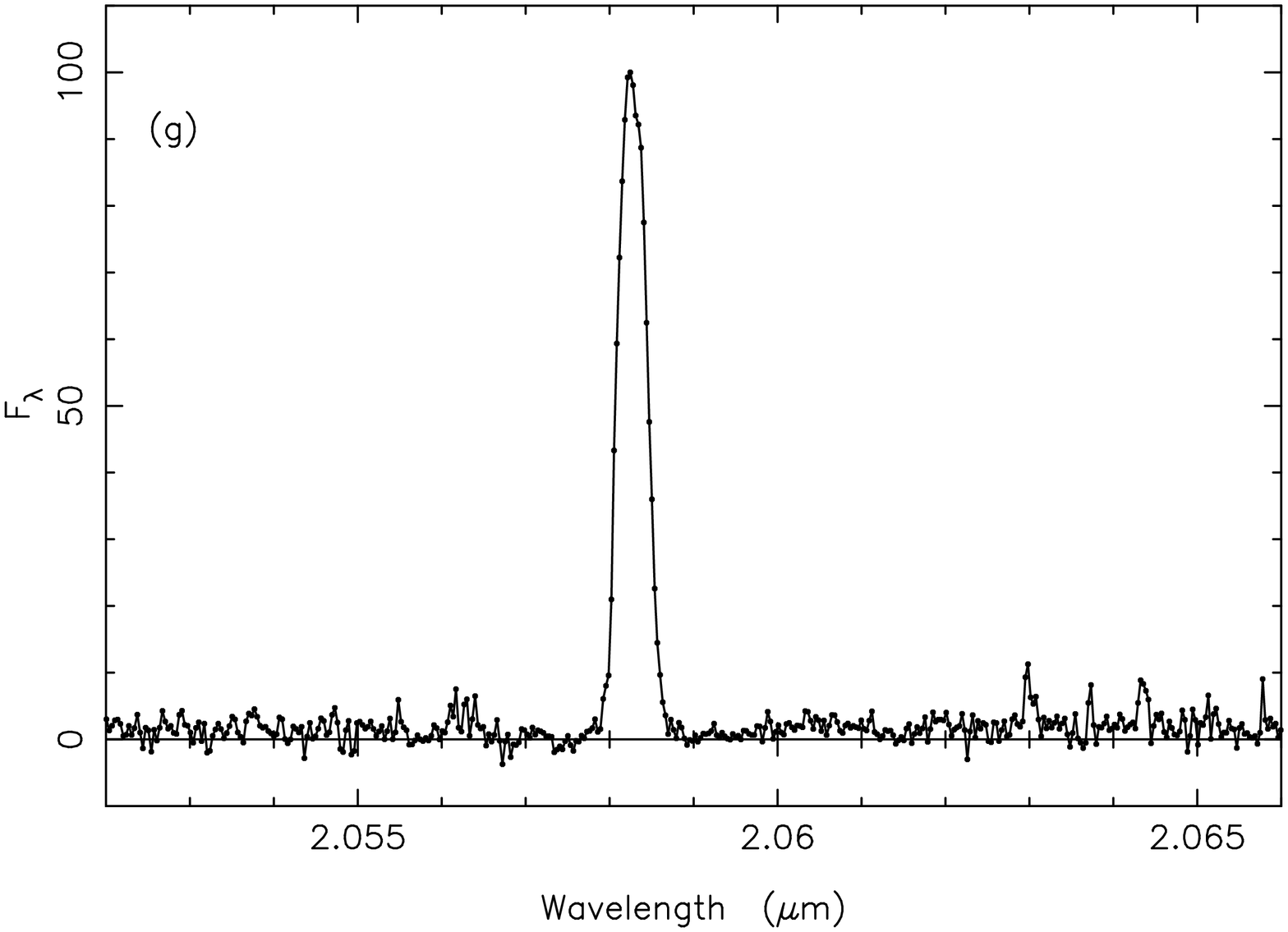,width=3in,angle=-90,clip=}
}\end{minipage}
\begin{minipage}{3.5in}{
\psfig{file=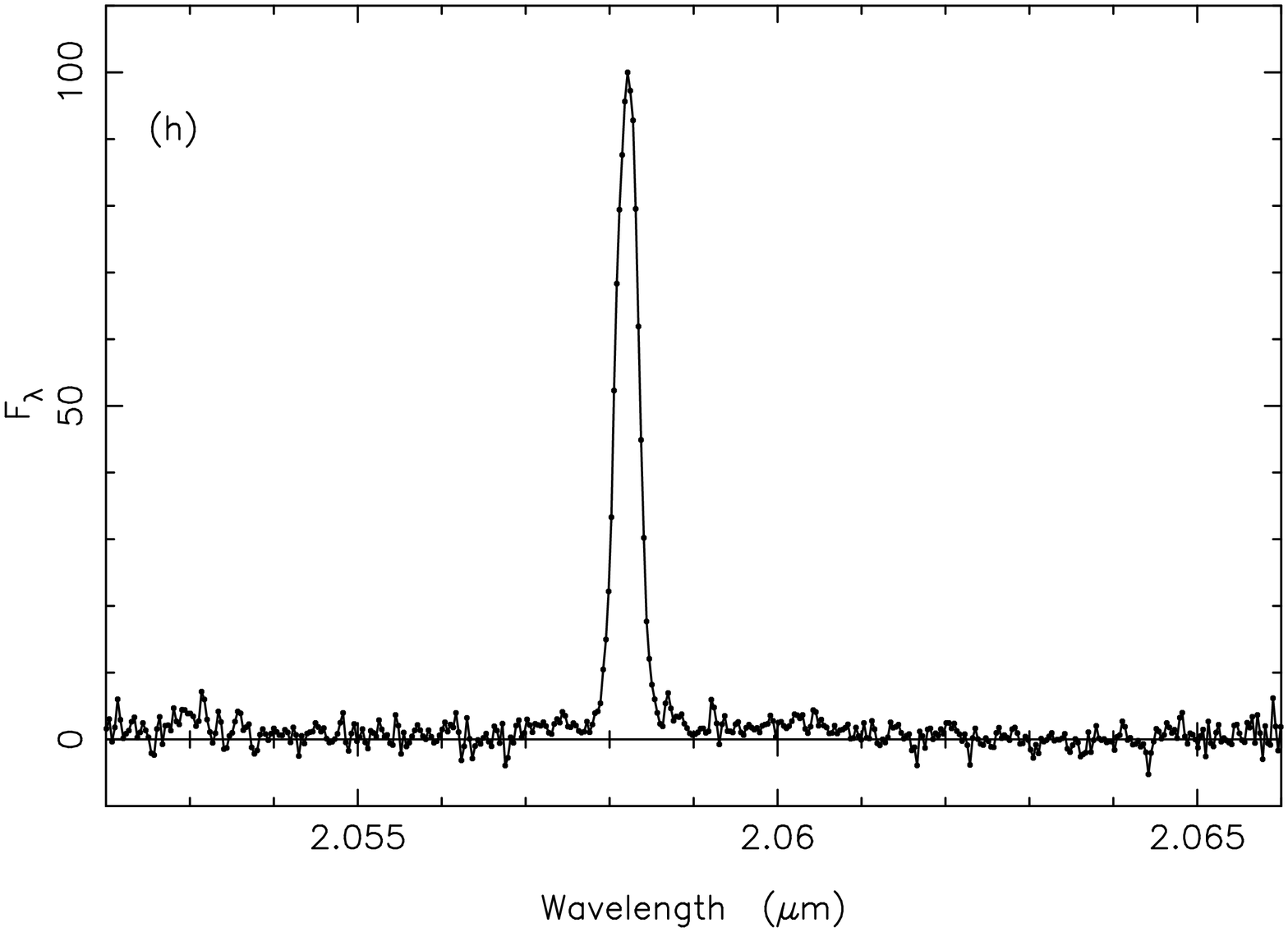,width=3in,angle=-90,clip=}
}\end{minipage}
\vspace*{-2mm}

\end{center}

\begin{center}
\begin{minipage}{3.5in}{
\psfig{file=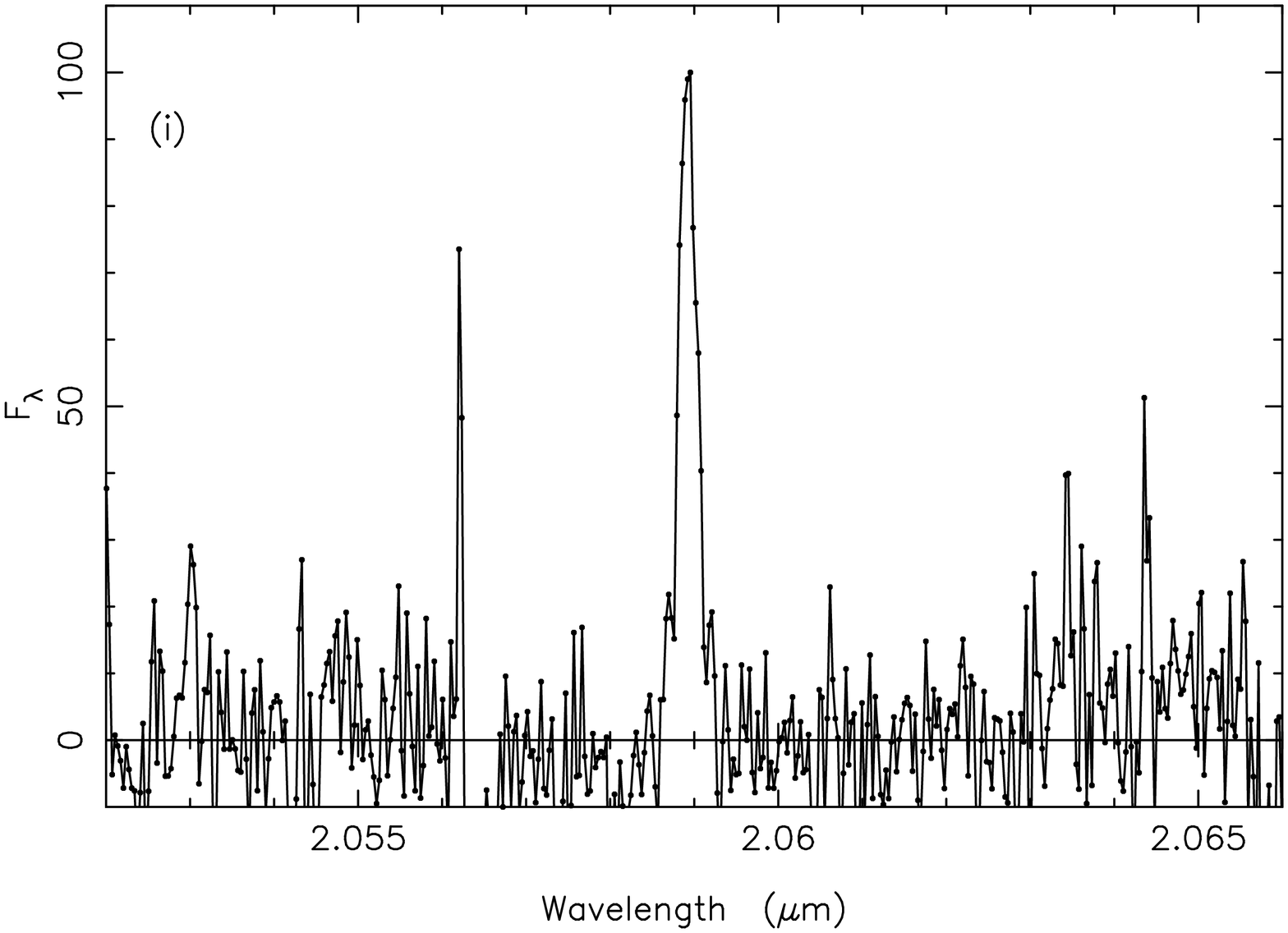,width=3in,angle=-90,clip=}
}\end{minipage}
\begin{minipage}{3.5in}{
\psfig{file=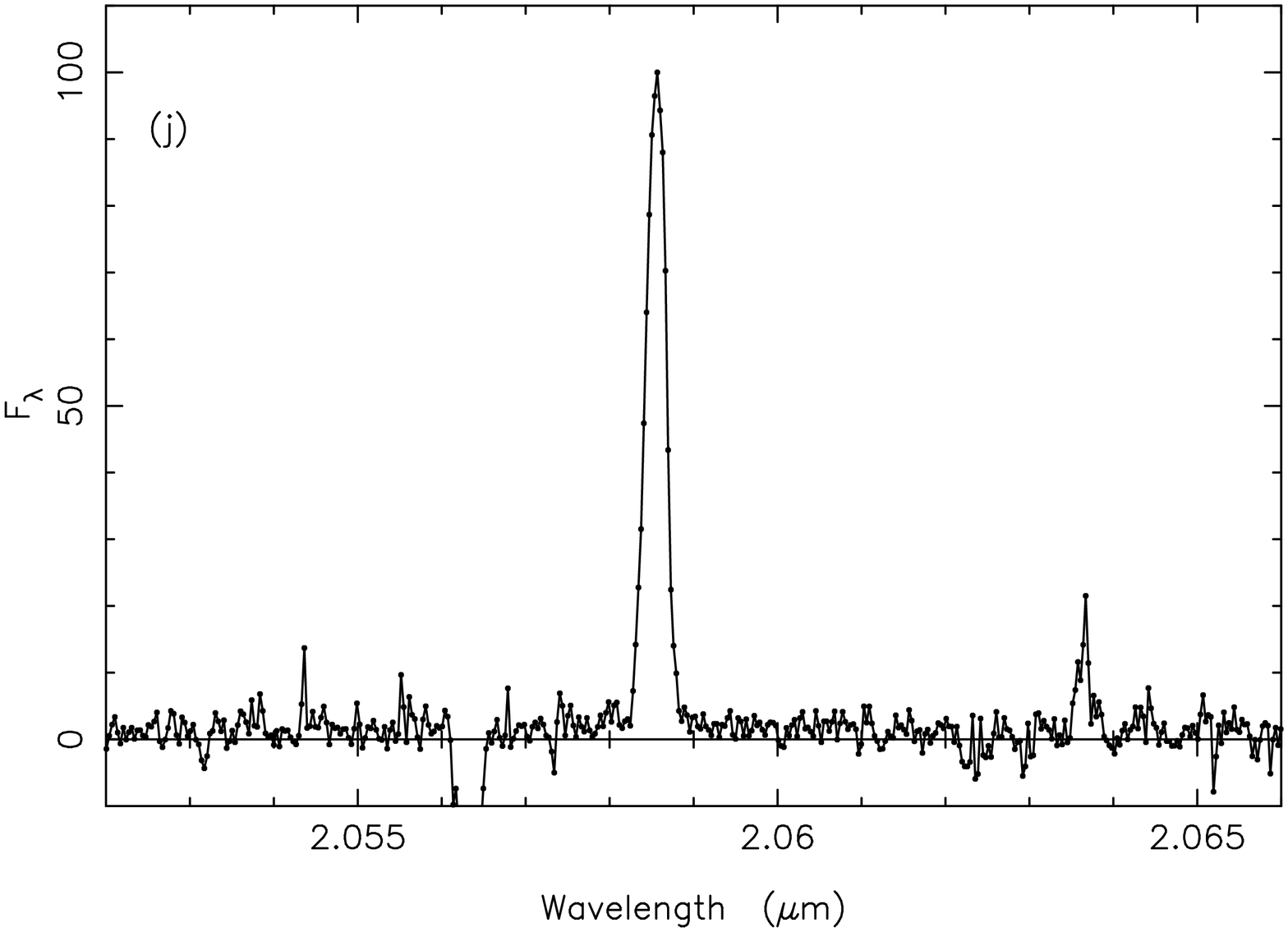,width=3in,angle=-90,clip=}
}\end{minipage}
\vspace*{-2mm}

\begin{minipage}{3.5in}{
\psfig{file=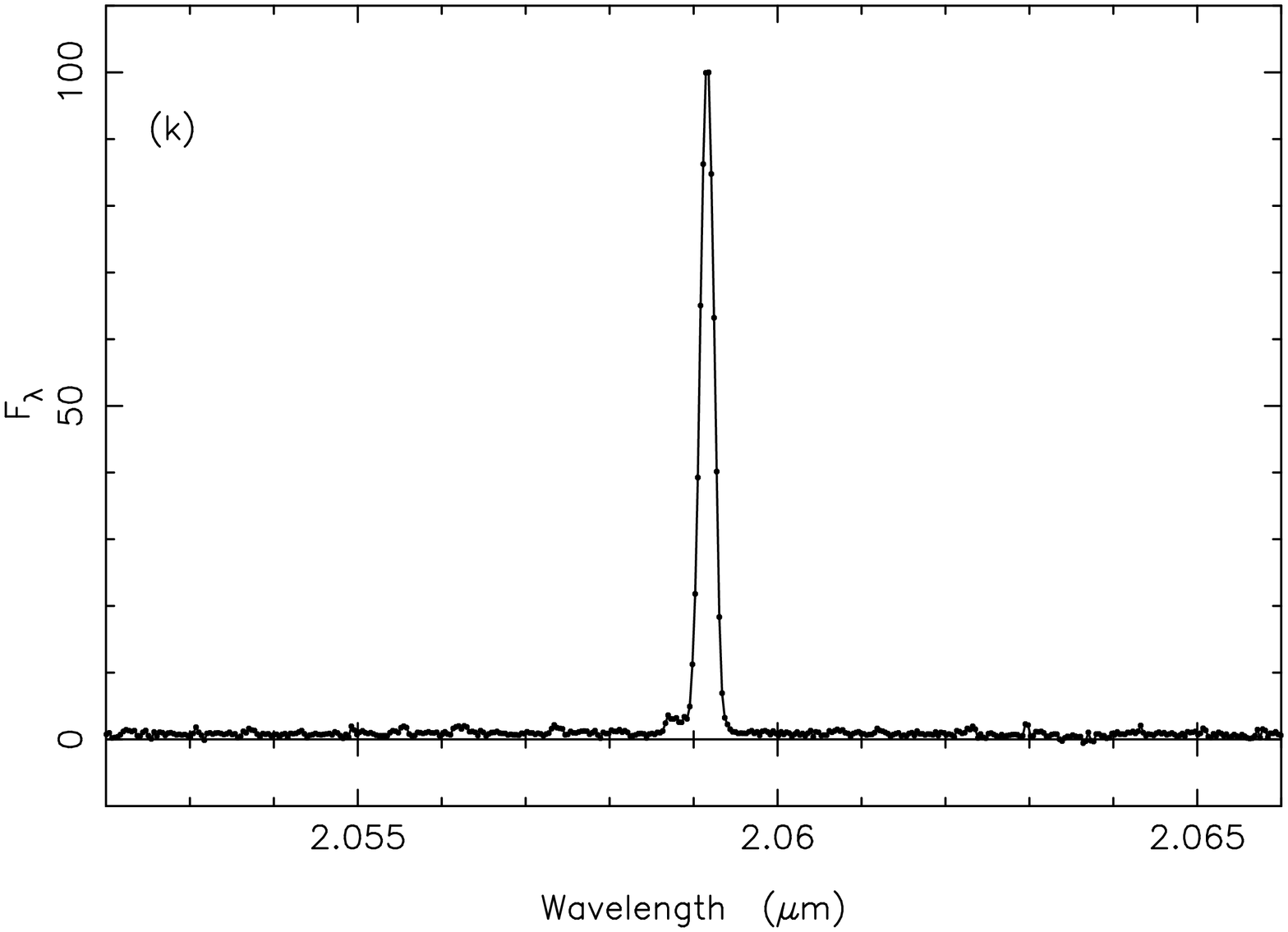,width=3in,angle=-90,clip=}
}\end{minipage}
\begin{minipage}{3.5in}{
\psfig{file=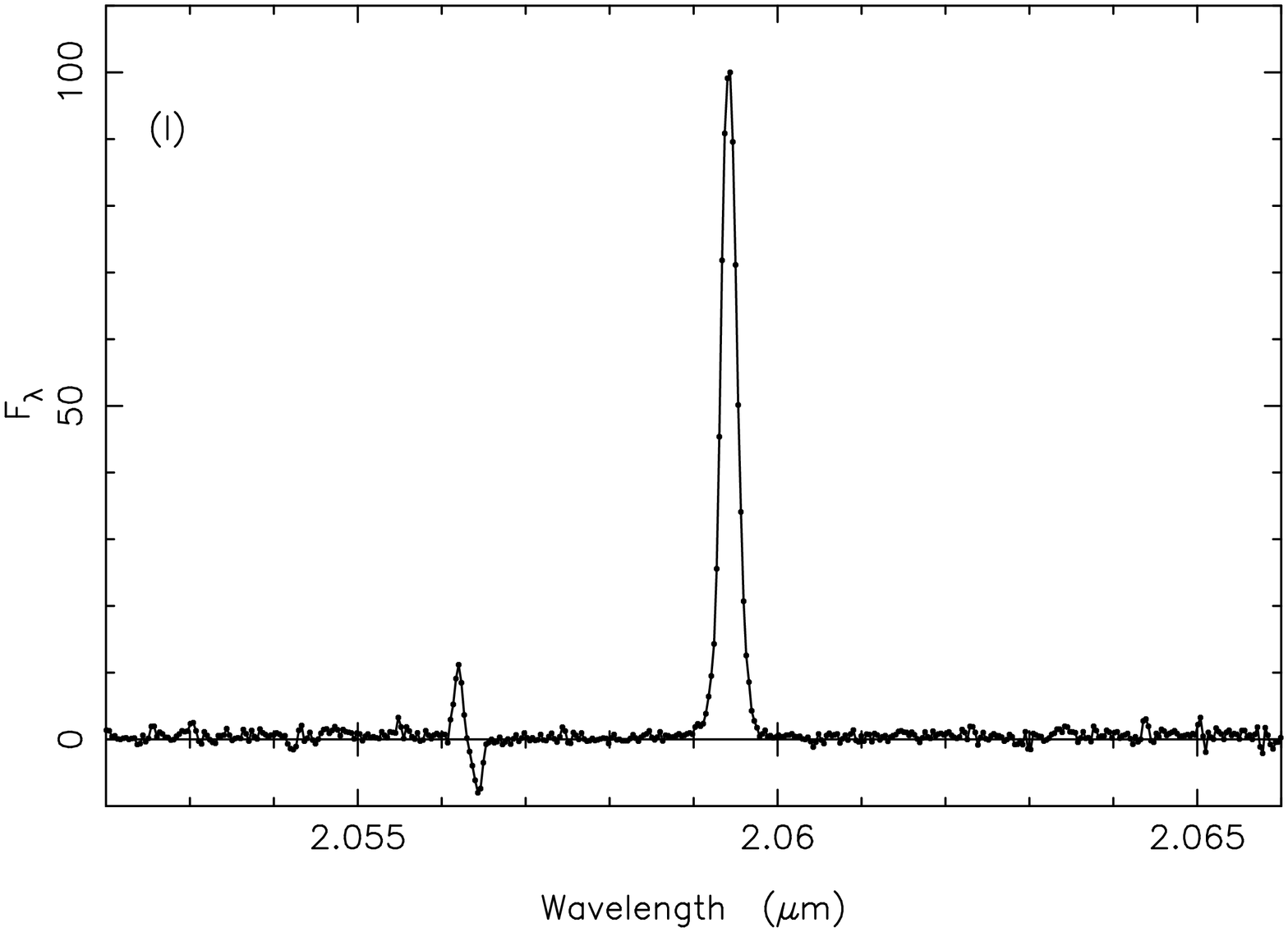,width=3in,angle=-90,clip=}
}\end{minipage}
\vspace*{-2mm}

\begin{minipage}{3.5in}{
\psfig{file=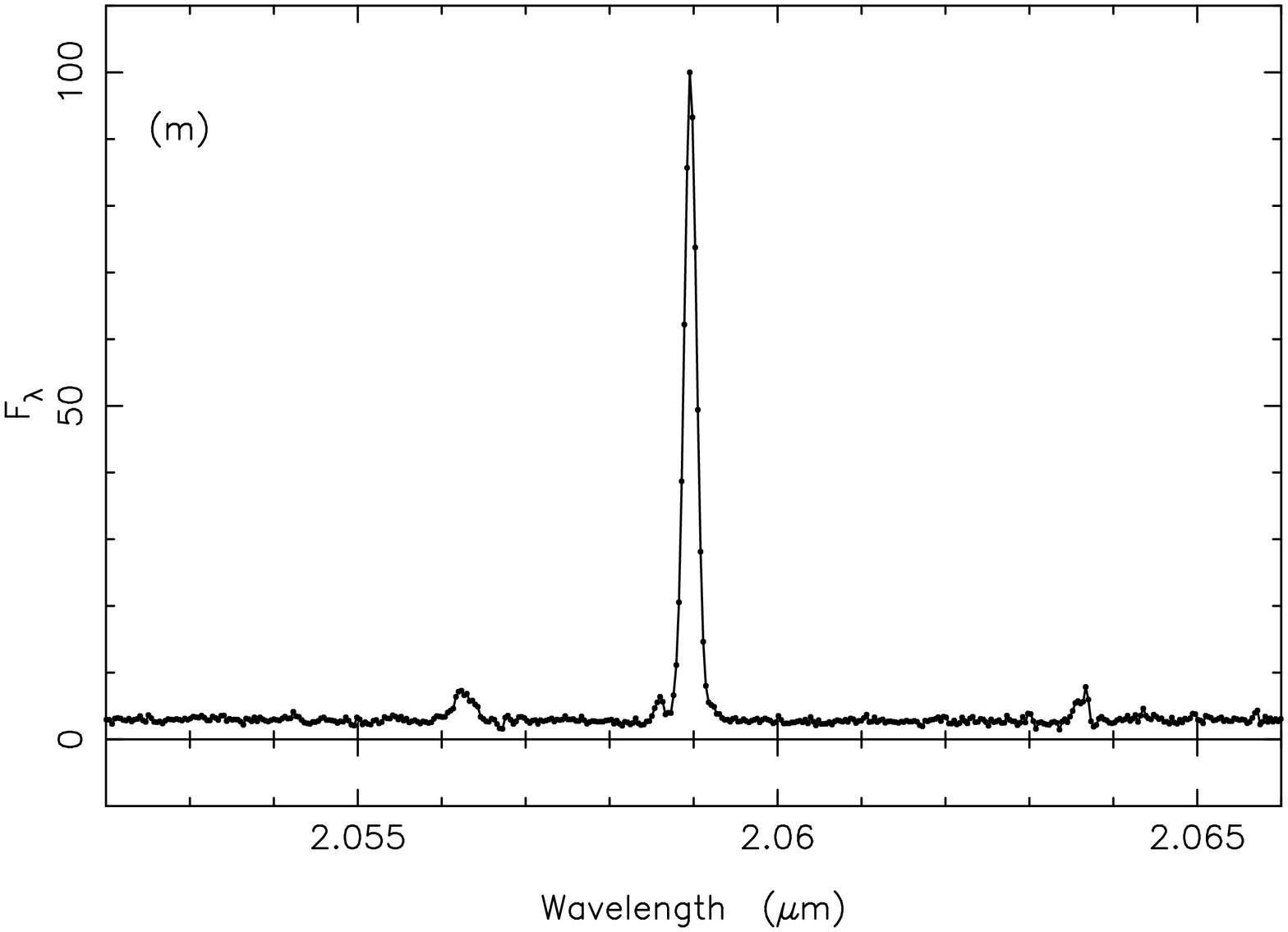,width=3in,angle=-90,clip=}
}\end{minipage}
\begin{minipage}{3.5in}{
\psfig{file=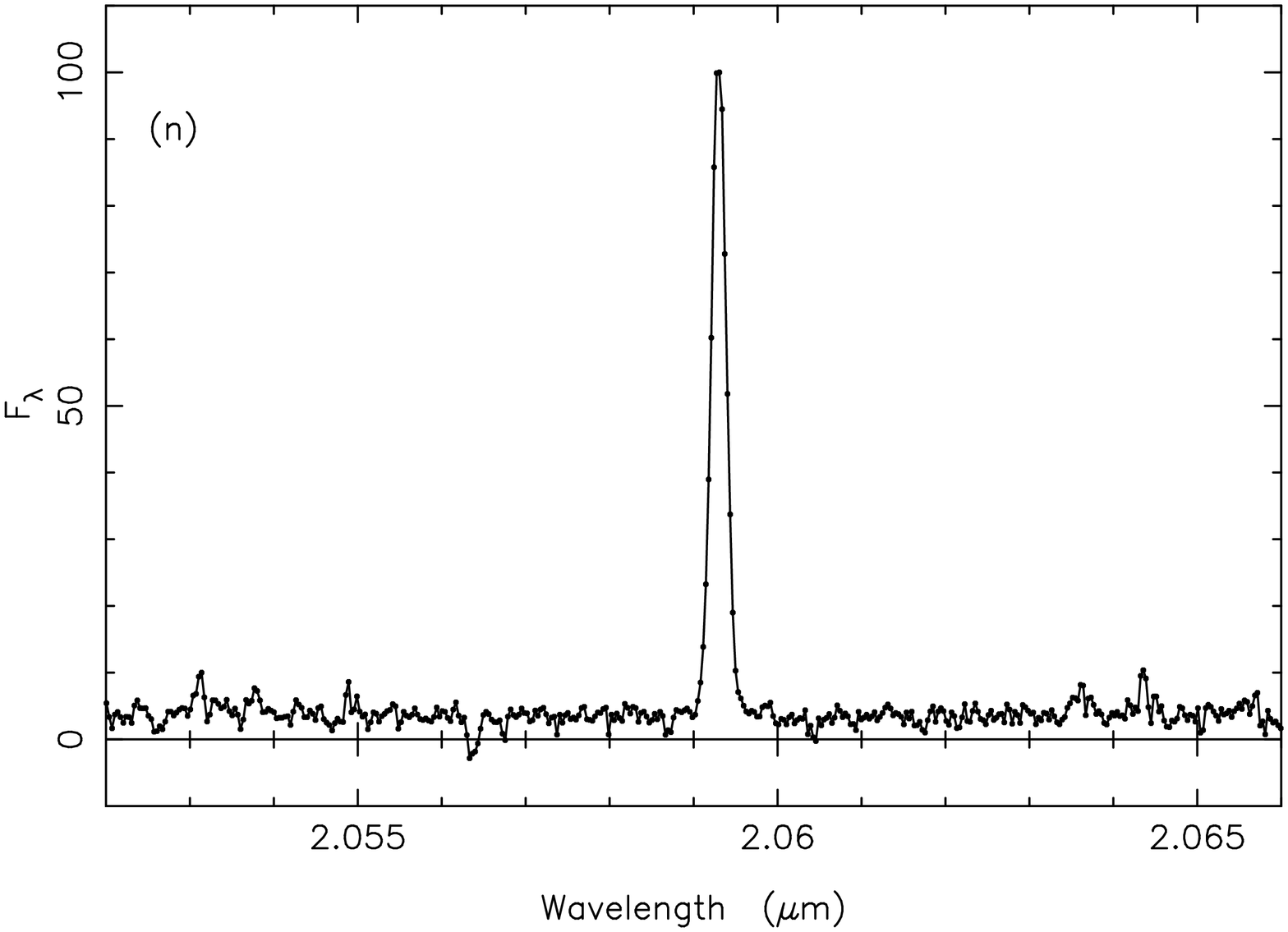,width=3in,angle=-90,clip=}
}\end{minipage}
\vspace*{-2mm}

\begin{minipage}{3.5in}{
\psfig{file=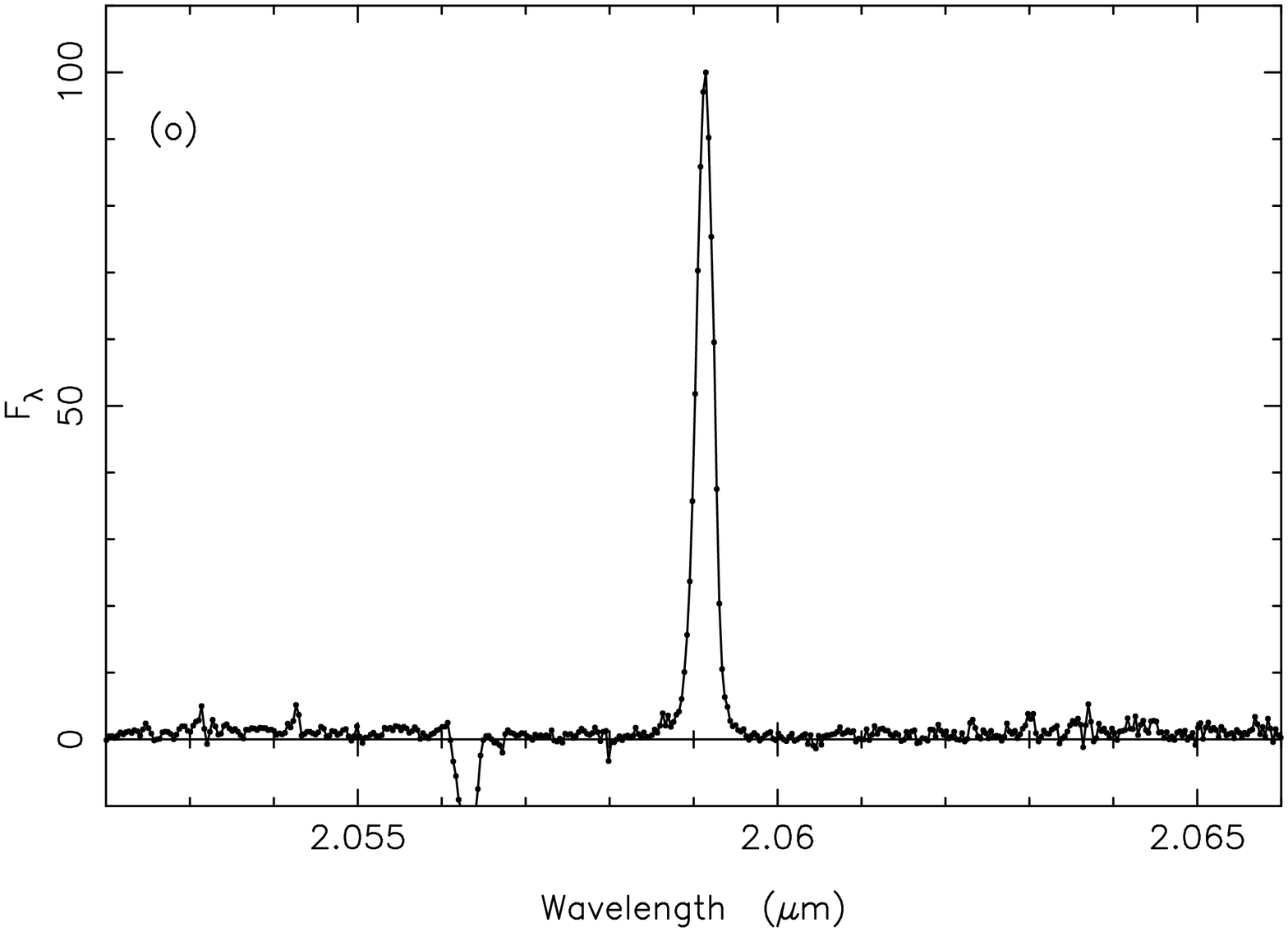,width=3in,angle=-90,clip=}
}\end{minipage}
\begin{minipage}{3.5in}{
\psfig{file=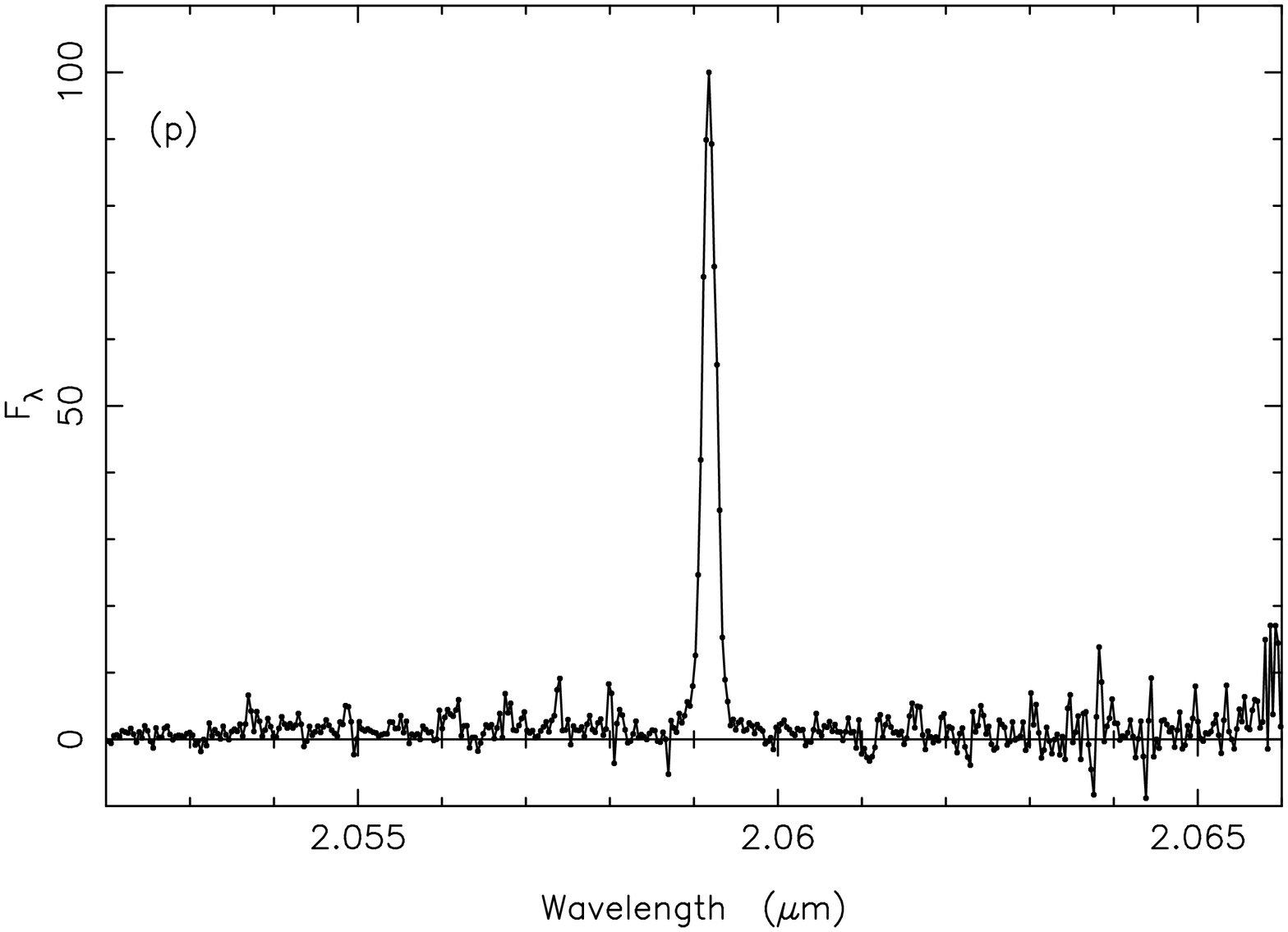,width=3in,angle=-90,clip=}
}\end{minipage}
\vspace*{-2mm}

\end{center}

\begin{center}

\begin{minipage}{3.5in}{
\psfig{file=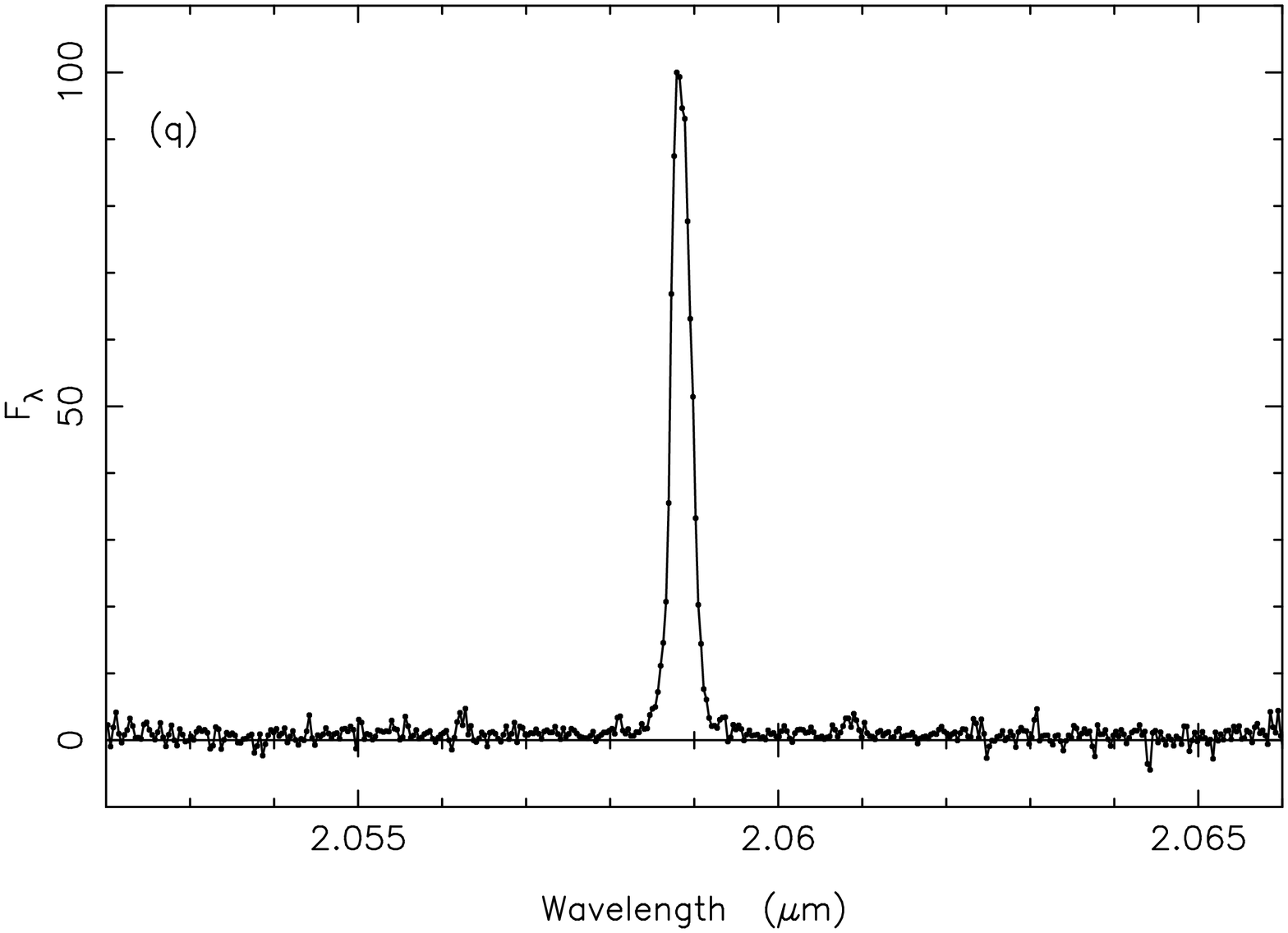,width=3in,angle=-90,clip=}
}\end{minipage}
\begin{minipage}{3.5in}{
\psfig{file=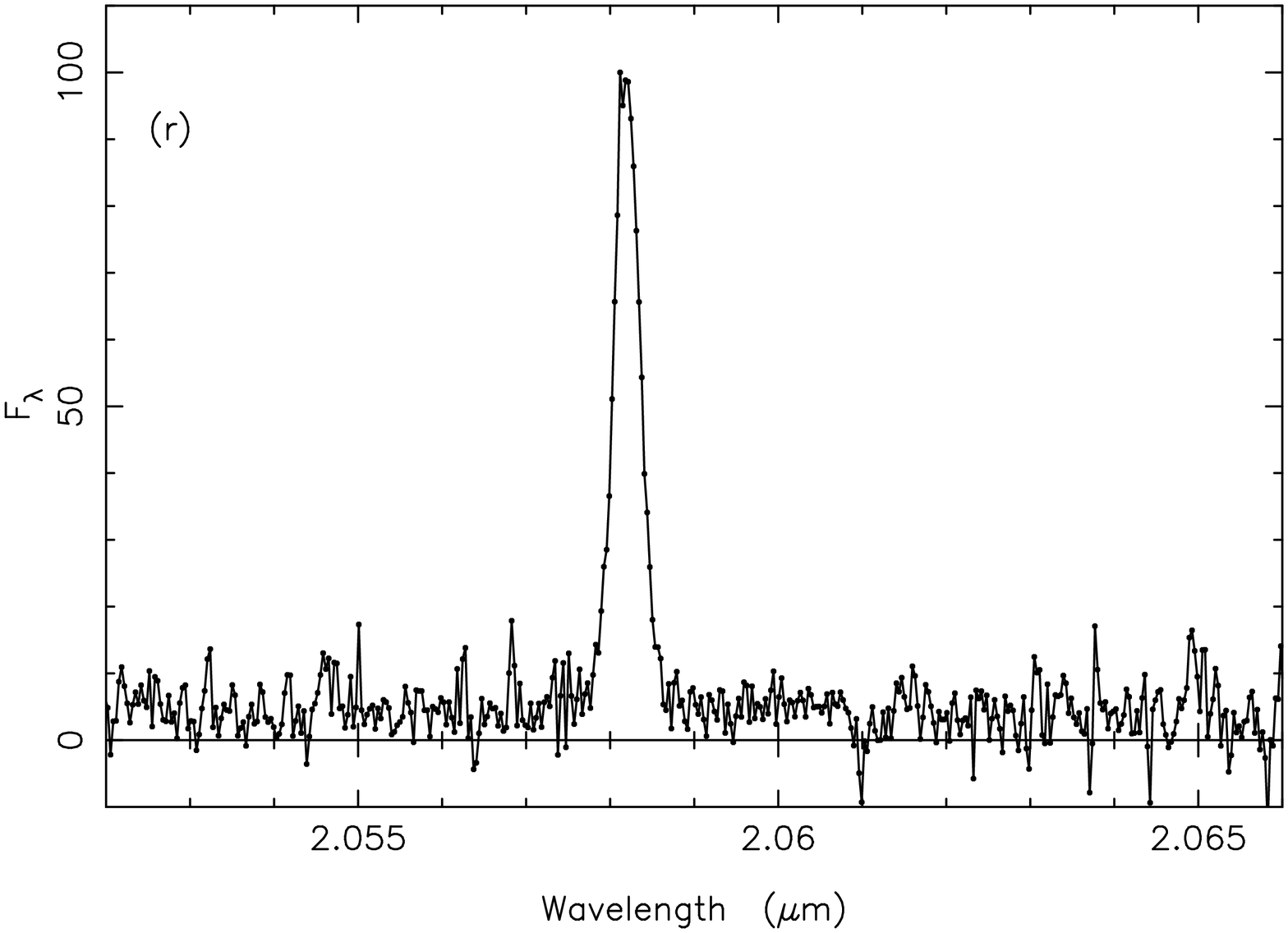,width=3in,angle=-90,clip=}
}\end{minipage}
\vspace*{-2mm}

\begin{minipage}{3.5in}{
\psfig{file=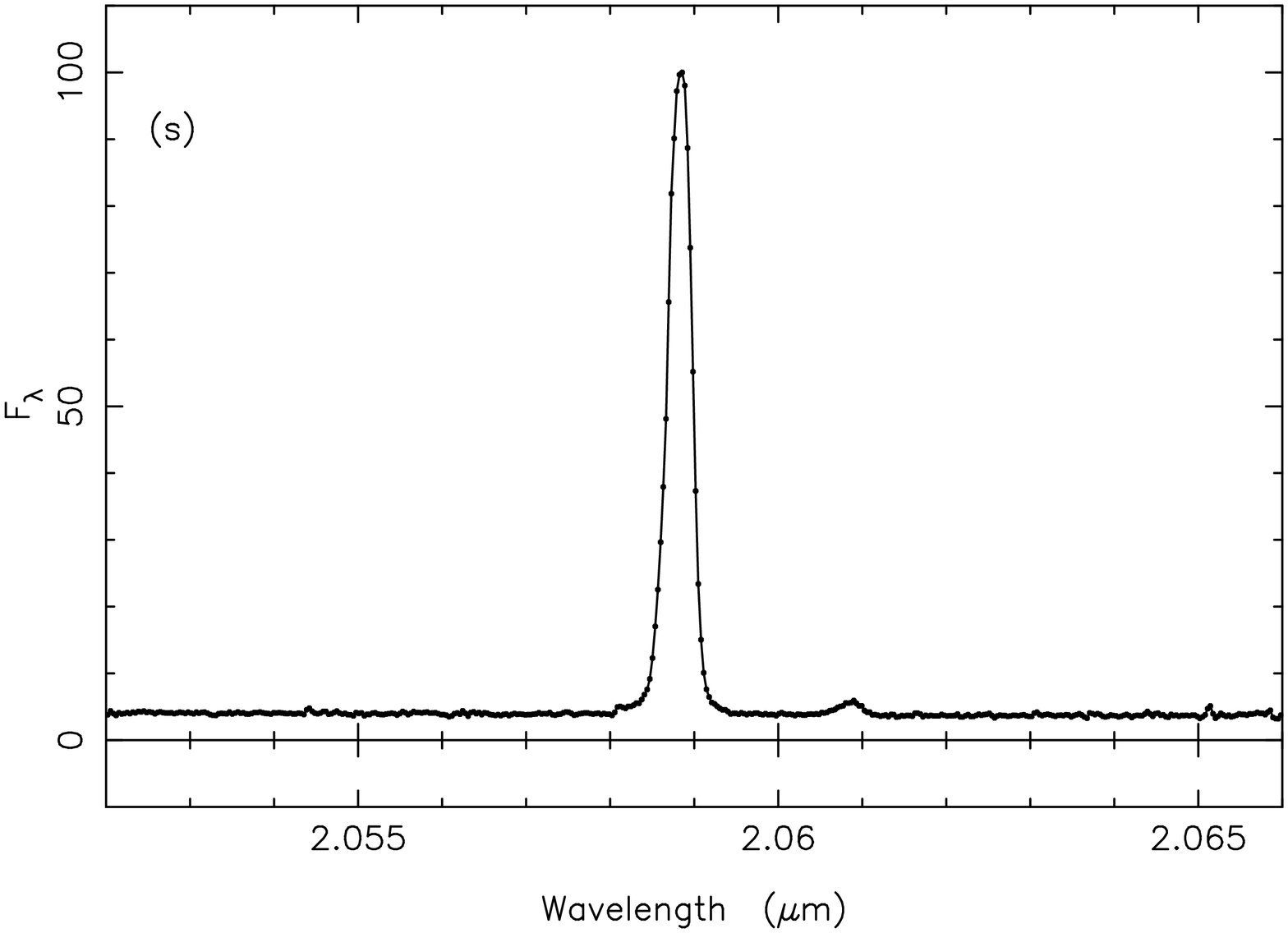,width=3in,angle=-90,clip=}
}\end{minipage}
\begin{minipage}{3.5in}{
\psfig{file=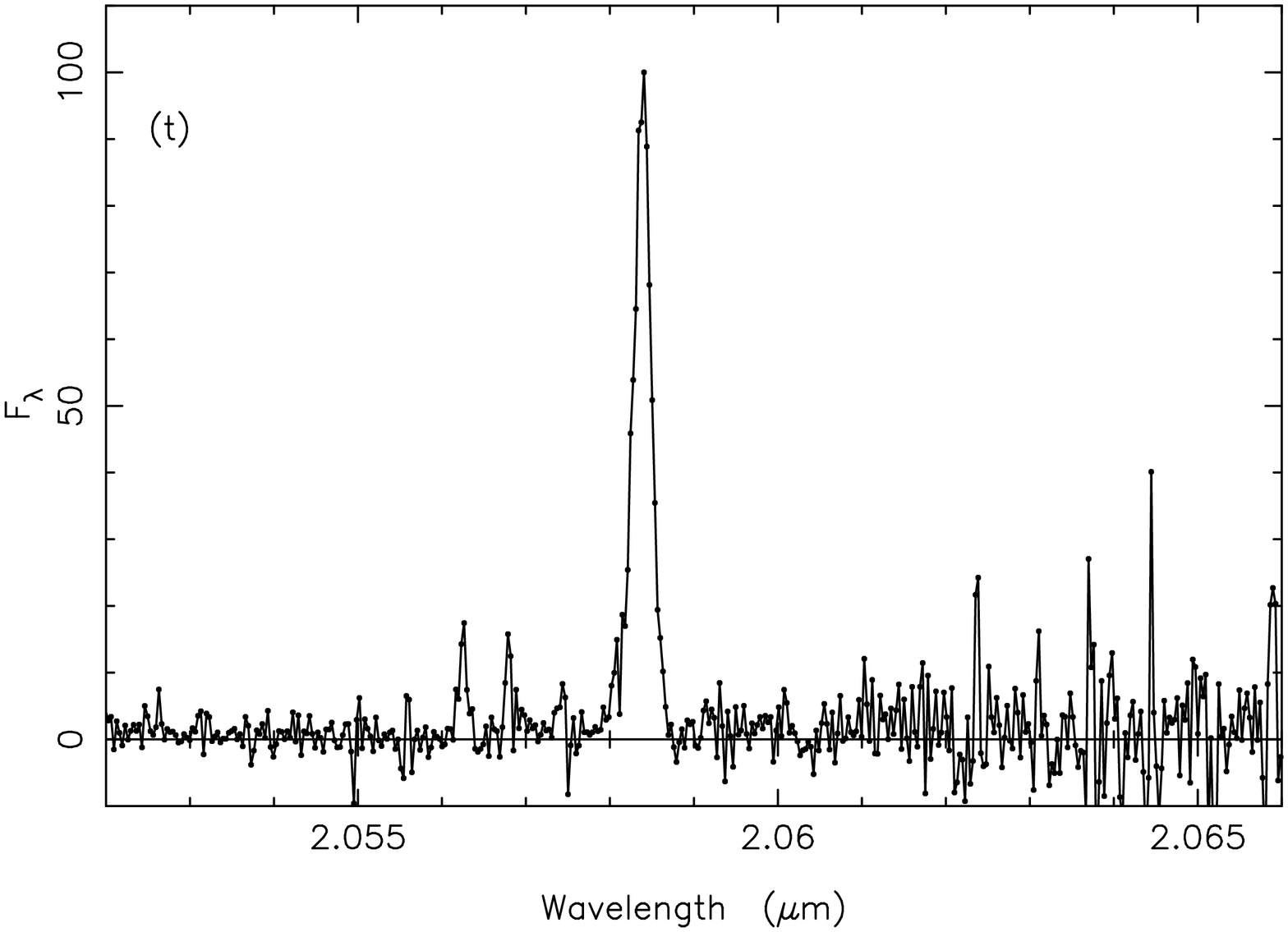,width=3in,angle=-90,clip=}
}\end{minipage}
\vspace*{-2mm}

\begin{minipage}{3.5in}{
\psfig{file=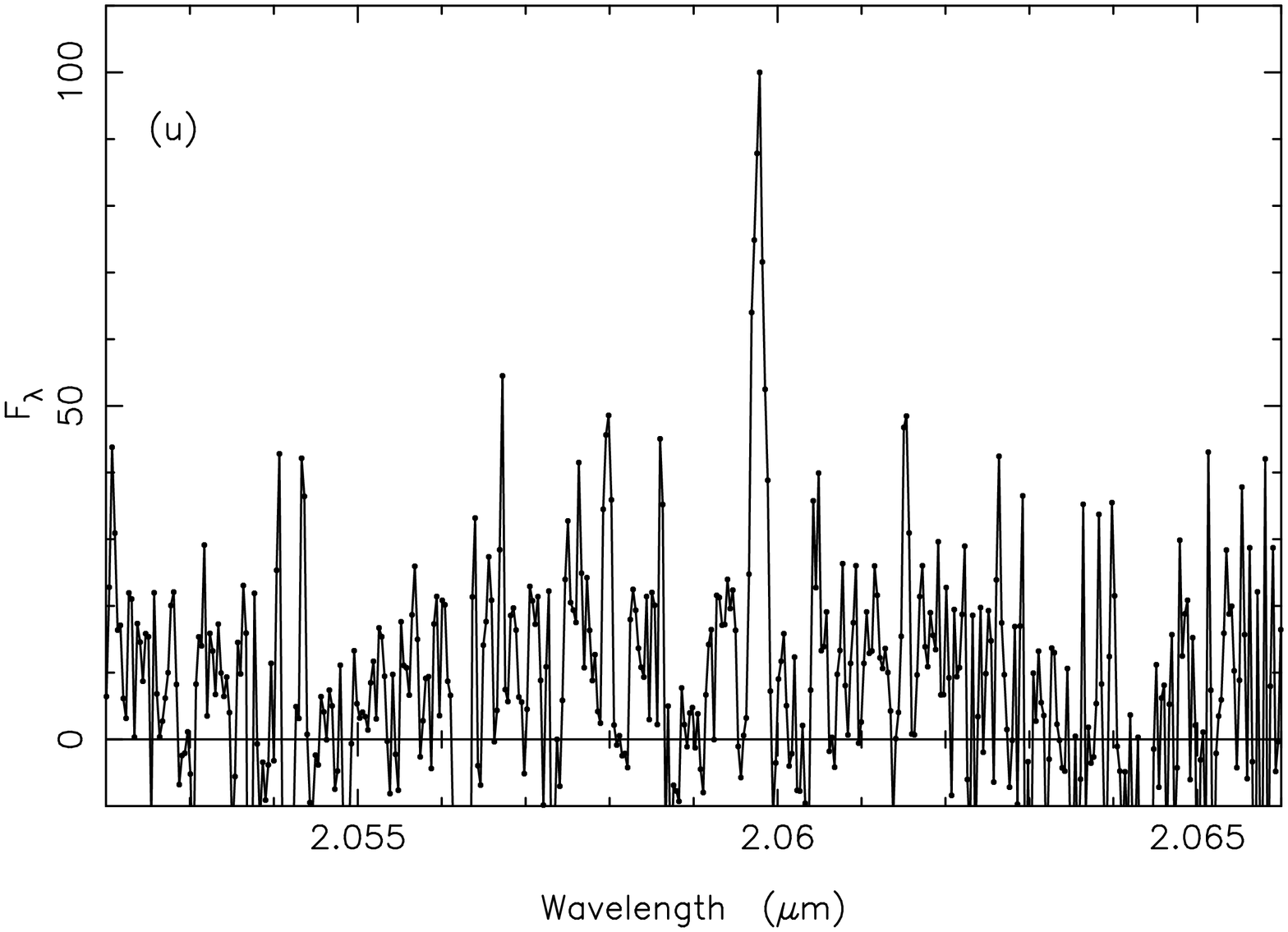,width=3in,angle=-90,clip=}
}\end{minipage}
\begin{minipage}{3.5in}{
\psfig{file=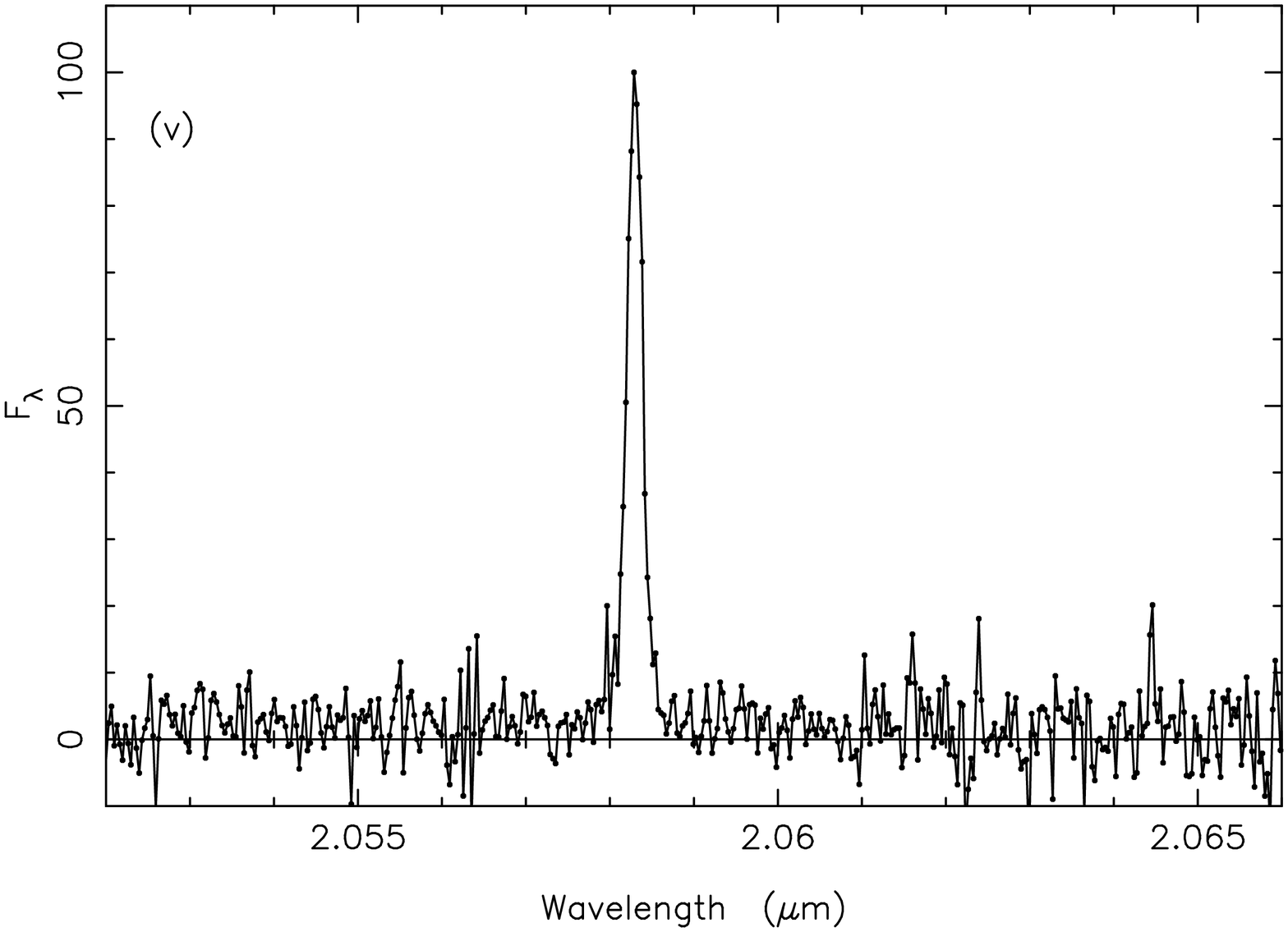,width=3in,angle=-90,clip=}
}\end{minipage}
\vspace*{-2mm}

\end{center}

{\bf Figure 2:} The HeI 2$^1$P--2$^1$S echelle spectra.  
The spectra are of: (a) BD+303639; (b) CRL 618; (c) DdDm~1; (d) Hu~1-2; 
(e) K~3-60; (f) K~3-62; (g) K~3-66; (h) K~3-67; (i) K 4-48; 
(j) M~1-4; (k) M~1-6; (l) M~1-9; (m) M~1-11;
(n) M~1-12; (o) M~1-14; (p) M~1-74; (q) M~1-78;
(r) NGC 7027; (s) PC~12; (t) SaSt~2-3; (u) Vy~1-1.   

\newpage

\begin{center}

\begin{minipage}{3.5in}{
\psfig{file=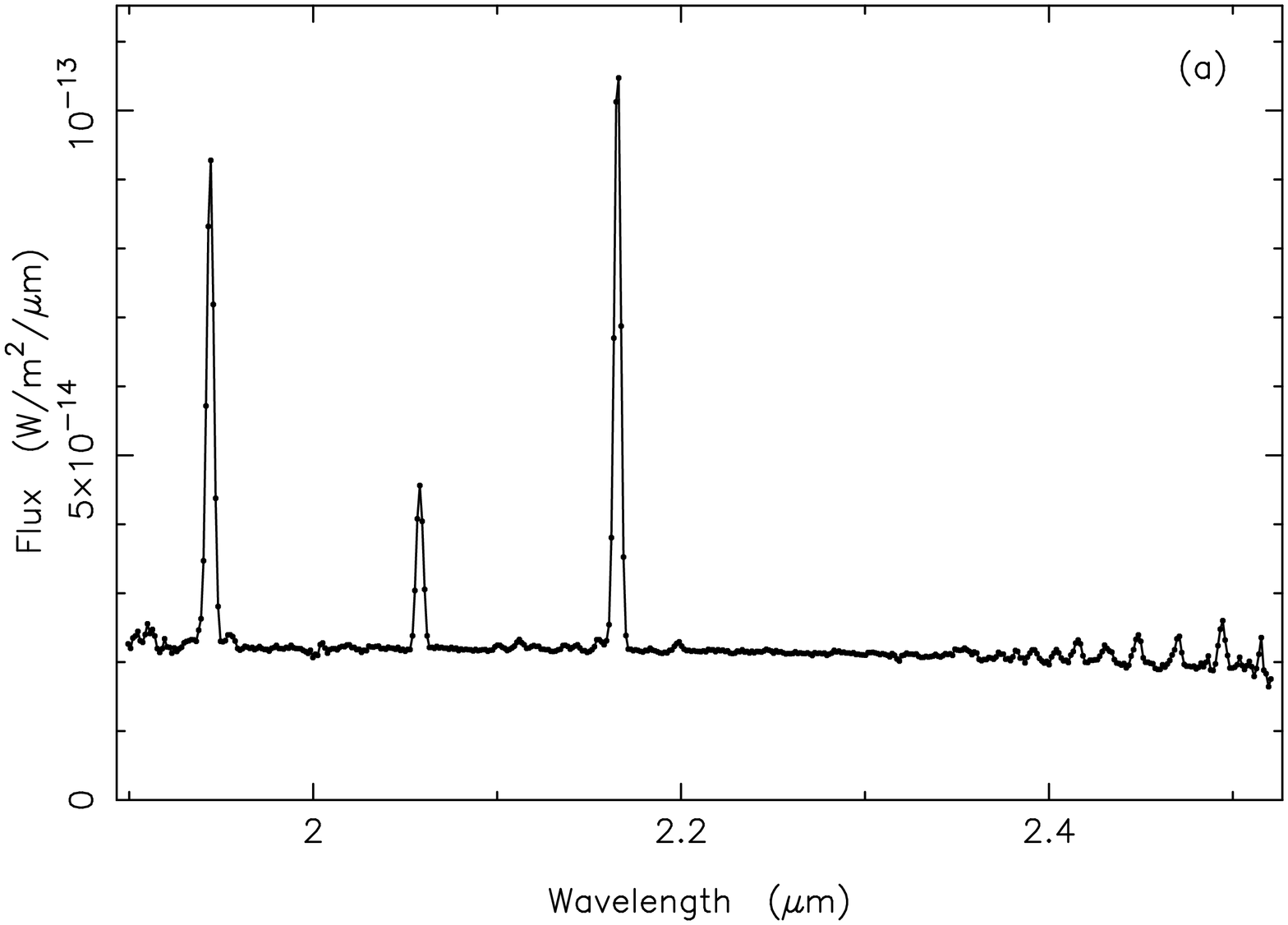,width=3in,angle=-90,clip=}
}\end{minipage}
\begin{minipage}{3.5in}{
\psfig{file=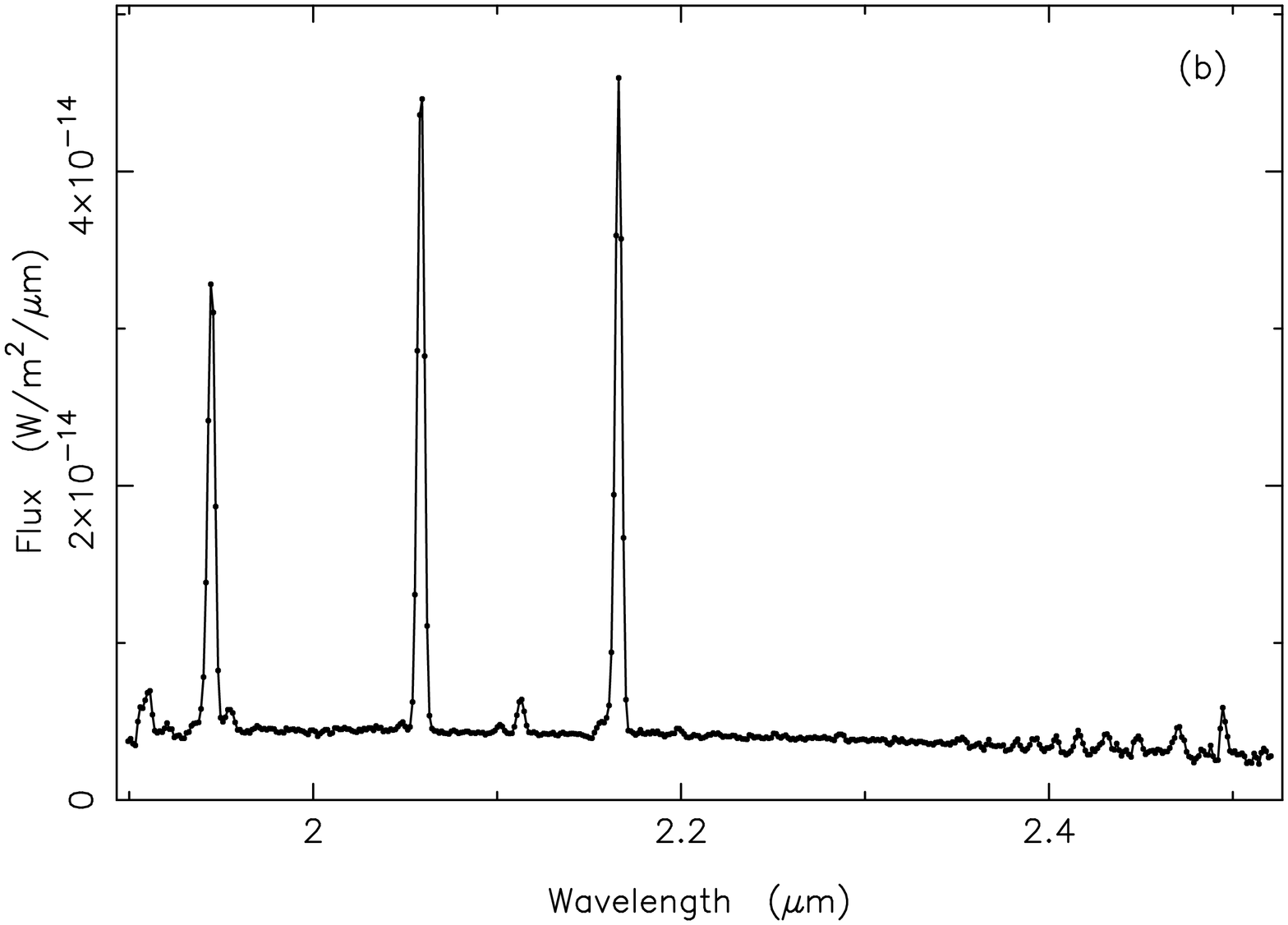,width=3in,angle=-90,clip=}
}\end{minipage}
\vspace*{-2mm}

\begin{minipage}{3.5in}{
\psfig{file=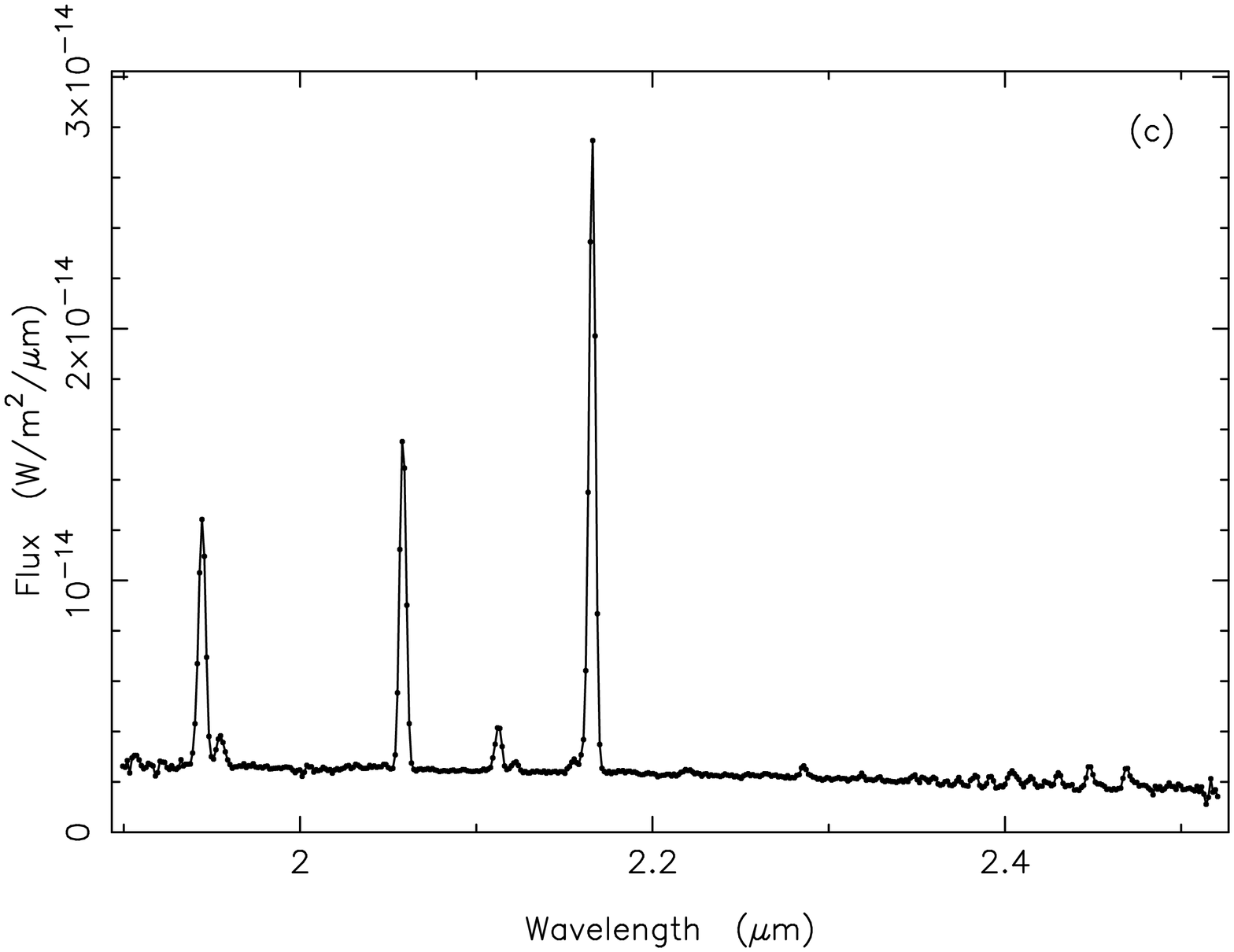,width=3in,angle=-90,clip=}
}\end{minipage}
\begin{minipage}{3.5in}{
\psfig{file=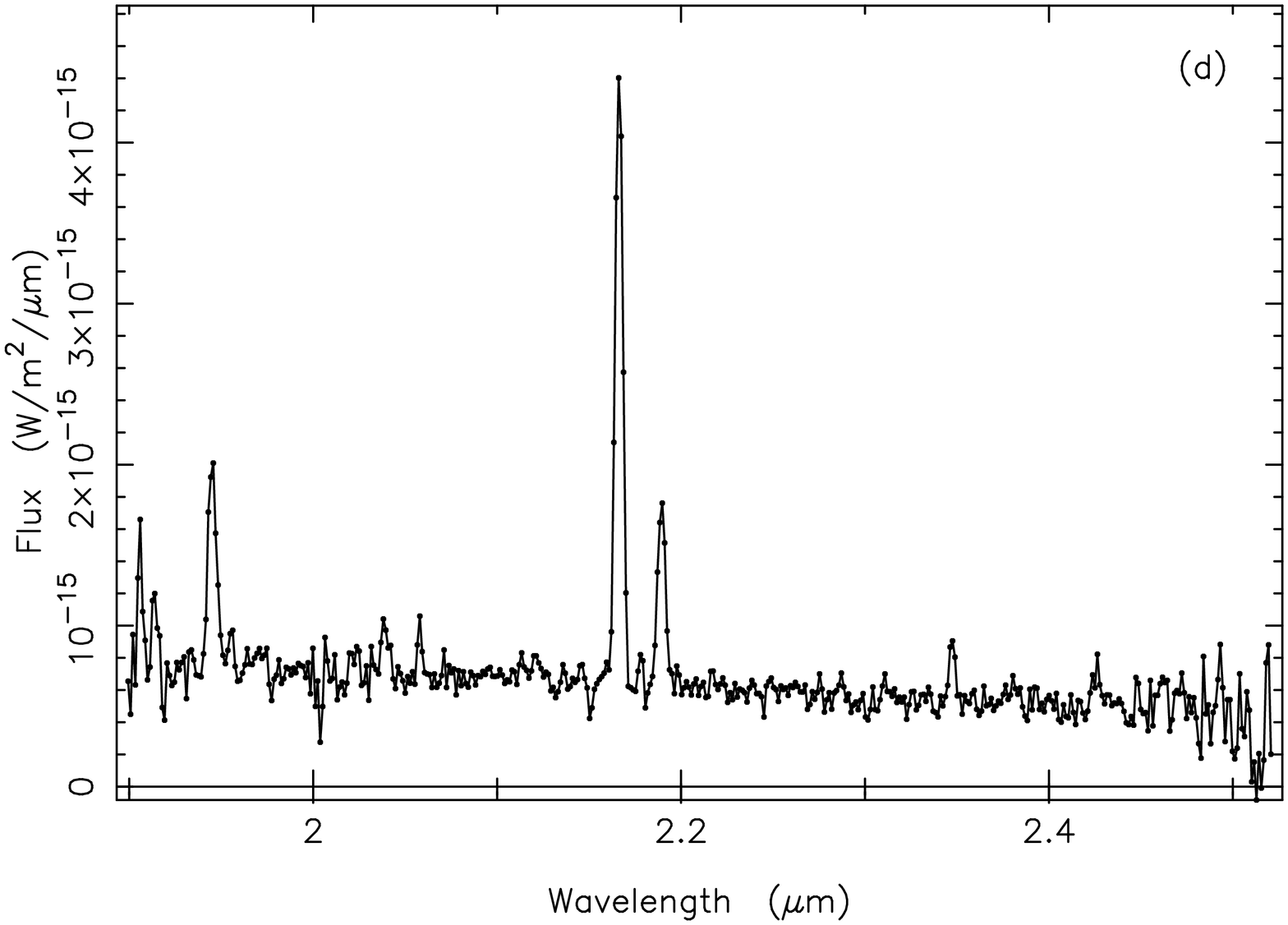,width=3in,angle=-90,clip=}
}\end{minipage}
\vspace*{-2mm}

\end{center}

{\bf Figure 3:} Typical low resolution spectra of the sample.  The objects
shown are (in order of increasing excitation) (a) M~1-12, (b) M~1-9,
(c) K~3-67 and (d) M~1-1.  The brightest lines are HI Br$\gamma$ and Br$\delta$
at 2.1661$\mu$m and 1.9451$\mu$m respectively, and HeI $2^{1}$S--$2^{1}$P
at 2.058$\mu$m.  The bright line to the red of Br$\gamma$ in M~1-1
is HeII 10--7 at 2.1891$\mu$m, indicative of the very high central star
temperature in this object.

\begin{center}

\begin{minipage}{3.5in}{(a)}\end{minipage}\begin{minipage}{3.3in}{(b)}\end{minipage}
\vspace*{-16mm}

\begin{minipage}{3.5in}{
\psfig{file=fig4a.ps,width=3in,angle=0,clip=}
}\end{minipage}
\begin{minipage}{3.5in}{
\psfig{file=fig4b.ps,width=3in,angle=0,clip=}
}\end{minipage}
\vspace*{-2mm}

\end{center}

{\bf Figure 4:} The observed ratios of (a) 2.166$\mu$m $7^{1,3}$G--$4^{3,1}$F
with Br$\gamma$ and (b) 2.058$\mu$m $2^{1}$S--$2^{1}$P
with Br$\gamma$ against 5007\AA\ [OIII] with H$\beta$.  Note the well
defined trend of the data in (a) against the much more complex behaviour
shown in (b).  This illustrates the difference between a HeI line arising 
purely from recombination and the $2^{1}$S--$2^{1}$P transition.

\begin{center}

\begin{minipage}{3.5in}{(a)}\end{minipage}\begin{minipage}{3.3in}{(b)}\end{minipage}
\vspace*{-16mm}

\begin{minipage}{3.5in}{
\psfig{file=fig5a.ps,width=3in,angle=0,clip=}
}\end{minipage}
\begin{minipage}{3.5in}{
\psfig{file=fig5b.ps,width=3in,angle=0,clip=}
}\end{minipage}
\vspace*{-2mm}

\end{center}

{\bf Figure 5:} The observed ratios of 6678\AA\ 3$^1$F--2$^1$D with
H$\beta$ plotted against (a) 2.166$\mu$m $7^{1,3}$G--$4^{3,1}$F
with Br$\gamma$ and (b) 2.113$\mu$m $4^{1,3}$S--$3^{3,1}$P
with Br$\gamma$.  Note how the data in (a) follow an essentially
linear relation, whereas in (b) there is significant scatter 
in the relation between the lines.  The latter is due to 
enhancement of the $4^{3}$S level of HeI either through collisions
or as a result of significant opacity in the $2^{3}$S--$n^{3}$P
series.    The solid lines in each figure represent the theoretical
bounds for the plotted ratios for the range of electron temperature and density
considered.

\begin{center}

\begin{minipage}{7.0in}{
\psfig{file=fig6.ps,width=6.5in,angle=0,clip=} }\end{minipage}

\end{center}

{\bf Figure 6:} The derived value of the full width at half maximum of the
integrated emission line using our simple expansion model as a function of
v$_{turb}$ and v$_{exp}$.  The solid curves represent (from bottom up) values
of v$_{turb}$ of 0, 5, 10, 15, 20, 25 and 30kms$^{-1}$.

\begin{center}
\begin{minipage}{3.5in}{
\psfig{file=fig7.ps,width=5in,angle=0,clip=}
}\end{minipage}
\vspace*{-2mm}
\end{center}

{\bf Figure 7:} The observed and predicted behaviour of the 5007\AA\ [OIII] to
H$\beta$ ratio as a function of effective temperature.  The solid lines are
generated using simple blackbodies for the model stellar atmospheres.  The
dashed lines are generated from Kurucz model atmospheres.  The crosses are the
data from Kaler \& Jacoby (1991), where we have used their tabulated Stoy
temperatures for T$_{eff}$.  The straight line is their best fit relation to
this data.  The two curved lines for each set of models are the extreme values
from our grid.  The upper line represents the prediction for a central star of
luminosity 7000\Lsolar, with electron density of 24000cm$^{-3}$, the lower a
star of luminosity 3000\Lsolar\ and electron density of 3000cm$^{-3}$.

\newpage

\begin{center}

\begin{minipage}{3.5in}{(a)}\end{minipage}\begin{minipage}{3.3in}{(b)}\end{minipage}
\vspace*{-16mm}

\begin{minipage}{3.5in}{
\psfig{file=fig8a.ps,width=3in,angle=0,clip=}
}\end{minipage}
\begin{minipage}{3.5in}{
\psfig{file=fig8b.ps,width=3in,angle=0,clip=}
}\end{minipage}
\vspace*{5mm}

\begin{minipage}{3.5in}{(c)}\end{minipage}\begin{minipage}{3.3in}{\ \phantom{(d)}}
\end{minipage}
\vspace*{-16mm}

\begin{minipage}{3.5in}{
\psfig{file=fig8c.ps,width=3in,angle=0,clip=}
}\end{minipage}\begin{minipage}{3.5in}
\phantom{\rule{3in}{0pt}}
\end{minipage}
\end{center}

{\bf Figure 8:} The predicted ratio of the 2.058$\mu$m HeI $2^{1}$S--$2^{1}$P
line with HI Br$\gamma$ as a function of stellar effective temperature.
In (a) we show the dependence on density, in (b) on turbulent velocity
and in (c) on central star luminosity.  In each case we held the other two
parameters constant at the values of $n_e=12000$cm$^{-3}$,
v$_{turb}=5$kms$^{-1}$ and L$_*=5000$\Lsolar.  
In (a) the curves correspond to $n_e=$3000, 6000, 12000, 24000 and
48000cm$^{-3}$ from lowest up.  In (b) the curves correspond to 
v$_{turb}=$0, 5, 10 and 15kms$^{-1}$ from highest down.  In (c) the
curves correspond to L$_*=$3000, 5000 and 7000\Lsolar\ from highest down.

\newpage
\begin{center}

\begin{minipage}{3.5in}{(a)}\end{minipage}\begin{minipage}{3.3in}{(b)}
\end{minipage}
\vspace*{-16mm}

\begin{minipage}{3.5in}{
\psfig{file=fig9a.ps,width=3in,angle=0,clip=}
}\end{minipage}
\begin{minipage}{3.5in}{
\psfig{file=fig9b.ps,width=3in,angle=0,clip=}
}\end{minipage}
\vspace*{-2mm}

\end{center}
{\bf Figure 9:} The observed and predicted behaviour of (a) the ratio of the
2.16475$\mu$m HeI $7^{1,3}$G--$4^{1,3}$F line with HI Br$\gamma$ and (b) the
6678\AA\ HeI 3$^1$D--2$^1$P with H$\beta$ as a function of effective
temperature.  The model plotted has L$_*=5000$\Lsolar and T$_e=10000$K.  There
is less than 5\% variation of the model ratios with either density or turbulent
velocity.

\begin{center}
\begin{minipage}{3.5in}{
\psfig{file=fig10.ps,width=5in,angle=0,clip=}
}\end{minipage}
\vspace*{-2mm}
\end{center}

{\bf Figure 10:} The observed and predicted behaviour of the ratio of the
2.058$\mu$m HeI $2^{1}$S--$2^{1}$P line with HI Br$\gamma$.  The models plotted
are for $n_e=48000$cm$^{-3}$ and v$_{turb}=0$kms$^{-1}$ (upper curve) and
$n_e=3000$cm$^{-3}$ and v$_{turb}=15$kms$^{-1}$ (lower curve) which are
essentially the extreme maxima and minima in our model grids.  In both cases
L$_*=5000$\Lsolar.  T$_{eff}$ is estimated from the observed 5007\AA\ [OIII] or
4686\AA\ HeII to H$\beta$ ratio as discussed in the text.

\begin{center}
\begin{minipage}{3.5in}{
\psfig{file=fig11.ps,width=5in,angle=0,clip=}
}\end{minipage}
\vspace*{-2mm}
\end{center}

{\bf Figure 11:} The observed and predicted behaviour of the ratio of the
2.058$\mu$m HeI $2^{1}$S--$2^{1}$P line with the 2.16475$\mu$m HeI
$7^{1,3}$G--$4^{1,3}$F line.  The models plotted are the same as in Figure 8.

\end{document}